\documentclass{aa}  \usepackage{txfonts}
\usepackage{natbib} \usepackage{graphicx} \usepackage{color}
\newcommand{\Lower}[1]{\smash{\lower 1.5ex \hbox{#1}}}
\newcommand\T{\rule{0pt}{2.6ex}}
\newcommand\B{\rule[-1.2ex]{0pt}{0pt}}
\begin{document}

\title{
Cosmic rays and the magnetic field in the nearby starburst galaxy
NGC\,253: II. The magnetic field structure }

\titlerunning{The magnetic field structure in NGC\,253}

\author{Volker Heesen\inst{1} \and Marita Krause\inst{2}
\and Rainer Beck\inst{2} \and Ralf-J\"urgen Dettmar\inst{3}}

\institute{Centre for Astrophysics Research, University of
  Hertfordshire, Hatfield AL10 9AB, UK \and Max-Planck-Institut f\"ur
  Radioastronomie, Auf dem H\"ugel 69, 53121 Bonn, Germany \and
  Astronomisches Institut der Ruhr-Universit\"at Bochum,
  Universit\"atsstr.\ 150, 44780 Bochum, Germany }

\date{Received 22 January 2009 / Accepted 31 July 2009}
\abstract {There are several edge-on galaxies with a known magnetic
 field structure in their halo. A vertical magnetic field
 significantly enhances the cosmic-ray transport from the disk into
 the halo. This could explain the existence of the observed radio
 halos. }
{We observed NGC\,253 that possesses one
 of the brightest radio halos discovered so far. Since this galaxy is
 not exactly edge-on ($i=78^\circ$) the disk magnetic field has to be
 modeled and subtracted from the observations in
 order to study the magnetic field in the halo.}
{We used radio continuum polarimetry with the VLA in D-configuration
 and the Effelsberg 100-m telescope. NGC\,253 has a very bright nuclear
 point-like source, so that we had to correct for instrumental
 polarization. We used appropriate Effelsberg beam
 patterns and developed a tailored polarization calibration to cope
 with the off-axis location of the nucleus in the VLA primary
 beams. Observations at $\lambda\lambda$ $6.2\,\rm cm$ and $3.6\,\rm
 cm$ were combined to calculate the RM distribution and to correct
 for Faraday rotation.}
{The large-scale magnetic field consists of a \emph{disk}
    $(r,\phi)$ and \emph{halo} $(r,z)$ component. The disk component
    can be described as an axisymmetric spiral field pointing inwards
    with a pitch angle of $25^\circ\pm5^\circ$ which is
    \emph{symmetric} with respect to the plane (even parity). This
  field dominates in the disk, so that the observed magnetic field
  orientation is disk parallel at small distances from the midplane.
  The halo field shows a prominent X-shape centered on the nucleus
  similar to that of other edge-on galaxies. We propose a model
    where the halo field lines are along a cone with an opening angle
    of $90\degr\pm30\degr$ and are pointing away from the disk in both
    the northern and southern halo (even parity). We can not exclude
    that the field points inwards in the northern halo (odd
    parity). The X-shaped halo field follows the lobes seen in
    H$\alpha$ and soft X-ray emission.}
{Dynamo action and a disk wind can explain the X-shaped halo field. The nuclear
 starburst-driven superwind may further amplify and align the halo
 field by compression of the lobes of the expanding superbubbles. The
 disk wind is a promising candidate for the origin of the gas in the
 halo and for the expulsion of small-scale helical fields as
   requested for efficient dynamo action.}
\maketitle
\setcounter{equation}{0}
\setcounter{figure}{0}
\setcounter{table}{0}
\setcounter{footnote}{0}
\setcounter{section}{0}
\setcounter{subsection}{0}
\section{Introduction}
\label{sec:mf_introduction}
There is increasing observational evidence for the existence of
gaseous halos around disk galaxies. They consist of different
interstellar medium (ISM) species, mainly the diffuse ionized gas,
dust, cosmic rays (CRs), and magnetic fields \citep[for a review, see e.g.][]{dettmar_92a}. The transport of gas from the disk into
the halo has been discussed in the past in terms of galactic chimneys
\citep{norman_89a} and superbubble blow outs \citep{maclow_99a}. These
models all include supernova explosions as the energy source that
drives the formation of the halo. The effects of star formation on the
ISM are dramatic: the heated gas in supernova remnants and accelerated
energetic particles are injected into the base of the halo. The
fundamental parameter, which determines the formation of a halo, is
therefore the energy input by supernova explosions. Thus, the halo
does not form above the entire disk, but only at radial distances
where the star formation takes place \citep{dahlem_95a}. Moreover, the
compactness of the star-forming regions determines the threshold
condition where the hot gas can break out against the pull of the
gravitational field \citep{dahlem_06a}. This corroborates the picture
of the hot gas injected by supernova explosions and stellar winds of
massive stars \citep[see e.g.][]{dettmar_06a}.
Direct evidence for supernova-heated galactic halos comes from imaging and
spectroscopy of diffuse soft X-ray emission. This has been impressively
demonstrated by XMM-Newton observations of NGC\,253 where the nuclear X-ray
plume can be explained only by star formation without the contribution of an
AGN \citep{pietsch_01a, bauer_08a}. Moreover, it clearly shows that the
galactic starburst must drive a thermal outflow, since there are strong
indications of collisionally excited oxygen and iron L line complexes in the
spectrum \citep{breitschwerdt_03a}.
In a pioneering work \citet{ipavich_75a} pointed out the importance of
CRs for the generation of a galactic wind. In the disk, the CR gas,
the magnetic field, and the hot gas contribute roughly the same amount of
pressure \citep{beck_96a}. In the halo, however, CRs and the magnetic
field dominate. The relativistic CR gas has a larger pressure
scaleheight than the hot gas,  because its adiabatic index is $4/3$
whereas it is $5/3$ for the hot gas; the magnetic field has an
even larger scaleheight \citep{beck_07a}.  Furthermore, the CR
nucleons, which contain the bulk of the energy, do not suffer from
strong radiative losses like the hot gas. The CR-driven galactic wind
is thus another scenario which needs consideration when discussing
gaseous halos.
The theoretical framework for the CR transport is given by the
combined diffusion-convection equation, which can be applied in the
local comoving coordinate system if the bulk speed of the background
medium is non-relativistic \citep{schlickeiser_02a}.
\citet{breitschwerdt_91a,breitschwerdt_93a} applied the transport
equation for different magnetic field configurations in the Milky
Way. They concluded that the CR transport can be split in an {\it
entirely diffusive} and an {\it entirely convective} regime,
depending only on the local magnetic field configuration. For the
lower Milky Way halo they assumed the magnetic field lines to be
turbulently excited by stochastic gas motions caused by expanding
and overlapping supernova remnants. Thus, there will be no preferred direction of
propagation of these magnetic fluctuations and hence no net
Alfv\'{e}nic drift. In the halo the magnetic field lines of the
superbubbles begin to overlap and form ``open'' field lines that
might be enhanced by magnetic reconnection \citep{parker_92a}.
Various polarization studies have shown that magnetic fields are a very
sensitive tracer of interaction in the ISM that is not visible at any other
wavelength. \citet{chyzy_04a} showed that the interacting pair of galaxies
NGC\,4038/39 (the Antennae) possesses a strong magnetic field of $20\,\mu \rm
G$, significantly stronger than non-interacting spirals. The polarized
emission in many of the Virgo cluster galaxies is shifted with respect to the
optical distribution \citep{wezgowiec_07a}. Simulations suggest that this
behavior can be explained by ram pressure stripping of galaxies moving through
the intracluster medium \citep{soida_06a,vollmer_08a}. But a close inspection
of the Virgo spiral NGC\,4254 showed that the observed magnetic field
structure requires additional MHD mechanisms other than ram pressure stripping
\citep{chyzy_07a, chyzy_08a}. The structure of the magnetic field is thus an
important tracer for MHD processes and the interaction between various ISM
components, which are expected to be present in galaxies with winds.
 From these considerations it is clear that there is a need for understanding
 the three-dimensional structure of the magnetic field in the halo of spiral
 galaxies. That restricts the observations to a few nearby edge-on galaxies
 that allow us to study the extra-planar magnetic field with high spatial
 resolution and sensitivity. Several edge-on galaxies have a known magnetic
 field structure \citep{golla_94a, dumke_98a, tuellmann_00a, krause_04a,
   dettmar_06a, krause_08a}. The nearby starburst galaxy NGC\,253 possesses
 one of the brightest radio halos discovered so far, but the inclination angle
 is only mildly edge-on ($i=78.5^\circ$).  We assumed a distance of $3.94\,\rm
 Mpc$ \citep{karachentsev_03a} where an angular resolution of $30\arcsec$
 corresponds to a spatial resolution of $600\,\rm pc$. Furthermore, we used
 $p.a.=52^\circ$ as the position angle of the major axis.
This paper is the successive paper of \citet{heesen_09a} (Paper~I
thereafter) where the CR distribution in NGC\,253 was found to be
consistent with a vertical CR transport from the disk into the
halo. In contradiction to this finding so far no vertical
magnetic field has been discovered in the halo of this galaxy. The most detailed
discussion of the magnetic field structure was
presented by \citet{beck_94a}, where a mainly disk-parallel magnetic
field was found in the disk and halo. This was explained by a strong
shearing of the magnetic field due to differential rotation. The study
presented in the present paper provides new sensitive observations of the
large-scale magnetic field in the halo of NGC\,253. We use the
polarimetry information of the observations presented in Paper~I at
different wavelengths which allow us to apply a correction for the
Faraday rotation and thus to determine the intrinsic magnetic field
orientation.
This paper is organized as follows: we start with the description of the
observations and explain especially the calibration of the polarization in
order to cope with instrumental polarization caused by the high dynamic range
(Sect.\,\ref{sec:mf_observations_and_data_reduction}). We present the
continuum maps of NGC\,253 along with the vectors of the intrinsic magnetic
field and briefly describe its morphology in
Sect.\,\ref{sec:mf_morphology}. Sect.\,\ref{sec:polarization} summarizes the
polarization properties. In Sect.\,\ref{sec:mf_structure} we investigate the
magnetic field structure and present a model for the large-scale magnetic
field. In Sect.\,\ref{sec:mf_discussion} we discuss the consequences of our
findings for the radio halo of NGC\,253 and the observed other phases of the
ISM. Finally we summarize our results and present the conclusions in
Sect.\,\ref{sec:mf_conclusions}.
\section{Observations and data reduction}
\label{sec:mf_observations_and_data_reduction}
\subsection{Effelsberg observations}
\label{subsec:mf_effelsberg}
In Paper~I we described our radio continuum observations of NGC\,253 at
$\lambda\lambda$ $6.2\,\rm cm$ and $3.6\,\rm cm$ with the 100-m Effelsberg
telescope.\footnote{The Effelsberg 100-m telescope is operated by the
  Max-Planck-Institut f\"ur Radioastronomie (MPIfR).} Here we only explain the
details important for the polarization measurements and for the general
calibration and data reduction we refer to Paper~I. The high dynamic range
$(\gtrsim 1000)$ due to the strong nuclear point-like source (hereafter for
simplicity called the nucleus) requires several additional steps in the data
reduction. In Paper~I we explained how we removed the sidelobes of the nucleus
via a H\"ogbom cleaning of the total power maps. But the high dynamic range
also influences the polarization via the leakage of total power emission to
the polarized intensity. The flux density of the so-called {\it instrumental
  polarization} is about $1.0\,\%$ for the Effelsberg telescope with respect
to the total power flux density \citep{heesen_08a}. Thus, it can be neglected
only for observations with a dynamic range significantly less than 100 which
is not fulfilled for our observations.
In order to apply a correction for the instrumental polarization we used beam
patterns for Stokes parameters $Q$ and $U$ obtained from deep observations of
the unpolarized point-like source 3C84. The cleaned total power emission was
convolved with the beam patterns and the computed instrumental contribution
was subtracted from the Stokes $Q$ and $U$ maps, respectively. Finally we
obtained the map of the polarized intensity from the corrected Stokes $Q$ and
$U$ maps using COMB (part of AIPS).\footnote{The Astronomical Image Processing
  System (AIPS) is distributed by the National Radio Astronomy Observatory
  (NRAO) as free software.} We applied a correction for the noise bias in
order to preserve the mean zero level of the polarized intensity. The rms
noise levels of the final maps are $200\,\mu\rm{Jy/beam}$ for $\lambda
6.2\,{\rm cm}$ at $144\arcsec$ resolution and $100\,{\rm \mu Jy/beam}$ for
$\lambda 3.6\,{\rm cm}$ at $84\arcsec$ resolution, respectively.\footnote{All
  angular resolutions in this paper are referred to as the Half Power Beam
  Width (HPBW).}
\begin{figure}[htbp]
\resizebox{\hsize}{!}
{\includegraphics{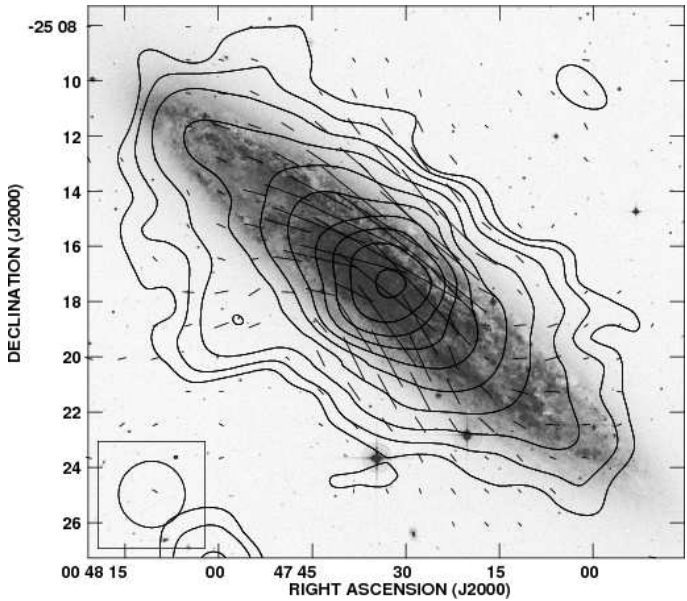}}
\caption{Total power radio continuum at $\lambda 6.2\,{\rm cm}$ from
  Effelsberg observations with $144\arcsec$ resolution. Contours are
  at 3, 6, 12, 24, 48, 96, 192, 384, 768, and 1536 $\times$ $1\,{\rm
    mJy/beam}$ (the rms noise level). The vectors show the orientation
  of the large-scale magnetic field (not corrected for Faraday
  rotation). A vector length of $1\arcsec$ is equivalent to
  $50\,{\mu\rm Jy/beam}$ polarized intensity. The circle in the
    lower left corner in this and the following figures indicates the
    size of the beam.}
\label{fig:n253cm6e_tpa_clean_dss_b144}
\end{figure}
\begin{figure}[t]
\resizebox{\hsize}{!}{
\includegraphics{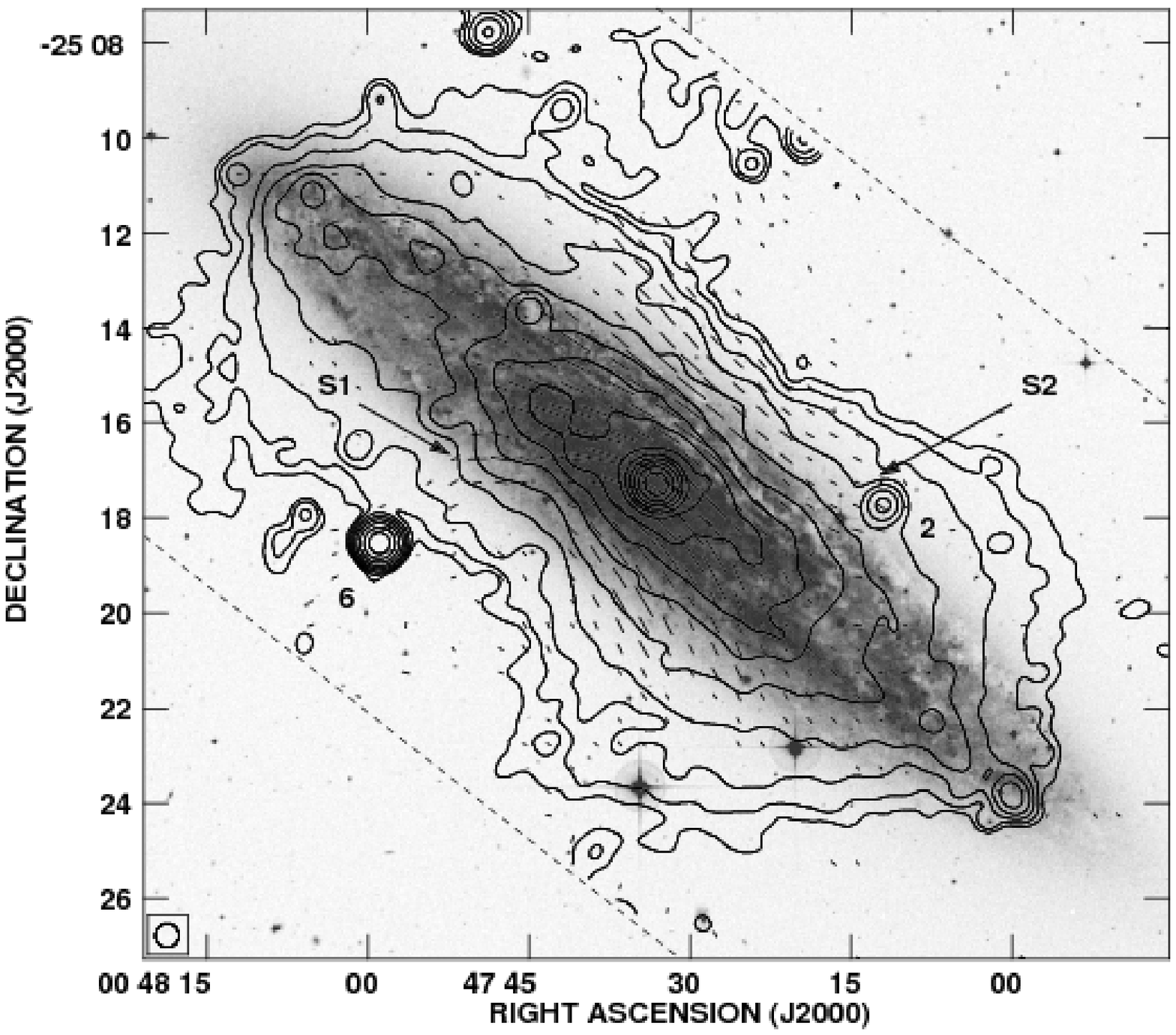}}
\caption{Total power radio continuum at $\lambda 6.2\,{\rm cm}$ from
 combined VLA + Effelsberg observations with $30\arcsec$
 resolution. Contours are at 3, 6, 12, 24, 48, 96, 192, 384, 768,
 1536, 3077, 6144, 12288, and 24576 $\times$ $30\,{\rm\mu Jy/beam}$
 (the rms noise level). The vectors show the orientation of the
 large-scale magnetic field (not corrected for Faraday rotation). A
 vector length of $1\arcsec$ is equivalent to $12.5\,{\mu\rm
   Jy/beam}$ polarized intensity. The spurs S1 and S2 and the
 point-like sources no.\,2 and no.\,6 are indicated.}
\label{fig:n253cm6ve_tpaf_dss_b30}
\end{figure}
\begin{figure}[thbp]
\resizebox{\hsize}{!}
{\includegraphics{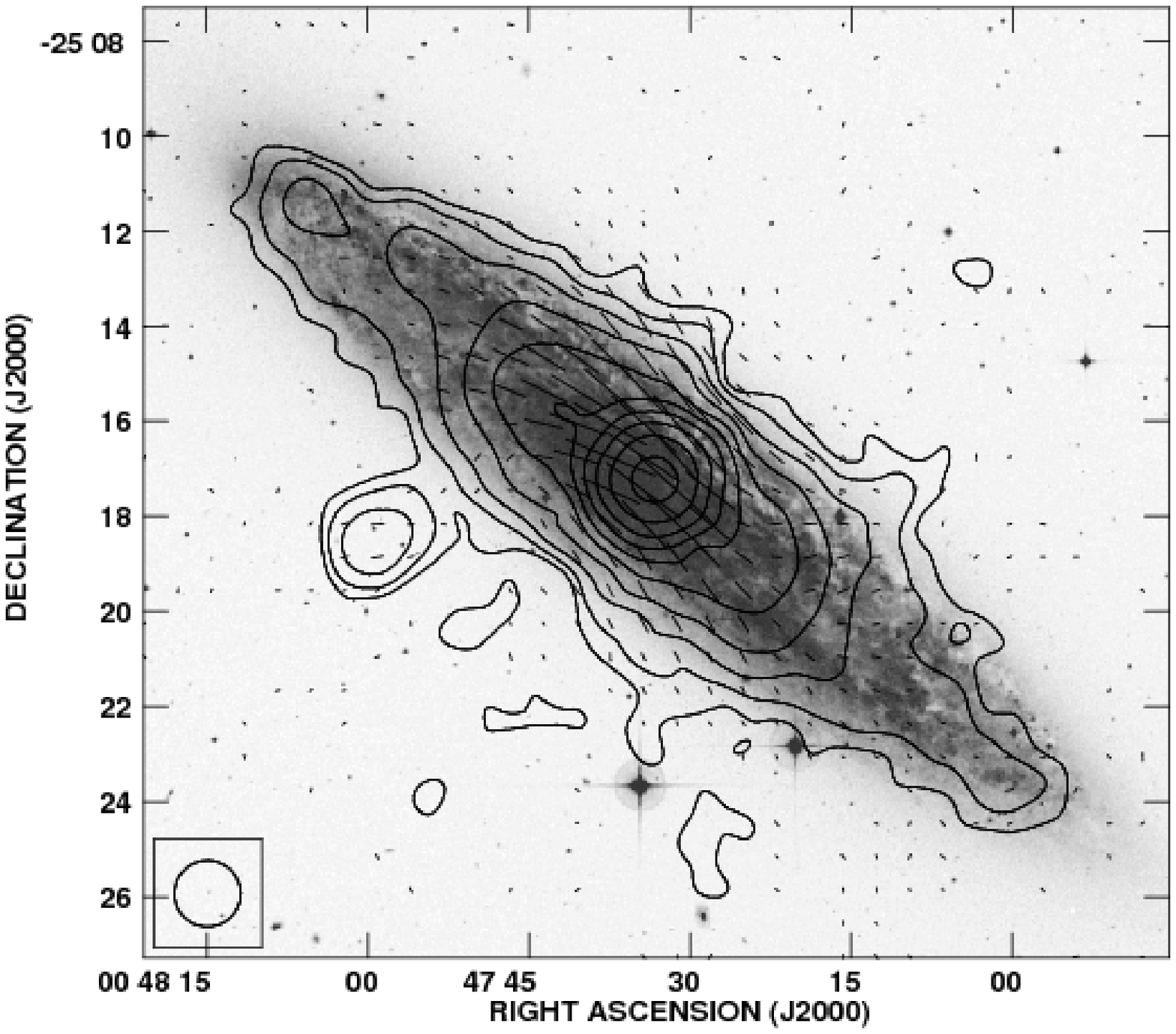}}
\caption{Total power radio continuum at $\lambda 3.6\,{\rm cm}$ from
 Effelsberg observations with $84\arcsec$ resolution. Contours are at
 3, 6, 12, 24, 48, 96, 192, 384, 768, and 1536 $\times$ $500\,{\rm\mu
   Jy/beam}$ (the rms noise level). The vectors show the
 orientation of the large-scale magnetic field (not corrected for Faraday
 rotation). A vector length of $1\arcsec$ is equivalent to
 $50\,{\mu\rm Jy/beam}$ polarized intensity.}
\label{fig:n253cm3e_tpaf_clean_dss_b84}
\end{figure}
\subsection{VLA observations}
\label{subsec:mf_vla}
For the polarization calibration of the VLA mosaic at $\lambda 6.2\,\rm cm$
(see Paper~I) we met the challenge that the nucleus is located at the edge of
the primary beam in some of our pointings.\footnote{The VLA (Very Large Array)
  is operated by the NRAO (National Radio Astronomy Observatory).} In the
center of the primary beam the instrumental polarization is less than 0.1\,\%
but at the edge of the primary beam it rises to more than 1\,\%. The
variability in time due to the apparent circular motion of the nucleus around
the center in the primary beam, caused by the alt-azimuthal mount of the VLA
antennas, complicates the situation even further. This prevents the removal of
sidelobes with the H\"ogbom clean algorithm as it works only for non-variable
sources. Thus, the standard polarization calibration technique applied in case
of NGC\,253 leads to a polarization map completely dominated by instrumental
polarization.
Hence, we developed a specially tailored technique for the pointings
with the nucleus located at the edge of the primary beam. The
polarization calibration PCAL (part of AIPS) was applied to the
nucleus itself instead of the secondary (phase) calibrator in order to
find a solution for the off-axis instrumental polarization. The
appropriate beam patterns in Stokes $Q$ and $U$ allowed us to subtract
the nucleus from the (u,v)-data of individual ``snapshots'', which
contained less than 10\,min of observing time. This effectively
removes the time-variable contribution of the instrumental
polarization caused by the nucleus. Thus we were able to use the full
set of (u,v)-data in order to get the best (u,v)-coverage which
resulted in the expected sensitivity of the extended emission. A more
detailed description of the polarization calibration can be found in
\citet{heesen_08a}.
The pointings that are not influenced by the nucleus were calibrated
in the standard way with the secondary calibrator using
PCAL. Inverting the (u,v)-data with natural weighting (i.e.\ Briggs
Robust=8) we produced maps of all pointings in Stokes $Q$ and $U$,
which were convolved with a Gaussian to $30\arcsec$ resolution. The
combination of the pointings was done with LTESS (part of AIPS) which
performs a linear superposition with a correction for the VLA primary
beam attenuation using information from each pointing out to the 7\,\%
level of the primary beam \citep{braun_88a}. The final maps are
essentially noise limited with a sensitivity of $30\,{\mu\rm Jy/beam}$
at $30\arcsec$ resolution in Stokes $Q$ and $U$. We used IMERG (part
of AIPS) for the combination of the VLA and Effelsberg Stokes $Q$ and
$U$ maps in order to fill in the missing zero-spacing flux; this was
done for the total power (Stokes $I$) map also. The map of the
polarized intensity was computed again using POLC with a correction
for the noise bias.
\begin{figure}
\resizebox{\hsize}{!}{
\includegraphics{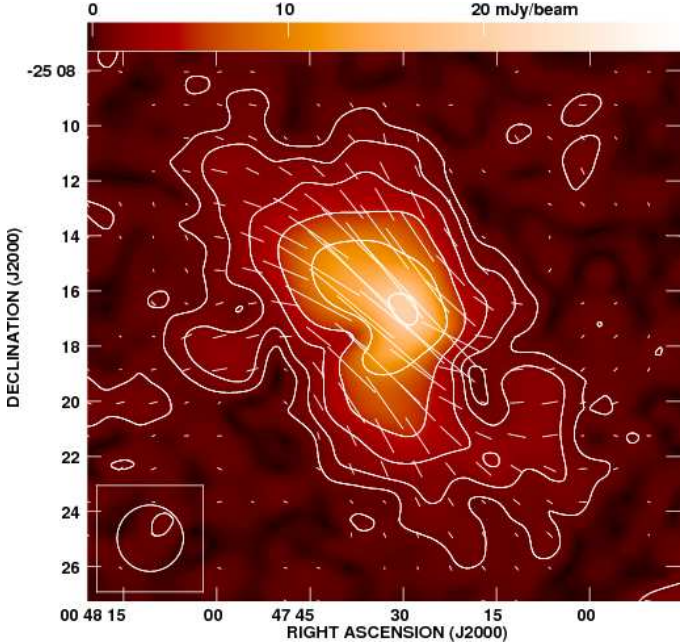}}
\caption{Polarized intensity at $\lambda 6.2\,{\rm cm}$ from
 Effelsberg observations with $144\arcsec$ resolution. Contours are
 at 3, 6, 12, 24, 48, and 96 times $200\,{\mu\rm Jy/beam}$ (the rms
 noise level). The vectors show the orientation of the Faraday
 corrected large-scale magnetic field. A vector length of $1\arcsec$
 is equivalent to $50\,{\mu\rm Jy/beam}$ polarized intensity.}
\label{fig:n253cm6e_pia_clean_dss_b144}
\end{figure}
\begin{figure}[htbp]
\resizebox{\hsize}{!}{
\includegraphics{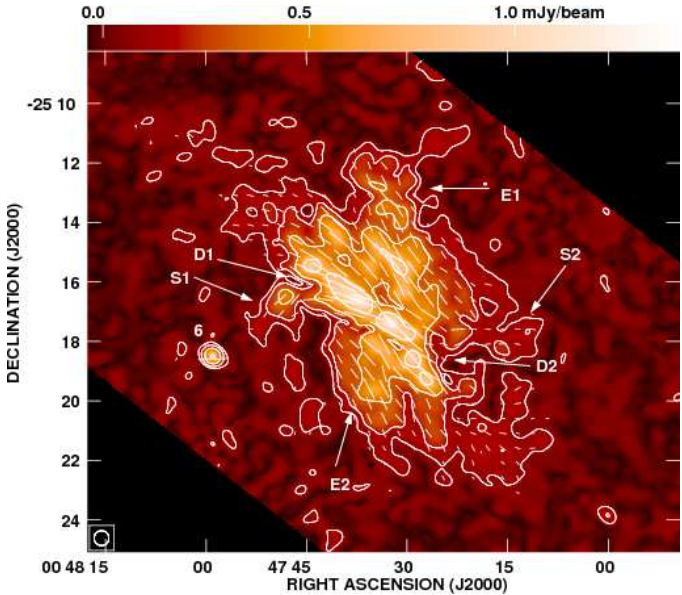}}
\caption{Polarized intensity at $\lambda 6.2\,{\rm cm}$ from combined VLA +
  Effelsberg observations with $30\arcsec$ resolution. Contours are at 3, 6,
  12, and 24 $\times$ $30\,{\mu\rm Jy/beam}$ (the rms noise level). The
  vectors show the Faraday corrected orientation of the large-scale magnetic
  field. A vector length of $1\arcsec$ is equivalent to $12.5\,{\mu\rm
    Jy/beam}$ polarized intensity. The extensions E1 and E2, the spurs S1 and
  S2, the relative minima D1 and D2, and the point-like source no.\,6 are
  indicated.}
\label{fig:n253cm6ve_piaf_dss_b30}
\end{figure}
\begin{figure}
\resizebox{\hsize}{!}{
\includegraphics{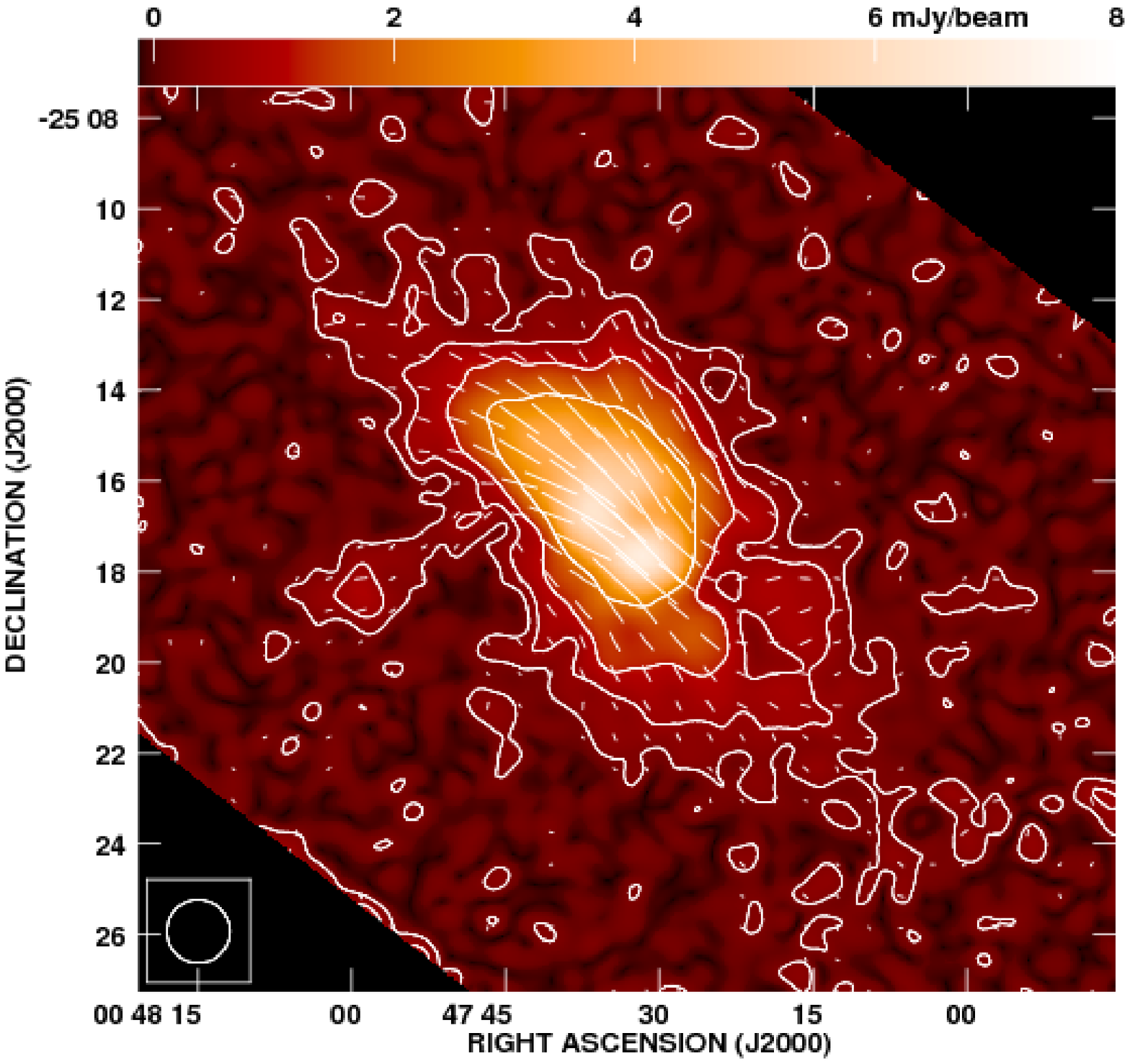}}
\caption{Polarized intensity at $\lambda 3.6\,{\rm cm}$ from
 Effelsberg observations with $84\arcsec$ resolution. Contours are at
 3, 6, 12, and 24 $\times$ $100\,{\mu\rm Jy/beam}$ (the rms noise
 level). The vectors show the Faraday corrected orientation
 of the large-scale magnetic field. A vector length of $1\arcsec$ is
 equivalent to $50\,{\mu\rm Jy/beam}$ polarized intensity.}
\label{fig:n253cm3e_pia_clean_dss_b84}
\end{figure}
\section{Morphology of the polarized emission}
\label{sec:mf_morphology}
%
\begin{table}[tbhp]
\caption{Maps of NGC\,253 presented in this paper of total power radio
 continuum (TP) and polarized intensity (PI).}
\begin{center}
\begin{tabular}{llrlll}
\hline\hline
$\lambda$ \T & Telescope & Resolution & TP rms & PI rms\\
$[\rm cm]$ \B & & & \multicolumn{2}{c}{$[\rm mJy/beam]$}\\\hline
$6.2$ \T & VLA + Effelsberg & $30\arcsec$ & $0.03$ & $0.03$\\
$6.2$ & Effelsberg & $144\arcsec$ & $1.00$ & $0.20$\\
$3.6$ \B & Effelsberg & $84\arcsec$ & $0.50$ & $0.10$\\\hline
\end{tabular}
\end{center}
\label{tab:mf_overview_observations}
\end{table}
In this section we discuss the morphology of the polarized emission and the
magnetic field orientation. In the figures, the first contours indicate
emission at three times the (apparent) rms noise level ($2^{m-1}\times 3\,
{\rm rms}$;\, $m=1,2,3,\ldots$). The polarization vectors (E-vectors) were
rotated by $90^\circ$ in order to display the orientation of the magnetic
field vectors (B-vectors). We have clipped the vectors below two times the rms
noise level in polarized intensity. The optical background image in
Figs.\,\ref{fig:n253cm6e_tpa_clean_dss_b144}-\ref{fig:n253cm3e_tpaf_clean_dss_b84}
is from the DSS (when the radio continuum map is smaller than the background
image, its boundary is indicated by dashed lines).\footnote{The compressed
  files of the ``Palomar Observatory -- Space Telescope Science Institute
  Digital Sky Survey'' of the northern sky, based on scans of the Second
  Palomar Sky Survey are \copyright 1993-1995 by the California Institute of
  Technology and are distributed herein by agreement.} A summary of the
polarization observations presented in this paper is given in
Table~\ref{tab:mf_overview_observations}.
In Fig.\,\ref{fig:n253cm6e_tpa_clean_dss_b144} we present the
distribution of the total power radio continuum emission together with
the B-vectors from the $\lambda 6.2\,{\rm cm}$ Effelsberg
observations. The low resolution of $144\arcsec$ shows the magnetic
field structure at the largest scales. The magnetic field is
disk parallel only in the disk plane. Further away from the disk, it
has a significant vertical component.
From the combined VLA + Effelsberg observations at $\lambda 6.2\,{\rm
 cm}$ we created the map shown in
Fig.\,\ref{fig:n253cm6ve_tpaf_dss_b30} with $30\arcsec$ resolution. At
the location S1 the magnetic field orientation is almost
perpendicular to the disk. This is an example of a ``radio spur''
where the large-scale magnetic field opens from a disk parallel to a
vertical field. At its end away from the disk the point-like source
no.\,6 is located; this polarized source likely is a background AGN
\citep{carilli_92a}. The second radio spur S2 has a magnetic field
orientation of $45^\circ$ with respect to the major axis; the
unpolarized point-like source no.\,2 is located at its end away from
the disk.
We show the $\lambda3.6\,{\rm cm}$ Effelsberg map in
Fig.\,\ref{fig:n253cm3e_tpaf_clean_dss_b84}. At a resolution of
$84\arcsec$ this map reveals no new details. But it is important
for the determination of the Faraday rotation as shown in
Sect.~\ref{subsec:mf_rm}.
The distribution of the polarized intensity from the $\lambda
6.2\,{\rm cm}$ Effelsberg observations is presented in
Fig.\,\ref{fig:n253cm6e_pia_clean_dss_b144}. It is very different from
the total power distribution that can be be described by a thick radio
disk with a vertical exponential
profile. Fig.\,\ref{fig:n253cm6e_pia_clean_dss_b144} shows a thick
disk formed by extensions E1, E2 apparent in the VLA + Effelsberg map
(Fig.\,\ref{fig:n253cm6ve_piaf_dss_b30}).
Figure~\ref{fig:n253cm6ve_piaf_dss_b30} shows the distribution of the
polarized intensity from the combined $\lambda 6.2\,{\rm cm}$ VLA +
Effelsberg observations. The bulk of the polarized emission arises in
the disk and extends into the halo. In contrast, in the outer parts of
the disk the polarized emission is weak at this resolution. The
polarized emission extends to large vertical heights above the inner
disk. The Effelsberg $\lambda 3.6\,{\rm cm}$ map of the polarized
intensity presented in Fig.\,\ref{fig:n253cm3e_pia_clean_dss_b84} is
similar to Fig.\,\ref{fig:n253cm6ve_piaf_dss_b30} where the emission
is slightly more concentrated to the disk than at $\lambda 6.2\,{\rm
 cm}$. This can be explained by the smaller scaleheight of the total
power emission at this shorter wavelength due to higher synchrotron
losses that influences also the polarized emission (see Paper~I).
The polarized intensity has a minimum between the radio spur S1 and
the extension E2. Another minimum of polarized intensity is located
near the radio spur S2. We will refer to these regions as
``depolarized regions''. The large-scale magnetic field is
disk parallel near the galactic midplane whereas at some locations we
find a significant vertical component. Hence, the large-scale magnetic
field may consist of two components, one disk parallel and one
vertical.
\begin{table}[t]
\caption[]{Integrated flux densities.}
\vfill
\begin{center}
\begin{tabular}{lrrrr}
\hline\hline $\lambda~{[\rm cm]}$ \T \B & Instrument & $S_{\rm e}~{\rm
[mJy]}$ & $S_{\rm p}~{[\rm mJy]}$ & $P~[\%]$\\\hline
6.2 \T & VLA + Eff. & $1440\pm70$ & $80\pm 12$ & $5.6\pm1.1$\\
6.2 & VLA & $1160\pm60$ & $64\pm 12$ & $5.5\pm1.3$\\
6.2 & Effelsberg & $1430\pm70$ & $87\pm 18$ & $6.1\pm1.6$\\
3.6 \B & Effelsberg & $680\pm40$ & $60\pm 10$ & $8.8\pm2.0$\\
\hline \end{tabular}
\end{center}
\label{tab:pi_flux}
\end{table}
%
%
\section{Polarization properties}
\label{sec:polarization}
\subsection{Degree of polarization}
\label{subsec:mf_total_polarized_flux}
Linearly polarized synchrotron emission is observed from CR electrons
spiraling in an ordered magnetic field. No linearly polarized signal is
detected from isotropic turbulent magnetic fields with randomly distributed
directions.\footnote{Note that the observed B-vectors of linearly polarized
  emission can trace either \emph{regular} magnetic fields (i.e.\ preserving
  their direction within the telescope beam) or \emph{anisotropic} fields
  (i.e.\ with multiple field reversals within the beam). Anisotropic fields
  can be generated from turbulent fields by shear or winds. To distinguish
  between these two components, additional Faraday rotation data is
  needed. The fields observed in polarization are called ``ordered''
  throughout this paper.} Hence, we can use the polarization degree as a
measure of the ratio between the ordered and the turbulent magnetic field.
The integrated polarized flux densities were obtained by integration in
ellipses which include the extra-planar emission as described in Paper~I (see
Table~\ref{tab:pi_flux}). The error of the flux densities was calculated as
the quadratic sum of a 5\,\% calibration error and the baselevel error. The
polarization degree $P$ is the ratio of the flux density of the extended total
power emission $S_{\rm e}$ (nucleus subtracted) to that of the polarized
emission $S_{\rm p}$. The polarization degree is between 6\,\% and 9\,\%,
which is similar to values found in other galaxies.
\begin{figure}[t]
\resizebox{\hsize}{!}{
\includegraphics{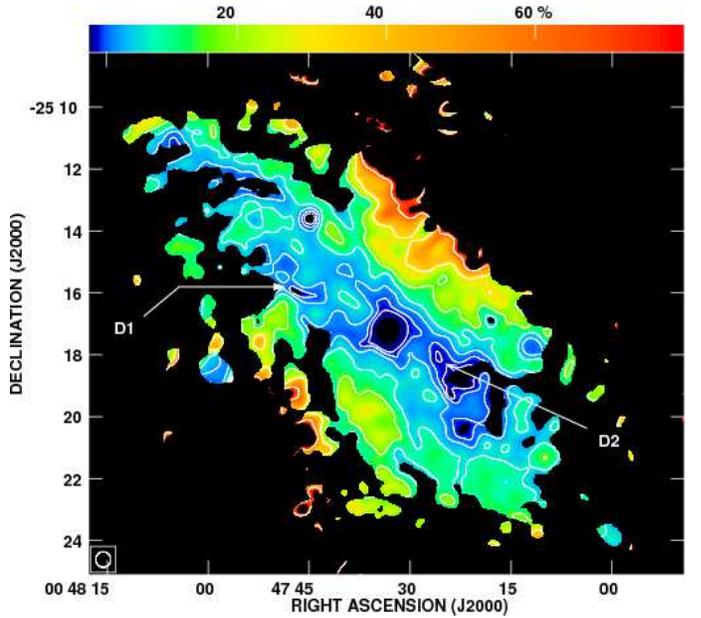}}
\vfill
\caption{Degree of polarization at $\lambda 6.2\,{\rm cm}$ from
 combined VLA + Effelsberg observations with $30\arcsec$
 resolution. Contours are at 1, 2, 4, 8, 16, 32, and 64 $\times$
 1\,\%. The depolarized regions D1 and D2 are indicated.}
\label{fig:n253cm6ve_pif_b30}
\end{figure}
%
%
%
In Fig.~\ref{fig:n253cm6ve_pif_b30} we present the distribution of the
degree of polarization at $\lambda 6.2\,{\rm cm}$. It has a minimum in
the midplane where it is smaller than $10\,\%$. Values larger than
$20\,\%$ are found at the radio spur S1 and and at the extensions E1
and E2. The depolarized regions D1 and D2 are prominently visible as
local minima.
%
\subsection{Distribution of the Rotation Measure}
\label{subsec:mf_rm}
\begin{figure}[t]
\resizebox{\hsize}{!}{
\includegraphics{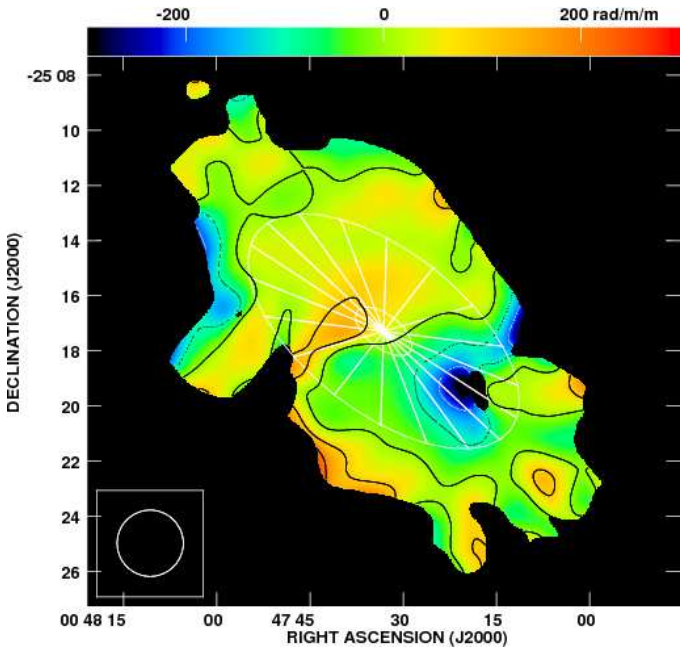}}
\vfill
\caption{RM distribution between $\lambda\lambda$ $6.2\,{\rm cm}$ and
  $3.6\,{\rm cm}$ from Effelsberg observations at $144\arcsec$
  resolution. Contours are at $-200$, $-100$, $0$, $100$, and $200$
  $\times$ $1\,{\rm rad\, m^{-2}}$. The vector maps at both
  wavelengths were clipped below $4\times$ the rms noise level in
  polarized intensity prior to the combination.  The sectors for
    the azimuthal RM variation are also shown (see
    Sect.\,\ref{subsec:the_disk_magnetic _field}).}
\label{fig:eff_rm}
\end{figure}

The polarization angle of the magnetic field vectors is changed by the Faraday
rotation which depends on the square of the wavelength. If the Faraday
depolarization is small, we can calculate the Rotation Measure (RM) from the
rotation $\Delta\chi$ of the polarization vector between two wavelengths:
\begin{equation}
\Delta\chi={\rm RM}\cdot(\lambda_1^{\phantom{1}2}-\lambda_2^{\phantom{2}2}).
\label{eq:mf_rm}
\end{equation}
In Fig.\,\ref{fig:eff_rm} we present the RM distribution between
$\lambda\lambda$ $6.2\,\rm cm$ and $3.6\,\rm cm$ at $144\arcsec$
resolution. We see a gradual change of positive RMs in the northeastern half
of the galaxy to negative RMs in the southwestern half. Strong negative RMs
$(<-200\,{\rm rad\, m^{-2}})$ in region D2 are an artifact due to strong
Faraday depolarization \citep{sokoloff_98a}.
According to the RM distribution the large-scale magnetic field has a
line-of-sight component that is pointing to the observer in the
northeastern half and pointing away from the observer in the
southwestern half. This agrees with a spiral magnetic field in the
disk as expected for a mean-field $\alpha$-$\Omega$ dynamo. For an
axisymmetric mode the magnetic field spirals in a uniform direction
\citep{baryshnikova_87a}. Rotation curves show that the northeastern
half is blueshifted whereas the southwestern half is redshifted
\citep{pence_81a, puche_91a}. Thus, the directions of the velocity and
magnetic field are opposite to each other. \citet{krause_89a} showed
that for such a case the spiral magnetic field is pointing inwards.
Using the RM distribution we corrected the magnetic field orientation for the
Faraday rotation. This was also done for the combined VLA + Effelsberg map at
$\lambda6.2\,\rm cm$ shown in Fig.\,\ref{fig:n253cm6ve_piaf_dss_b30}, although
we do not have a RM map with sufficient resolution. We corrected the
polarization angles for angular scales larger than $84\arcsec$ using a
combined VLA + Effelsberg RM map (not shown). The Effelsberg polarization maps
at $\lambda\lambda$ $6.2\,\rm cm$
(Fig.\,\ref{fig:n253cm6e_pia_clean_dss_b144}) and $3.6\,\rm cm$
(Fig.\,\ref{fig:n253cm3e_pia_clean_dss_b84}) could be corrected without loss
of resolution.
The effect of the Faraday correction is especially important at the
the radio spur S1: the magnetic field vectors are almost perpendicular
to the disk after applying the Faraday correction (a RM of $100\,\rm
rad\,m^{-2}$ corresponds to a rotation of $15^\circ$). In general the
Faraday correction made the magnetic field orientation more
disk parallel in locations near the disk plane (compare
Figs.\,\ref{fig:n253cm6ve_tpaf_dss_b30} and
\ref{fig:n253cm6ve_piaf_dss_b30}).
%
\section{The magnetic field structure}
\label{sec:mf_structure}
\begin{figure*}[htbp]
\resizebox{\hsize}{!}{
\Large\input{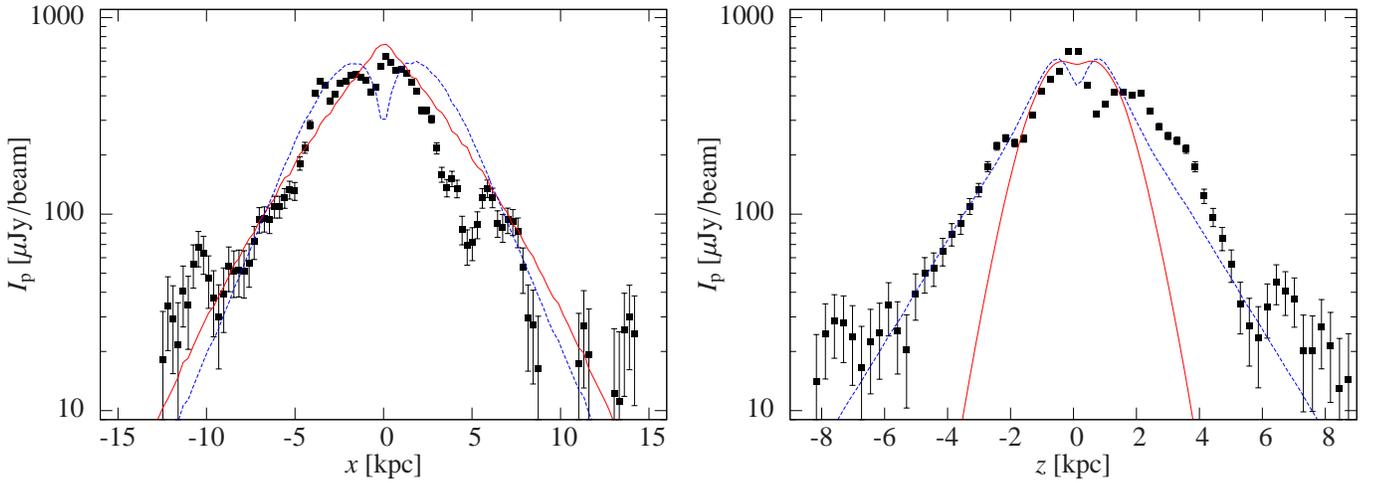}}
\vfill
\caption{Profiles of the polarized intensity along the major (left)
 and minor axis (right). The black symbols show the measured
 intensities from combined VLA + Effelsberg observations at $\lambda
 6.2\,\rm cm$. The solid (red) lines show the distribution derived from the
 disk magnetic field model. The dashed (blue) lines show the model of
 the combined disk and halo magnetic field (see
 Sect.\,\ref{subsec:the_disk_magnetic _field}).}
\label{fig:tor}
\end{figure*}
\subsection{Axisymmetric model for the disk magnetic field}
\label{subsec:mf_dynamo}
\begin{figure}[htbp]
\resizebox{\hsize}{!}{
\includegraphics{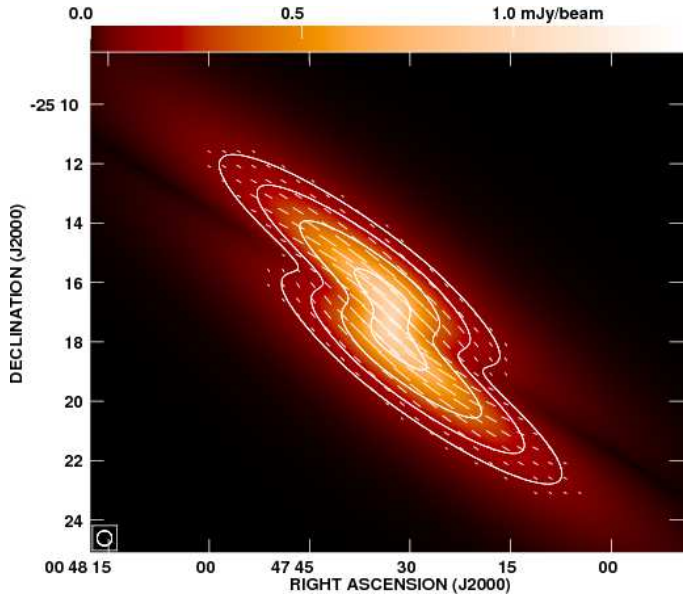}}
\vfill
\caption{Modeled polarized intensity at $\lambda 6.2\,\rm cm$ of the
  axisymmetric magnetic field (ASS) of the disk at $30\arcsec$
 resolution. Contours are at 3, 6, 12, and 24
 $\times$ $30\,{\mu\rm Jy/beam}$ of polarized intensity. The vectors
 indicate the magnetic field orientation where the length of the
 vectors of $1\arcsec$ is equivalent to $12.5\,{\mu\rm
   Jy/beam}$ polarized intensity.}
\label{fig:pa25_g700_pia}
\end{figure}

In Sect.\,\ref{sec:mf_morphology} we suggested that the observed
large-scale magnetic field is the superposition of a disk parallel
and a vertical component. We will refer to the disk-parallel
magnetic field with radial and azimuthal $(r,\phi)$ components as
the \emph{disk} magnetic field. The vertical magnetic field with
radial and vertical $(r,z)$ components we will refer to
as the \emph{halo} magnetic field.
As a model for the disk magnetic field we use a spiral magnetic field
corresponding to an axisymmetric spiral (ASS) mode of a galactic dynamo, which
is symmetric with respect to the plane (even parity). The inclination angle of
the disk was prescribed as $i=78.5^\circ$, which is that of the optical
disk. For a pitch angle of $25^\circ\pm5^\circ$ the polarized intensity
of our model resembles the observed distribution (the optical pitch angle is
$20^\circ$ \citep{pence_81a}). We fitted the observed polarized intensity
along the major axis with a Gaussian distribution consisting of an inner disk
($\rm FWHM =6.5\, kpc$) and an outer disk ($\rm FWHM =13\, kpc$). For the
vertical polarized emission profile we chose the exponential scaleheight of
the synchrotron emission from the thin disk of $0.4\,\rm kpc$ at $\lambda
6.2\,\rm cm$ (Paper~I). We assumed energy equipartition between CRs
and the magnetic field (where the CR number density is $n_{\rm CR}\propto B^2$),
so that the Gaussian FWHM and the exponential scaleheight of the magnetic
field are two and four times larger, respectively, than that of the
synchrotron emission. The magnetic field has a FWHM of $13\,\rm kpc$ and
$26\,\rm kpc$ in the inner and outer disk, respectively, and a vertical
scaleheight of $1.6\,\rm kpc$. The profiles of the polarized flux density of
the model and the observations are presented in Fig.\,\ref{fig:tor}.  A
summary of the model parameters can be found in
Table~\ref{tab:model_parameters}.
\begin{table}[t]
\caption[]{Parameters used for the ASS model of the disk magnetic field.}
\vfill
\begin{center}
\begin{tabular}{lcr}
\hline\hline Parameter \T \B & Value & Notes\\\hline
inclination angle $i$ \T & $78.5\degr$ & optical disk\\
pitch angle & $25\degr$ & fitted to PI\\
magnetic field FWHM & 13\,kpc / 26\,kpc & fitted to PI profile\\
magnetic field scaleheight & 1.6\,kpc & thin radio disk\\
ordered magnetic field & $3.0\,\mu\rm G$ & fitted to RM\\
electron FWHM & 13\,kpc & fitted to H$\alpha$ profile\\
electron scaleheight & 1.4\,kpc & fitted to H$\alpha$ profile\\
electron density $n_{\rm e}$ \B & $0.05\,\rm cm^{-3}$ & from H$\alpha$\\
\hline \end{tabular}
\end{center}
\label{tab:model_parameters}
\end{table}
We produced maps
of Stokes $Q$ and $U$ by integrating along the line-of-sight
\begin{equation}
Q \propto n_{\rm CR} B_\perp^2 \cdot \cos(2\psi) \quad U \propto n_{\rm CR}
B_\perp^2 \cdot \sin(2\psi),
\label{eq:qu_integration}
\end{equation}
where $B_\perp$ is the component of the magnetic field perpendicular to the
line-of-sight and $\psi$ is the orientation angle of the local polarization
vector. The maps were used to calculate the polarized emission and the
orientation of the magnetic field. At this point we do not take any Faraday
rotation into account but we will do so in
Sect.\,\ref{subsec:the_disk_magnetic _field}. The ASS model for the disk
magnetic field is shown in Fig.\,\ref{fig:pa25_g700_pia}.
The model of the disk magnetic field clearly shows why the polarized emission
is not symmetric with respect to the major and minor axis: the locations of
the maxima of the polarized intensity are shifted due to the pitch angle in
counterclockwise direction from the minor axis. Hence, the polarized intensity
distribution resembles an S-shape, as observed.  The depolarized regions D1
and D2 in the observed map of the polarized intensity
(Fig.\,\ref{fig:n253cm6ve_piaf_dss_b30}) can be explained by the model,
too. They are located where different components of the magnetic field occur
in one beam and thus cancel each other. We note that the magnetic field
orientation is mainly disk parallel as expected for a spiral magnetic
field. The good agreement between the model and the observed distribution in
the disk justifies the choice of the model.
The comparison of the profiles of the polarized emission between the
observations and the model in Fig.\,\ref{fig:tor} shows that along the
minor axis the halo shows up as additional emission. The halo
magnetic field is investigated in the next section.
\begin{figure}[bhtp]
\resizebox{\hsize}{!}{
\includegraphics{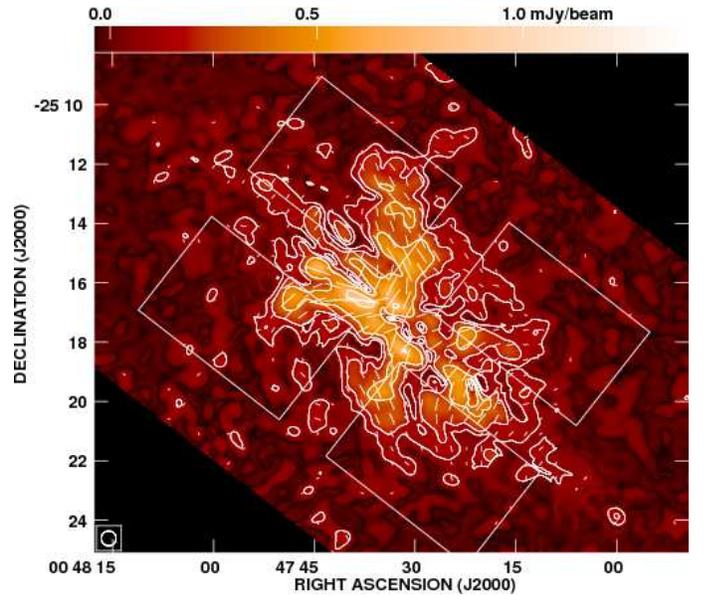}}
\vfill
\caption{Polarized intensity at $\lambda 6.2\,\rm cm$ of the halo
  magnetic field with $30\arcsec$ resolution after subtraction of the
  disk field model. Contours and vectors are identical to
  Fig.\,\ref{fig:pa25_g700_pia}. The boxes for the integration of the
  orientation angle $\widehat\psi$ are also shown.}
\label{fig:pa25_g700_sub_b180_180}
\end{figure}
\subsection{The halo magnetic field}
\label{subsec:mf_halo}
The good agreement in the disk region of the simple model shown in
Fig.\,\ref{fig:pa25_g700_pia} with the observations shows
that the disk field dominates in the disk. We note that the two
radio spurs are located near the two depolarized regions (D1 and
  D2 in Fig.\,\ref{fig:n253cm6ve_piaf_dss_b30}), where the
projected disk field has a minimum. Elsewhere the visibility of
the halo field is reduced by the dominating disk field.
The observed polarized emission is the superposition of the disk
and halo component with $Q=Q_{\rm d}+Q_{\rm h}$ and $U=U_{\rm
 d}+U_{\rm h}$. We subtracted the disk magnetic field model
from the maps of Stokes $Q$ and $U$ of the combined VLA +
Effelsberg observations at $\lambda 6.2\,\rm cm$. The polarized
intensity and the orientation of the halo magnetic field was
calculated from the maps of $Q_{\rm h}$ and $U_{\rm h}$ and is
presented in Fig.\,\ref{fig:pa25_g700_sub_b180_180}.
The halo magnetic field clearly resembles an X-shape centered on
the nucleus. Its orientation is mirror symmetric both to the major and
minor axis. X-shaped halo fields have been observed in several edge-on
galaxies with inclination angles $i\ge 85^\circ$, almost edge-on
\citep{krause_06a, krause_07a}. This is the first galaxy with an X-shaped field which is only mildly edge-on ($i=78.5^\circ$),
so that the emission from the disk and halo are superimposed.
In order to quantify the halo magnetic field we determined the
mean orientation angle $\widehat\psi$ in the four boxes shown in
Fig.\,\ref{fig:pa25_g700_sub_b180_180}. We integrated Stokes $Q$ and
$U$ and calculated the orientation angle of the magnetic field by
\begin{equation}
\widehat\psi=\left|\frac{1}{2}\arctan\left(\frac{U}{Q}\right)-p.a. +
90^\circ \right|,
\label{eq:mf_inclination}
\end{equation}
where $p.a.=52^\circ$ is the position angle of the major axis and
$+90^\circ$ was included in order to convert the electric polarization
angle into the magnetic field orientation. We
find a mean orientation angle of $\widehat\psi=46^\circ\pm15^\circ$
with respect to the major axis.
%
\subsection{Modeling the Rotation Measure distribution}
\label{subsec:the_disk_magnetic _field}
\begin{figure}[tpbh]
\resizebox{\hsize}{!}{\Large
 \input{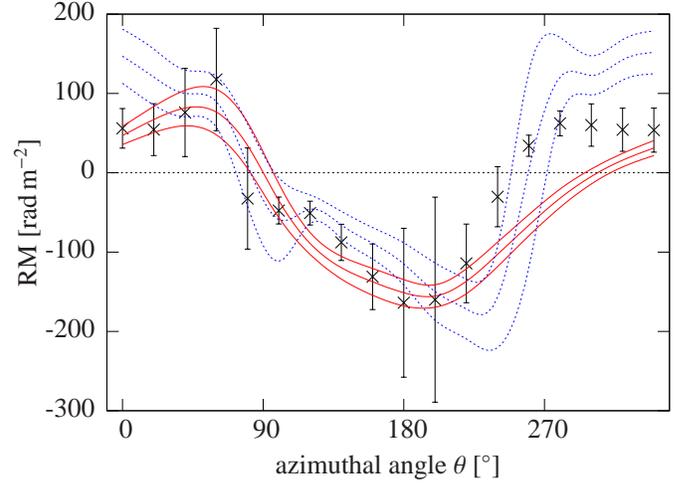}}
\vfill
\caption{RM variation as a function of the azimuthal angle
  $\theta$. The crosses give the observed RMs averaged in sectors as
  shown in Fig.\,\ref{fig:eff_rm} and their errors. The solid (red)
  lines represents the ASS model of the disk magnetic field together
  with its error interval. The dashed (blue) lines shows the model of
  the combined disk and halo magnetic field (and its error interval)
  which are both of even parity.}
 \label{fig:RM_sec_b144}
\end{figure}

\begin{figure}[bhpt]
\resizebox{\hsize}{!}{
  \includegraphics{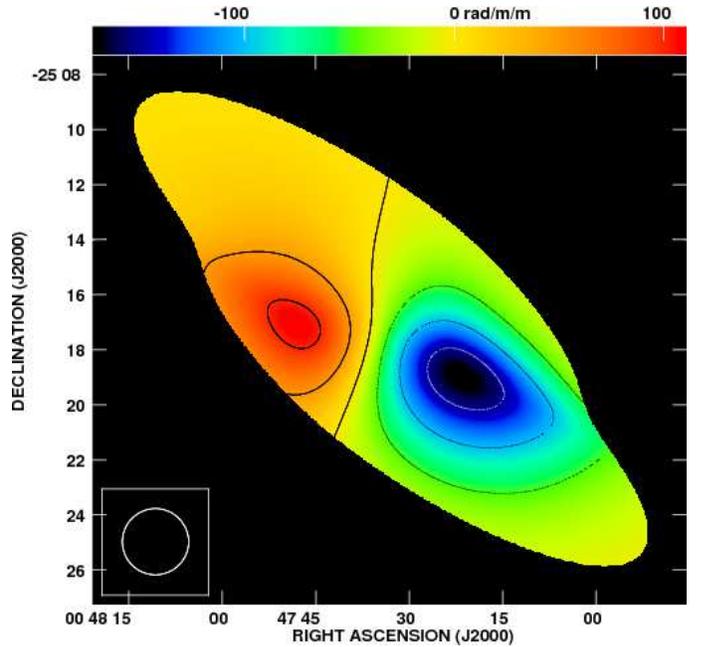}} \vfill 
\caption{RM distribution of the axisymmetric spiral (ASS) of the disk magnetic
  field at $144\arcsec$ resolution. Contours are at $-150$, $-100$, $-50$,
  $0$, $50$, and $100$ $\times$ $1\,\rm rad\,m^{-2}$. The map was clipped at
  $0.2\,\rm mJy/beam$ in polarized intensity.}
\label{fig:pa25_g700_rm_b180}
\end{figure}

An analysis of the RMs, averaged in sectors, as a function of the azimuthal
angle $\theta$ gives further information about the structure of the
large-scale magnetic field. The sector integration was applied between a
galactocentric radius of $1.4\,{\rm kpc}$ and $6.7\,{\rm kpc}$ with a spacing
of $20^\circ$ in the azimuthal angle (see Fig.\,\ref{fig:eff_rm}). An
effective inclination of $60\degr$ describes the observed distribution of the
polarized emission in the disk and halo at $\lambda 6.2\,\rm cm$ with
$144\arcsec$ resolution (Fig.\,\ref{fig:n253cm6e_pia_clean_dss_b144}). We
chose the RM distribution at $144\arcsec$ resolution, because the low
resolution favors the study of the large-scale RM structure.
The crosses in Fig.\,\ref{fig:RM_sec_b144} give the azimuthal RM variation
together with their errors. There is a broad maximum between $270\degr$ and
$60\degr$ and the minimum is near $\theta=180^\circ$. This confirms the result
previously obtained from the morphology of the RM distribution: the
northeastern half of the galaxy contains positive RMs
($-90^\circ\lesssim\theta\lesssim90^\circ$) whereas the southwestern half of
the galaxy contains negative RMs ($90^\circ\lesssim\theta\lesssim270^\circ$).
The disk magnetic field is dominating the polarized emission. Therefore we
will use the ASS model of the disk magnetic field in order to make a model for
the RM distribution. We integrated the polarization vector along the
line-of-sight, so that we take Faraday depth effects into account
\citep{sokoloff_98a}. From the Emission Measure $EM=10\,\rm cm^{-6}\,pc$
\citep{hoopes_96a} we derived an electron density of $n_{\rm e}=0.05\,\rm
cm^{-3}$, where we assumed a pathlength of $6.5\,\rm kpc$. From the H$\alpha$
distribution we derived a Gaussian distribution of the electron density along
the major axis with a FWHM of 13\,kpc and a vertical scaleheight of 1.4\, kpc.
The RM distribution of the ASS model is presented in
Fig.\,\ref{fig:pa25_g700_rm_b180}. It is notably asymmetric where the minimum
has a larger amplitude than the maximum.
Averaging Stokes $Q$ and $U$ of our ASS magnetic field model in sectors
provides the azimuthal RM variation shown as solid line in
Fig.\,\ref{fig:RM_sec_b144}. We found reasonable agreement with the observed
variation for an ordered magnetic field strength of $3.0\pm0.5\,\mu \rm
G$. From pulsar RMs we obtain a foreground of $RM_{\rm for}=10\pm5\,\rm
rad\,m^{-2}$ at the position of NGC\,253, which is near to the Galactic south
pole \citep{noutsos_08a}. Compared with the amplitude of the RM variation the
foreground RM is $\approx 10\,\%$, which we further neglect in our analysis.
The azimuthal RM variation of a spiral ASS field in an RM screen or a
homogeneous, mildly inclined emitting layer can be described by a simple
cosine, with a phase shift $\Delta\theta$ equal to the pitch angle of the
magnetic field spiral \citep{krause_89a}. However, the strongly inclined disk
of NGC\,253 requires to take into account the distribution of Faraday depths
along the line-of-sight.
Our model can explain the observations much better than the cosine variation
of an RM screen. For the screen the amplitudes of the maximum and the minimum
are equal, whereas our model can reproduce both the maximum with $RM\approx
120\,\rm rad\,m^{-2}$ and the minimum $RM\approx -180\,\rm
rad\,m^{-2}$. Furthermore, the RM screen predicts the maximum at
$\Delta\theta$ but our model shows that the maximum is very broad, in
agreement with the observations. This can be explained by the thickness of the
disk (not a thin emitting layer) and the inclination of the disk together with
the smearing by the observation beam. The difference in amplitude between the
maximum and minimum is a strong function of inclination.
The model of the disk magnetic field alone fits already to the observed RM
variation (reduced $\chi^2=6.5$). But so far we took not into account the halo
magnetic field that we also observe and which contributes significantly to the
polarized emission. With observations at only two wavelengths there is no
direct way to decompose the observed RM of the disk and halo magnetic field,
because they are overlapping along the line-of-sight. Still, we can produce a
model including the halo and disk magnetic fields.
For the halo magnetic field we propose a model where the field lines are along
a cone above and below the galactic plane, so that in projection to the plane
of the sky the field forms an X-shape. From the orientation angle of the halo
magnetic field as derived in Sect.\,\ref{subsec:mf_halo} we deduce an opening
angle of the cone of $180^\circ-2 \times \widehat\psi=90\degr\pm30\degr$. Note
that the contributions to the RM do not cancel if the line-of-sight components
of the field change sign along the line-of-sight, as is the case in our cone
model, because of different pathlengths towards the cone's far and near
sides. Hence, we see the RM mainly from the front half of the cone.
\begin{figure}[t]
\begin{center}
\resizebox{0.9\hsize}{!}{
\includegraphics{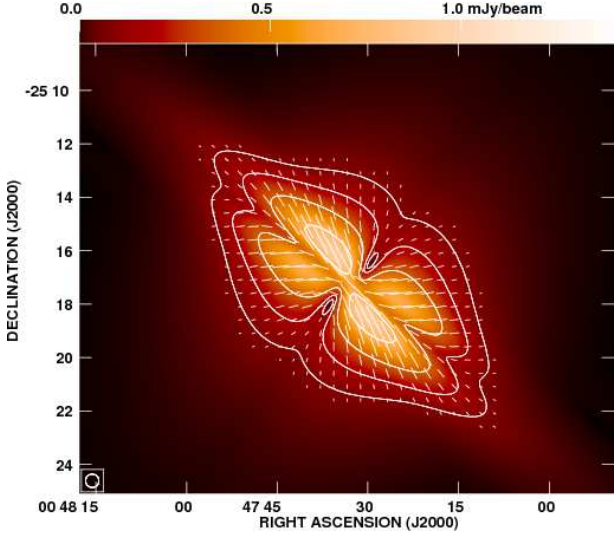}}
\end{center}
\vfill \caption{Modeled polarized intensity at $\lambda 6.2\,\rm cm$ of the
  combined even disk and even halo magnetic field at $30\arcsec$
  resolution. Contours are at 3, 6, 12, and 24 $\times$ $30\,{\mu\rm Jy/beam}$
  of polarized intensity. The vectors indicate the magnetic field orientation
  (corrected for Faraday rotation) where the length of the vectors of
  $1\arcsec$ is equivalent to $12.5\,{\mu\rm Jy/beam}$ polarized
  intensity. This Figure corresponds to the model shown in
  Fig.\,\ref{fig:even_even}.}
\label{fig:disk_and_halo}
\end{figure}
\begin{figure}[tbp]
\begin{center}
\resizebox{0.9\hsize}{!}{
\includegraphics{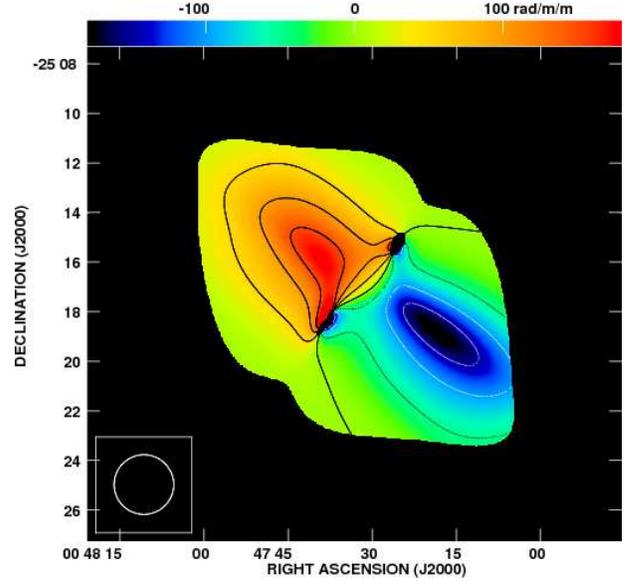}}
\end{center}
\vfill \caption{Model RM distribution for the combined even disk and
  even halo magnetic field at $144\arcsec$ resolution. Contours are at
  $-150$, $-100$, $-50$, $0$, $50$, $100$ and $150$ $\times$ $1\,\rm
  rad\,m^{-2}$. This Figure is identical with the Fig.\,\ref{fig:even_even}c.}
\label{fig:disk_and_halo_rm}
\end{figure}
\begin{figure}[htbp]
\begin{center}
\resizebox{0.7\hsize}{!}{
\includegraphics{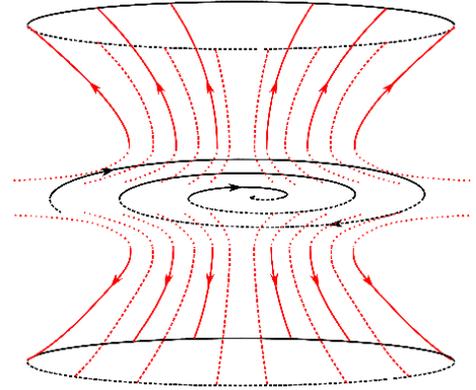}}
\end{center}
\vfill \caption{Sketch of the observable components of the large-scale
  magnetic field structure. In the
  disk, the ASS spiral magnetic field is even and pointing inwards. In
  the halo, the magnetic field is also even and pointing
  outwards. Dashed lines indicate components on the rear side. The
  dotted part of the halo magnetic field is not observed.}
\label{fig:disk}
\end{figure}
For the same reason we can only see the field direction in the southern halo,
because it lies in front of the bright disk and acts therefore as a Faraday
screen. Our models show that the magnetic field points away from the disk in
the southern halo. In case that the field points also away from the disk in
the northern halo, the halo field has \emph{even} parity. Otherwise, if the
field points towards the disk in the northern halo, the halo field has
\emph{odd} parity. Because the field direction in the northern halo has only a
small influence on the RM distribution the difference of the azimuthal RM
variation is also small. We tested all possible combinations of disk and halo
magnetic field configurations. The result is summarized in a catalog in
Appendix\,\ref{app:models}. The disk magnetic field is of even parity, where
the field is pointing in the same direction above and below the galactic
midplane. An odd disk magnetic field in combination with any halo magnetic
field leads to strong local gradients in the RM distribution which are not
observed. The even disk field together with an even halo field fits slightly
better to the observations with a reduced $\chi^2=4.9$ (see
Fig.\,\ref{fig:RM_sec_b144}) than with an odd halo field with $\chi^2=13$ (see
Fig.\,\ref{fig:even_odd}). The polarized intensity and the RM distribution of
our favored model for the combined even disk and even halo magnetic field is
shown in Figs.\,\ref{fig:disk_and_halo} and \ref{fig:disk_and_halo_rm}. A
sketch of the magnetic field lines is presented in
Fig.\,\ref{fig:disk}.\footnote{The magnetic field lines are schematically
  shown for the observed components only. We miss structures that are smaller
  than our resolution element and regions where not enough CR electrons are
  present. Therefore, all field lines are closed by components that we do not
  observe.}
We investigated also the possibility of an azimuthal halo magnetic field
component which is predicted by some halo models
\citep{dallavecchia_08a,wang_09a}. It was included into the conical magnetic
field model and the RM variation of the combined disk and halo model was
studied. It turned out that the azimuthal component -- if there is any -- must
point in the opposite direction to that of the disk magnetic field. As an
upper limit of the azimuthal component we estimate $60\,\%$ of the field
strength ($\chi^2=3.9$). We note that the shape and direction of the model
halo magnetic field is still consistent with the observations, i.e.\ is not
very sensitive to the relative strength of the azimuthal component.
\subsection{Magnetic field strengths}
The magnetic field strength can be calculated using the energy equipartition
assumption as described in \citet{beck_05a}. As in Paper~I we use a pathlength
through the total power emission of 6.5\,kpc and a CR proton to electron ratio
of $K=100$, and a nonthermal radio spectral index of $1.0$.  With a typical
nonthermal total power flux density of 8\,mJy/beam at $\lambda$6.2\,cm with
30\arcsec resolution we find a total magnetic field strength of $15\pm
1.5\,\mu\rm G$ in the disk. At a polarization degree of $5\,\%$ this
corresponds to $4.4\pm 0.5\,\mu\rm G$ for the ordered magnetic field in the
disk. The strength of the halo magnetic field was estimated by fitting
profiles of the polarized emission to the observations (see
Fig.\,\ref{fig:tor}). We found $4.4\pm0.5\,\mu \rm G$ for the halo magnetic
field.
We now compare the magnetic field strengths of the ordered field calculated
from the equipartition and from the RM analysis. The equipartition value can
be larger than that of the RM analysis, because the anisotropic turbulent
magnetic field emits polarized emission but does not or only weakly contribute
to the observed RM \citep{beck_05b}. From the azimuthal RM variation of our
modeled ASS disk field we found an ordered magnetic field component of
$3.0\pm0.5\,\rm \mu G$ for the disk. In the halo, the magnetic field strength
is similar with $3.0\pm0.5\,\rm \mu G$.  There may be some anisotropic field
component both in the disk and halo because the equipartition strength of the
ordered magnetic field is about $50\,\%$ larger than that of the RM
analysis. The anisotropic field component can contribute significantly to the
polarized emission in galaxies as shown for NGC\,1097 \citep{beck_05b} and M\,51 \citep{fletcher_09a}.
\section{Discussion}
\label{sec:mf_discussion}
\subsection{The superwind model}
\begin{figure}[tbhp]
\resizebox{\hsize}{!}{
\includegraphics{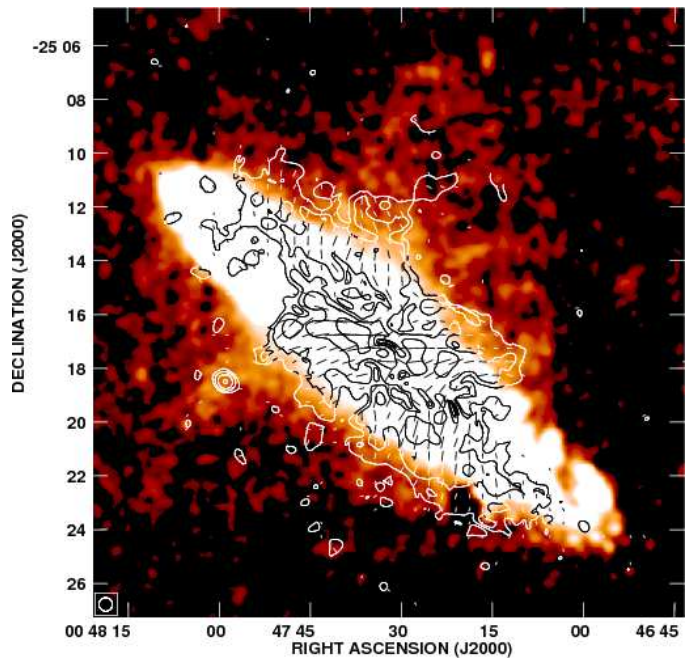}}
\vfill
\caption{Halo magnetic field overlaid onto diffuse H$\alpha$ emission as
  greyscale. Contours are at 3, 6, 12, and 24 $\times$ $30\,{\mu\rm Jy/beam}$
  (the rms noise level). A vector length of $1\arcsec$ is equivalent to
  $12.5\,{\mu\rm Jy/beam}$ polarized intensity. The H$\alpha$ map from
  \citet{hoopes_96a} was convolved with a Gaussian to $20\arcsec$ resolution
  in order to show the weak diffuse emission in the halo.}
\label{fig:pa25_g700_pia_Ha}
\end{figure}

\begin{figure}[tbhp]
\resizebox{\hsize}{!}{
\includegraphics{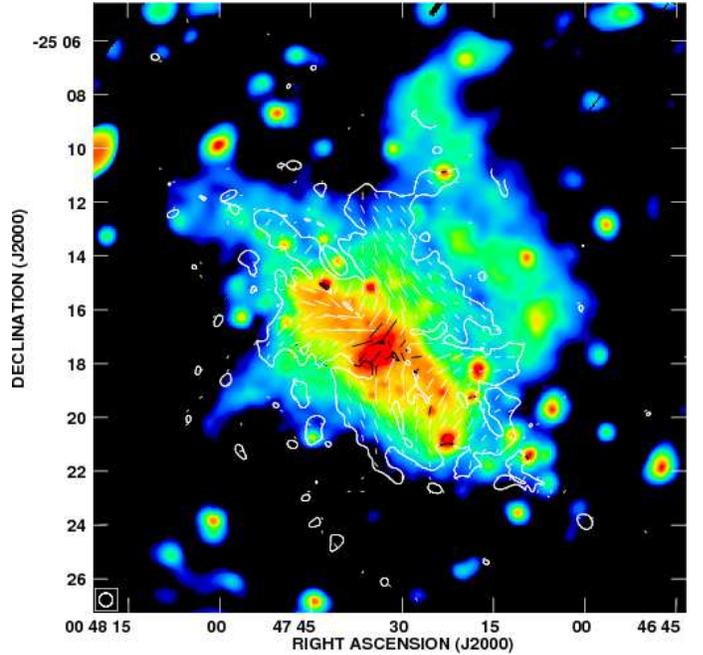}}
\vfill
\caption{Halo magnetic field overlaid onto diffuse X-ray emission. The contour
  is at 3 $\times$ $30\,{\mu\rm Jy/beam}$ (the rms noise
  level). A vector length of $1\arcsec$ is equivalent to $12.5\,{\mu\rm
    Jy/beam}$ polarized intensity. The X-ray map is from XMM-Newton
observations in the energy band $0.5-1.0\,\rm keV$ \citep{bauer_08a}.}
\label{fig:n253_pol_XMM2}
\end{figure}

NGC\,253 is a prototypical nuclear starburst galaxy, which
is the source of an outflow of hot X-ray emitting gas
\citep{strickland_00a, bauer_07a}. Spectroscopic measurements of
H$\alpha$-emitting gas in the southern nuclear outflow cone by
\citet{schulz_92a} give an outflow velocity of $390\,{\rm km\,
 s^{-1}}$. Apart from the nuclear outflow, huge lobes of
diffuse X-ray emission in the halo are extending up to $8\,\rm kpc$ away from
the disk. These lobes are thought to be the walls of two huge
bubbles containing a hot gas with a low density \citep{pietsch_00a,
 strickland_02a}. Sensitive XMM-Newton observations revealed that
indeed the entire bubbles are filled with X-ray emitting gas
\citep{bauer_08a}. Similar structures are found in other galaxies with
intense star formation, with M\,82 as the most prominent example,
and are now known as \emph{superwinds} \citep{heckman_00a}.
Figure~\ref{fig:pa25_g700_pia_Ha} shows the diffuse H$\alpha$ emission in
greyscale. There are several H$\alpha$ plumes extending from the disk into the
halo. We note that the large plume east of the nucleus corresponds to the
radio spur S1 (see Fig.\,\ref{fig:n253cm6ve_piaf_dss_b30}). The diffuse X-ray
emission in Fig.\,\ref{fig:n253_pol_XMM2} shows a similar structure with the
lobe at the extension E1 extending furthest into the halo. The abundance of
heated gas has an asymmetry with respect to the minor axis: the northeastern
half contains significantly more diffuse H$\alpha$ and soft X-ray emission
than the southwestern half which cannot be explained by a symmetric
superwind. This asymmetry is also observed in \ion{H}{I} emission
\citep{boomsma_05a} which surrounds the superbubbles. An explanation is
attempted in Sect.~\ref{subsec:gas}.
\subsection{The disk wind model}
\label{sec:disc_magnetic_field_structure}
The halo magnetic field allows CRs to stream along the field lines from the
disk into the halo. In Paper~I we determined the vertical CR bulk speed as
$\upsilon_{\rm CR}=300\pm30\,\rm km\,s^{-1}$ which is surprisingly constant
over the entire extent of the disk. This shows the existence of a disk wind in
NGC\,253. CRs cannot stream faster than with Alfv\'{e}n speed with respect to
the magnetic field, because they are scattered at self-excited Alfv\'{e}n
waves via the so-called streaming instability \citep{kulsrud_69a}. A typical
magnetic field strength of $B\approx 15\,\rm \mu G$ and a density of the warm
gas of $n\approx0.05\,\rm cm^{-3}$ (Sect.\,\ref{subsec:the_disk_magnetic
  _field}) leads to an Alfv\'{e}n speed of $\upsilon_{\rm
  A}=B/\sqrt{4\pi\rho}\approx\, 150\,\rm km\,s^{-1}$. The super-Alfv\'{e}nic
CR bulk speed requires that the CRs and the magnetic field are advectively
transported together in the disk wind. CRs stream with Alfv\'{e}n speed with
respect to the magnetic field that is frozen into the thermal gas of the
wind. The measured CR bulk speed is the superposition $\upsilon_{\rm
  CR}=\upsilon_{\rm W} + \upsilon_{\rm A}$ of the outflow speed of the thermal
gas $\upsilon_{\rm W}$ in the wind and the Alfv\'{e}n speed $\upsilon_{\rm A}$
\citep{breitschwerdt_02a}. Adopting this picture the wind speed at the
galactic midplane in NGC\,253 is $\upsilon_{\rm w}\approx 150\,\rm
km\,s^{-1}$.
The Alfv\'{e}n speed is difficult to estimate, so that the uncertainty in the
actual value of the wind speed is large as well.  Furthermore, we could
measure only the average wind speed between 1\,kpc and 5\,kpc above the
plane. Therefore, the bulk speed near the galactic midplane may be lower than
$150\,\rm km\,s^{-1}$.  Studies of CR transport in a galactic wind by
\citet{breitschwerdt_91a} and follow-up works assumed a wind speed of only a
few $10\,\rm km\,s^{-1}$ at the starting point in the galactic midplane which
accelerates to the escape velocity at larger distances to the disk
\citep{breitschwerdt_93a, zirakashvili_96a,breitschwerdt_02a}. On the other
hand, \citet{everett_08a} found a high wind speed of $170\,\rm km\,s^{-1}$ in
the plane of the Milky Way in order to explain the soft X-ray emission in the
halo. Our value measured in NGC\,253 does not exclude any of these models. It
will be left to future observations to provide an accurate estimate of the
wind speed at different heights above the galactic plane.
If the large-scale magnetic field is frozen into the thermal gas in the wind
the halo magnetic field has an azimuthal component and winds up in a
spiral. The field configuration is similar to that of the sun with the Parker
spiral magnetic field \citep{parker_58a,weber_67a}. The azimuthal component
increases with height, so that above a certain height the magnetic field is
almost purely azimuthal. Up to the Alfv\'{e}nic point
the stiff magnetic field lines corotate with the underlying disk. If the
azimuthal component of the halo magnetic field is small, the observable radio
halo is within the corotating regime and the Alfv\'{e}n radius is larger than
the vertical extent of the halo. As an azimuthal halo component cannot be
excluded (Sect.\,\ref{subsec:the_disk_magnetic _field}), the Alfv\'{e}nic
radius may be smaller than $2\,\rm kpc$, which is the scaleheight of the
polarized emission in the halo.
The existence of a disk wind has important consequences for the transport of
angular momentum. As we have shown in Paper~I the CR bulk speed is fairly
constant over the extent of the disk, i.e.\ it does not depend on the
galactocentric radius. If this is also true for the speed of the disk wind,
angular momentum is effectively transported to large galactocentric radii
\citep{zirakashvili_96a}.  Moreover, the angular momentum of the wind per unit
wind mass is proportional to the Alfv\'{e}n radius as shown by
\citet{zirakashvili_96a}. An Alfv\'{e}n radius of $\approx 2\,\rm kpc$ agrees
with their models for which they found a significant loss of angular momentum
over the lifetime of a galaxy.  We note that the disk wind can account for
larger angular momentum losses than the superwind in the center (at small
galactocentric radii).
\begin{figure}[t]
\resizebox{\hsize}{!}{
\includegraphics{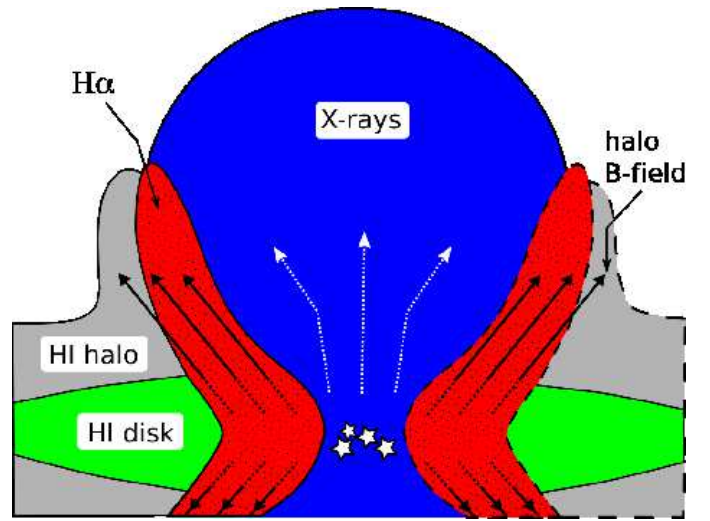}}
\vfill
\caption{Halo structure of NGC\,253. Reproduced from
 \citet{boomsma_05a} and extended. The superbubble, filled with soft
 X-rays emitting gas, expands into the surrounding medium (indicated
 by dotted lines with arrows). The halo magnetic field is aligned
 with the walls of the superbubble. Dashed lines denote components
 that are not (or only weakly) detected in the southwestern half of
 NGC\,253.}
\label{fig:superbubble_new}
\end{figure}
\subsection{The origin of X-shaped halo magnetic fields}
\label{subsec:x_shaped_halo_magnetic_fields}
The distribution of the halo magnetic field is X-shaped in both
orientation and intensity. As the intensity of the polarized emission
depends on the perpendicular component of the ordered field, a
possible explanation of the intensity distribution is limb brightening
in the conical halo magnetic field. We modeled this effect using a
halo magnetic field lying on a cone with an opening angle of $90\degr$ and
the inclination angle of NGC\,253. The model can reproduce indeed an
increased vertical extension along the limbs.
The halo magnetic field follows the lobes of the heated gas that are
regarded as walls of huge bubbles expanding into the surrounding
medium. The large-scale magnetic field may be compressed and aligned
by shock waves in the walls. These shock waves are also able to heat a
pre-existing cold halo gas, so that the gas becomes visible as
H$\alpha$ and soft X-ray emission. The cold halo gas seen in
\ion{H}{I} emission surrounds the superbubbles and shows the same
asymmetric distribution as expected if this gas is the source for the
heated gas in the halo. A cartoon of the halo structure including the
large-scale magnetic field is shown in
Fig.\,\ref{fig:superbubble_new}.
Both the disk wind model and the superwind model, in conjunction with
a large-scale dynamo action, can explain the X-shaped halo magnetic
field. We should note, however, that such magnetic field structures
are observed in several edge-on galaxies. These galaxies show a very
different level of star formation \citep{krause_08a}. Most of them do
not possess a starburst in the center and hence they have no
superwind. While the classical $\alpha$-$\Omega$ dynamo can not
explain such an X-shaped halo field, model calculations of a galactic
disk within a spherical halo including a galactic wind showed a
similar magnetic field configuration
\citep{brandenburg_93a}. Hydrodynamical simulations show that the wind
in spiral galaxies has a significant radial component due to the
radial gradient of the gravitational potential. The wind flow reveals
an X-shape when the galaxy is seen edge-on
\citep{dallavecchia_08a}. New MHD simulations of disk galaxies
including a galactic wind are in progress and may explain the halo
X-shaped field \citep{gressel_08b, hanasz_09a, hanasz_09b}. The first
global galactic-scale MHD simulations of a CR-driven dynamo
give very promising results showing directly that magnetic flux is
transported from the disk into the halo \citep{hanasz_09c}. Therefore
we suggest that in NGC\,253 the X-shaped halo field is connected
rather to the disk wind than to the superwind. The superwind gas flow
may be collimated by the halo magnetic field.
The transport of CRs and magnetic fields has important
consequences for the possibility of a working dynamo in NGC\,253.
The $\alpha$-$\Omega$ dynamo relies on the differential rotation of
the galactic disk ($\Omega$-effect) and the cyclonic motions of the
ionized gas ($\alpha$-effect). The latter one can be generated by
subsequent supernova explosions \citep{ferriere_92a}. The
amplification of magnetic fields by an $\alpha$-$\Omega$ dynamo
requires expulsion of small-scale helical fields \citep[see
 e.g.\ ][]{brandenburg_05a}. \citet{sur_07a} showed that galactic
winds can transport the helicity flux by advection.  A galactic wind
may be thus vital for a working dynamo.
\subsection{The origin of the gas in the halo}
\label{subsec:gas}
The cold gas in the halo of NGC\,253 is necessary to explain the halo
structure with the superwind model. Its origin, however, is yet
unknown. \citet{boomsma_05a} discussed a gas infall in NGC\,253 by a
minor merger to explain the asymmetric \ion{H}{I} halo distribution
with respect to the minor axis. But they noted that a minor merger
event is unlikely because the distribution of \ion{H}{I} is very
symmetric with respect to the major axis. Could the disk wind be the
origin of the cold gas in the halo?  The disk wind is asymmetric with
respect to the minor axis. The convective northeastern half indicates
a strong disk wind which transports gas from the disk into the halo
while in the diffusive southwestern half the wind is weaker (Paper~I).
We can only speculate about the reason why the two halo parts are so
different. Star formation in the disk most likely plays an important
role. The northeastern spiral arm contains significantly more
\ion{H}{II} regions than the southwestern one as is visible from
Fig.\,4b in \citet{hoopes_96a}. Moreover, the amount of total radio
continuum emission indicates a higher star-formation rate in this
part of the disk. Since the energy input of star formation is
essential for the injection of thermal and CR gas, the disk wind can
be more easily driven in the northeastern half. Thus, the disk wind
is a good candidate for the origin of the cold gas in the halo.
%
%
%
\section{Summary and conclusions}
\label{sec:mf_conclusions}
The three-dimensional magnetic field structure can be
investigated using sensitive radio continuum polarimetry. Our main
results are:
\begin{enumerate}
\item A disk-parallel magnetic field exists along the midplane of
 the disk. The magnetic field lines are slowly opening further away
 from the midplane. The vertical magnetic field component is most
 prominent at the edge of the inner disk where we find two ``radio
 spurs'', one previously known east of the nucleus and one newly
 discovered west of the nucleus. The magnetic field configuration can
 be described as an X-shaped pattern, as in other edge-on galaxies.
\item The large-scale magnetic field can be decomposed into a disk $(r,\phi)$
  and halo $(r,z)$ component.
\item The disk magnetic field of NGC\,253 can be described by an axisymmetric
  spiral (ASS) magnetic field with a constant pitch angle of
  $25^\circ\pm5^\circ$ which is symmetric with respect to the plane. This
  model shows a high resemblance to the observed magnetic field in the disk.
\item The distribution of the polarized intensity and the
orientation of the halo magnetic field shows a distinct X-shaped
pattern centered on the nucleus. Our model of the disk magnetic
field was subtracted to construct a map of the halo magnetic field.
We propose a model where the halo field lines are along a cone with
an opening angle of $90\degr \pm 30\degr$ and are pointing away from
the disk (even parity). An odd halo magnetic field is also possible,
because we can not reliably determine the magnetic field direction in
the northern halo.
\item The distribution of the halo magnetic field coincides in shape
 with the extra-planar heated gas traced by H$\alpha$ and soft X-ray
 emission. Possible explanations are limb brightening and compression
 of the halo magnetic field in the walls of the expanding
 superbubbles.
\item A disk wind plus dynamo action is a promising scenario for the origin of
  the gas in the halo and for the expulsion of small-scale helical fields as
  requested for efficient dynamo action. The disk wind can also account for
  large angular momentum and magnetic flux losses over galactic time scales.
\end{enumerate}
Future investigations should include polarimetry at low frequencies,
to take advantage of studying the magnetic structure far away from the
disk illuminated by an aged population of CR electrons. This should
become feasible with the upcoming class of telescopes operating at low
frequencies like LOFAR and the low-frequency SKA array. RM Synthesis
\citep{brentjens_05a, heald_09a} applied to multi-channel polarization
observations is required to separate the RM contributions from the
disk and the halo of strongly inclined galaxies.
\begin{acknowledgements}
  VH acknowledges the funding by the Graduiertenkolleg GRK\,787 and the
  Sonderforschungsbereich SFB\,591 during the course of his PhD.  The GRK\,787
  ``Galaxy groups as laboratories for baryonic and dark matter'' and the
  SFB\,591 ``Universal properties of non-equilibrium plasmas'' are funded by
  the Deutsche Forschungsgemeinschaft (DFG). RJD is supported by DFG in the
  framework of the research unit FOR\,1048.\\ We thank Dieter Breitschwerdt
  and Andrew Fletcher for many fruitful discussions. Moreover, we would like
  to thank Charles Hoopes for kindly providing us his H$\alpha$ map of
  NGC\,253. We thank Michael Bauer for providing us the XMM-Newton map. We are
  grateful to Elly Berkhuijsen and Matthias Ehle for carefully reading the
  manuscript and suggesting many improvements to the paper.
\end{acknowledgements}

\bibliography{bib2}

This paper has been typeset from a \TeX/\LaTeX file prepared by the author. 

\appendix

\section{Magnetic field models}
\label{app:models}
\begin{figure*}[tbhp]
  \begin{minipage}[b]{0.25\textwidth}
    \resizebox{\hsize}{!}{
      \includegraphics{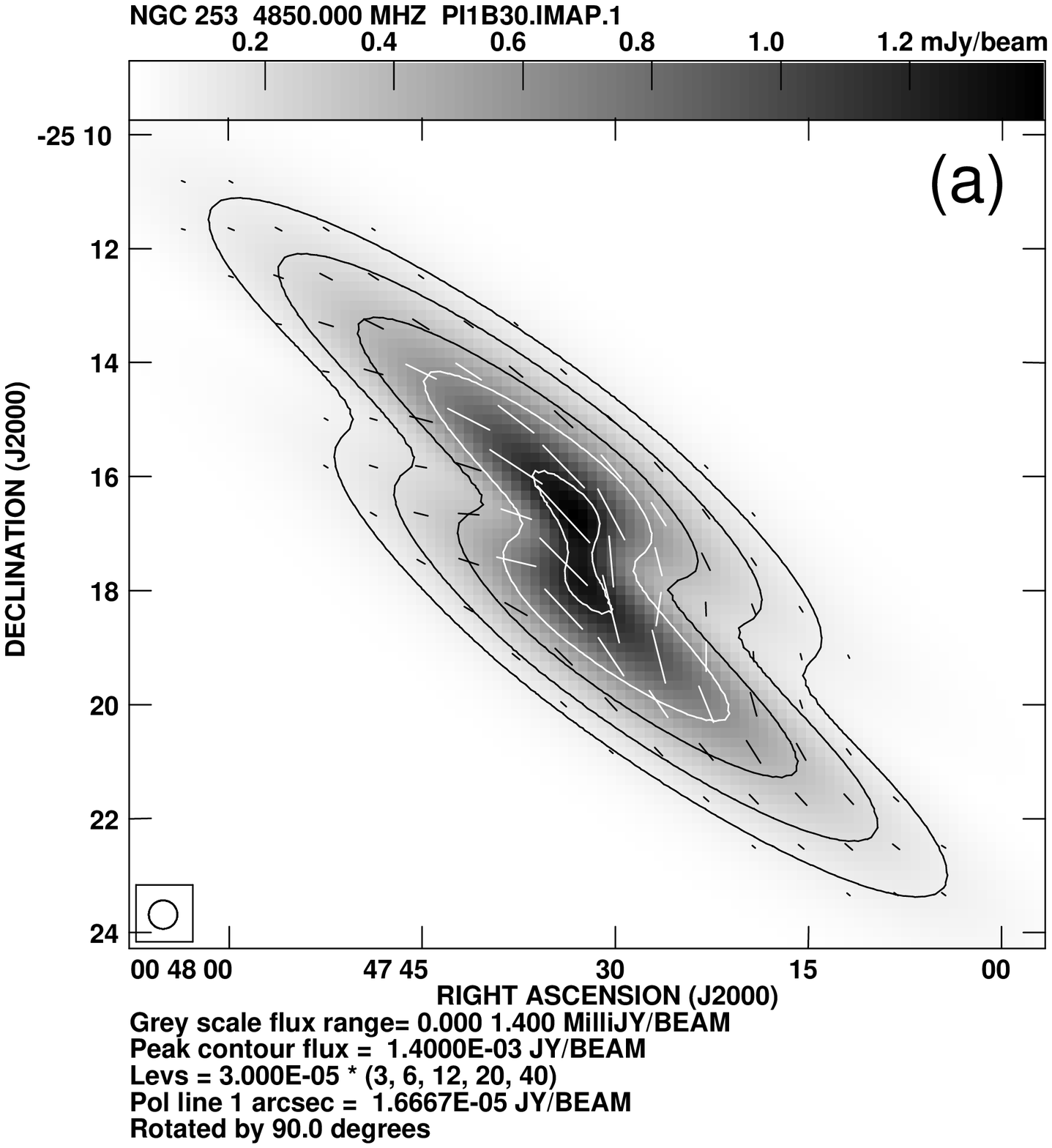}}
  \end{minipage}
  \begin{minipage}[b]{0.25\textwidth}
    \resizebox{\hsize}{!}{\includegraphics{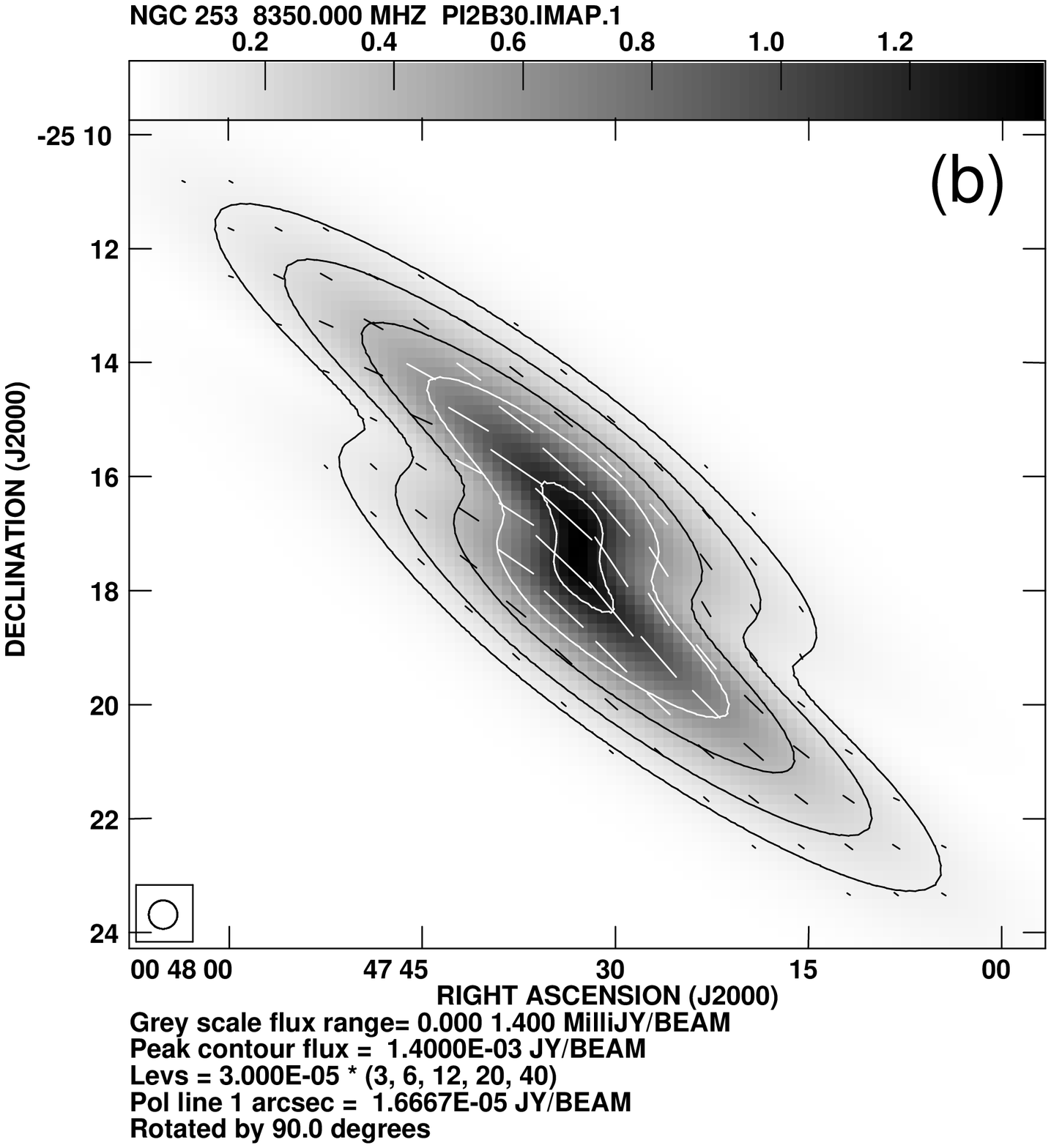}}
  \end{minipage}
  \begin{minipage}[b]{0.25\textwidth}
    \resizebox{\hsize}{!}{\includegraphics{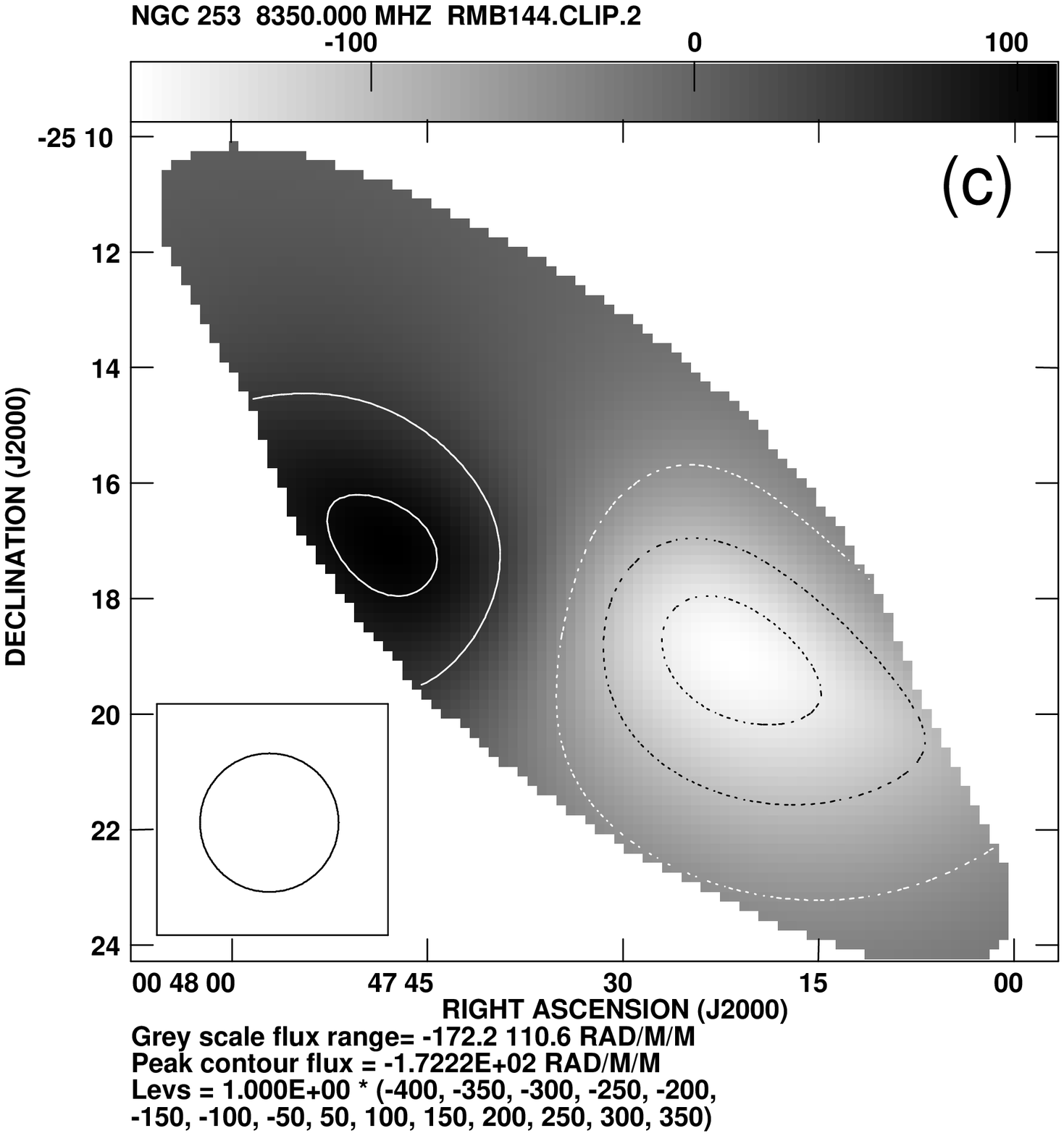}}
  \end{minipage}
  \begin{minipage}[b]{0.24\textwidth}
    \centering\resizebox{0.8\hsize}{!}{\includegraphics{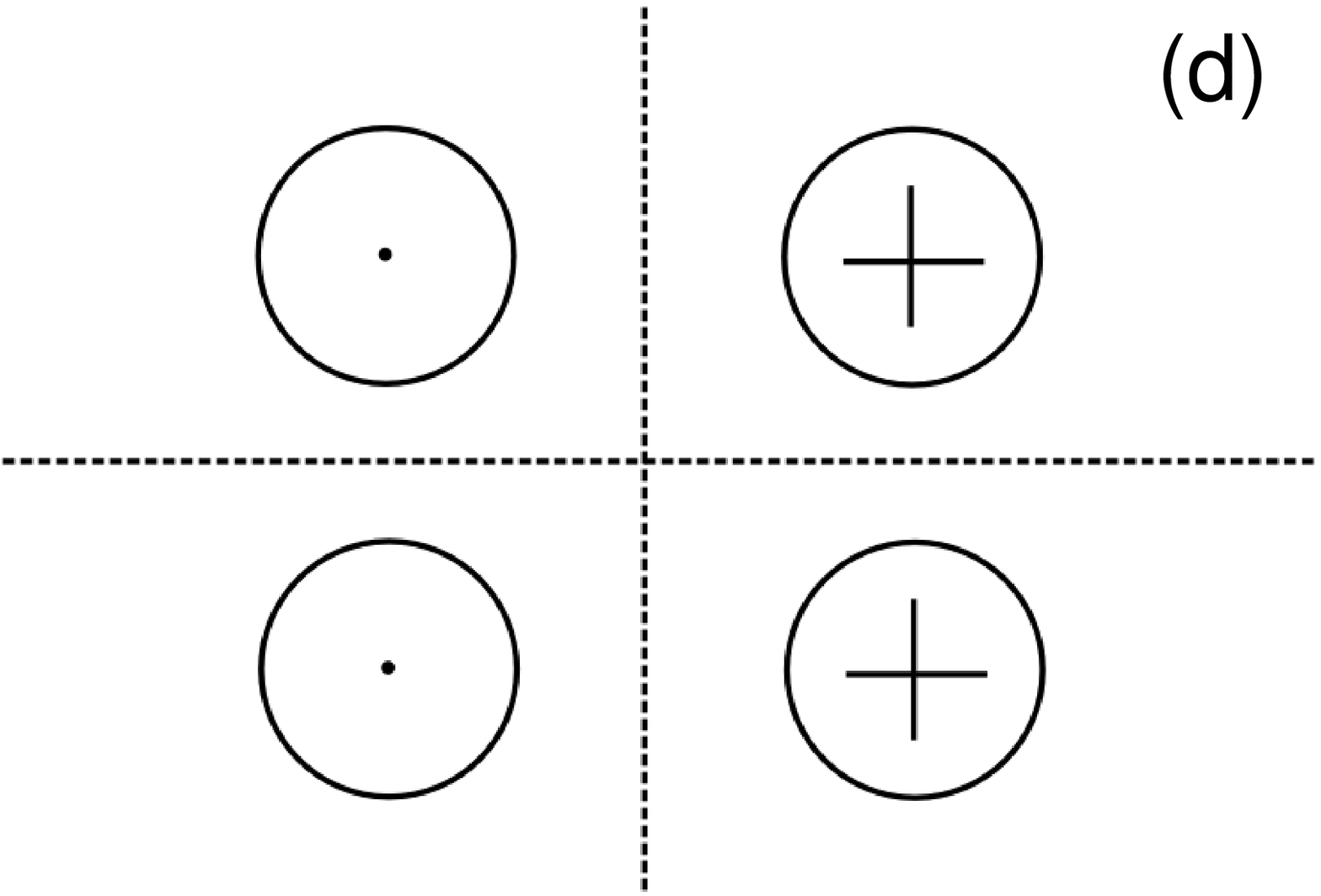}}
    \vspace{1.8cm}
  \end{minipage}
  \caption{Even disk magnetic field. For
    Figs.\,\ref{fig:disk_even}-\ref{fig:odd_odd_n}: a) polarized flux density
    and magnetic field orientation at $\lambda 6.2\,\rm cm$ with $30\arcsec$
    resolution. b) likewise but for $\lambda 3.6\,\rm cm$. c) RM distribution
    between $\lambda\lambda$ $6.2\,\rm cm$ and $3.6\,\rm cm$ with $144\arcsec$
    resolution. d) sketch of the magnetic field direction.}
\label{fig:disk_even}
\end{figure*}

\begin{figure*}[tbhp]
  \begin{minipage}[b]{0.25\textwidth}
    \resizebox{\hsize}{!}{\includegraphics{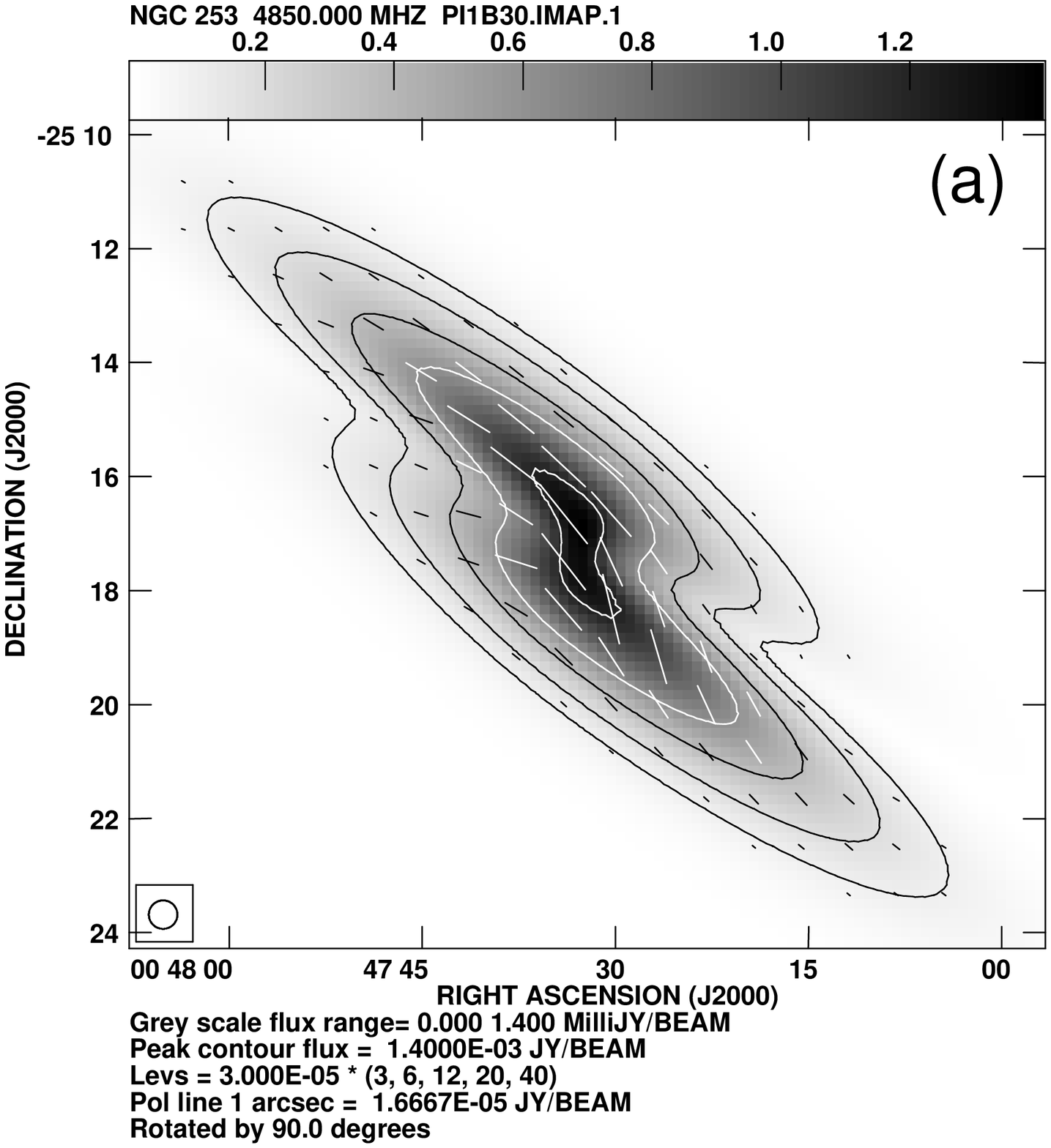}}
  \end{minipage}
  \begin{minipage}[b]{0.25\textwidth}
    \resizebox{\hsize}{!}{\includegraphics{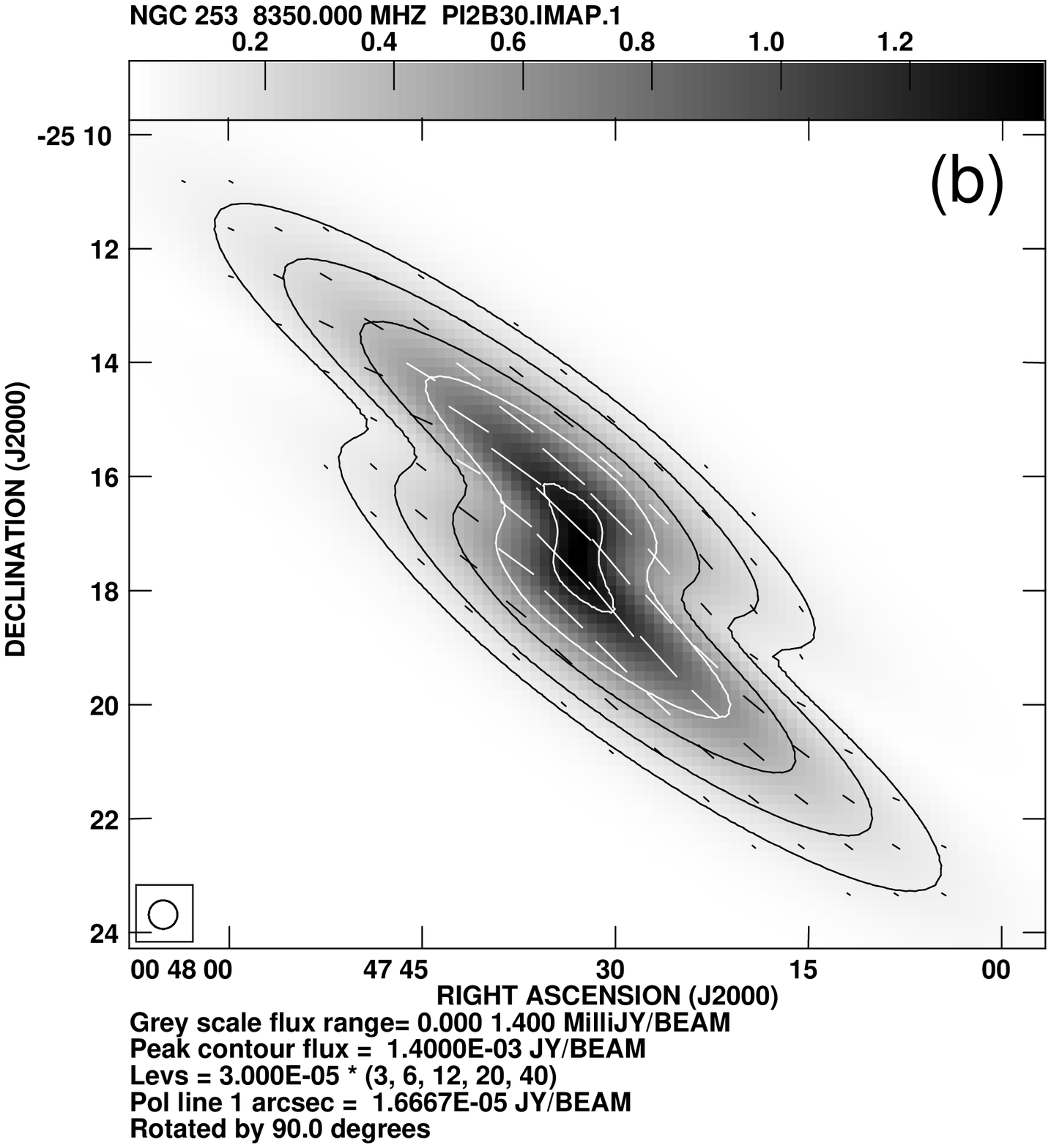}}
  \end{minipage}
  \begin{minipage}[b]{0.25\textwidth}
    \resizebox{\hsize}{!}{\includegraphics{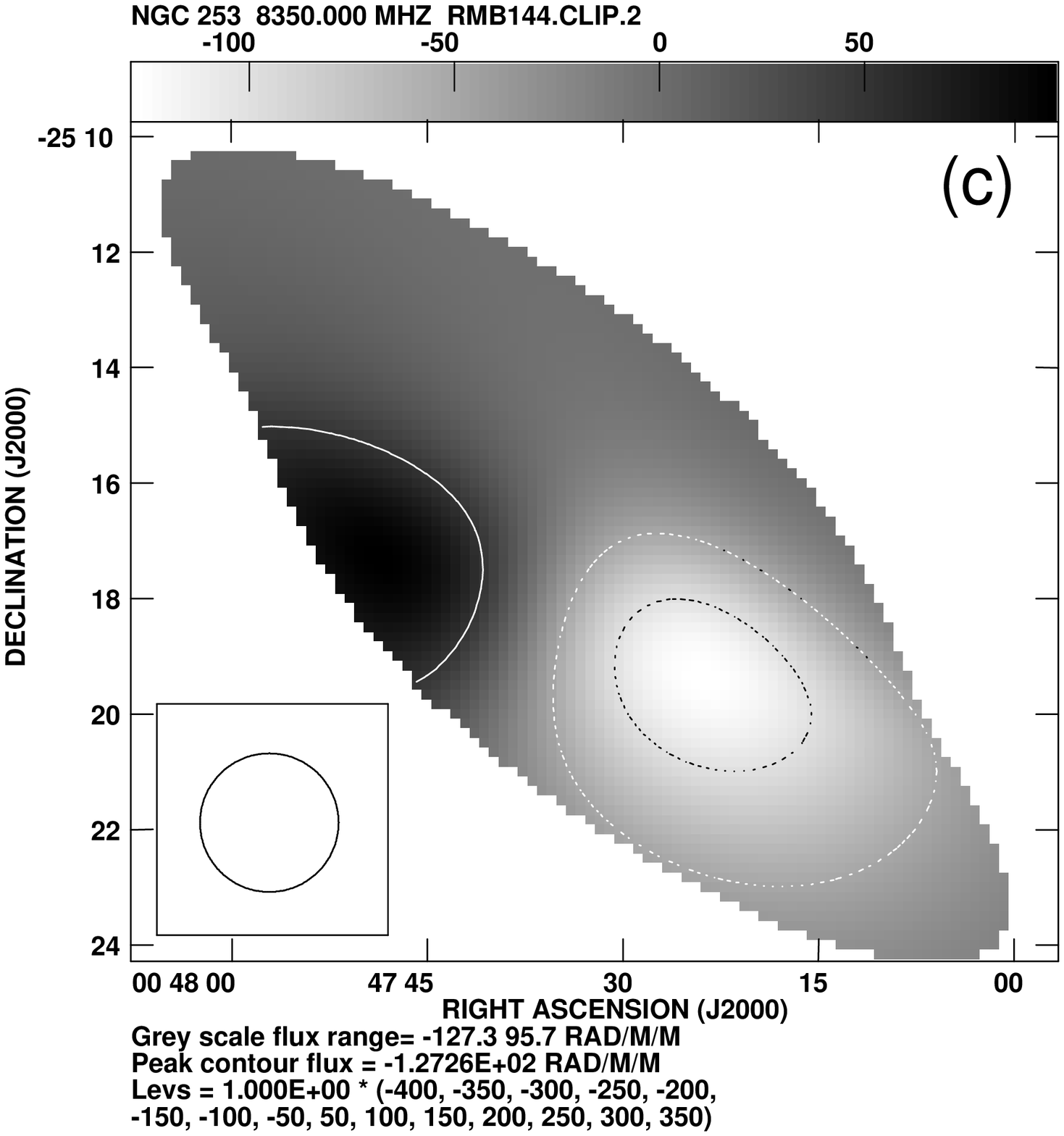}}
  \end{minipage}
  \begin{minipage}[b]{0.24\textwidth}
    \centering\resizebox{0.8\hsize}{!}{\includegraphics{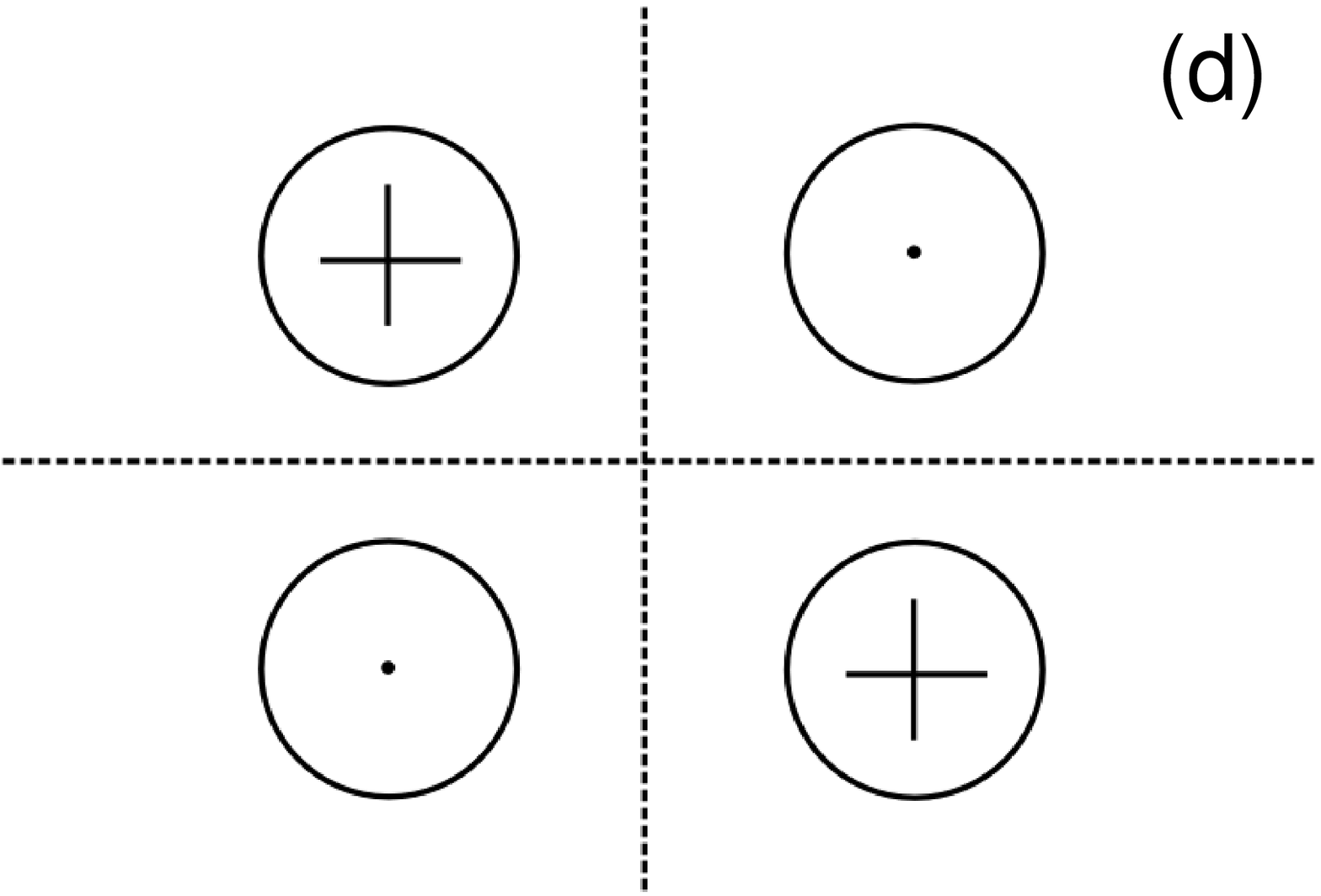}}
    \vspace{1.8cm}
  \end{minipage}
\caption{Odd disk magnetic field.}
\label{fig:disk_odd}
\end{figure*}

\begin{figure*}[tbhp]
  \begin{minipage}[b]{0.25\textwidth}
    \resizebox{\hsize}{!}{\includegraphics{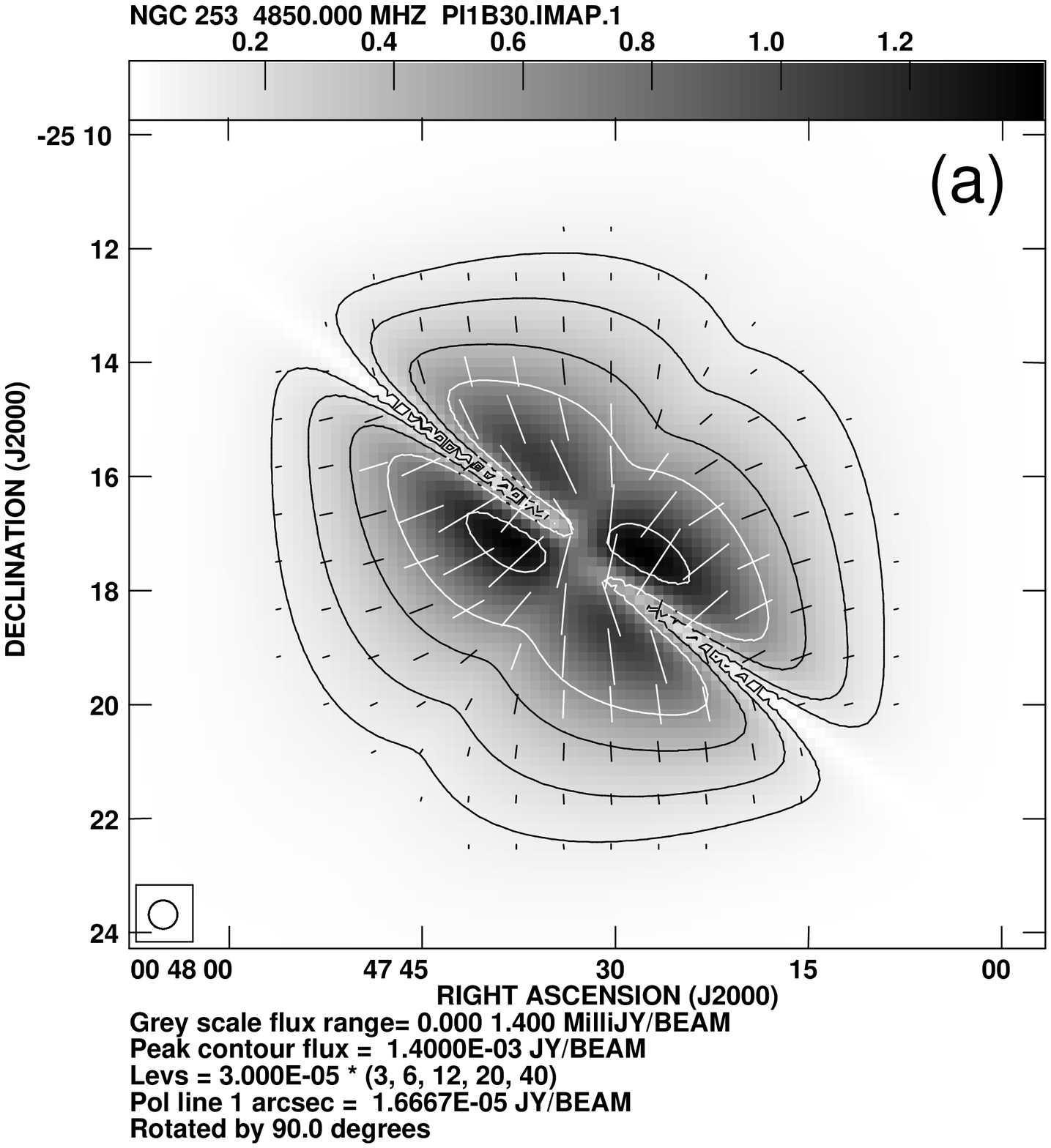}}
  \end{minipage}
  \begin{minipage}[b]{0.25\textwidth}
    \resizebox{\hsize}{!}{\includegraphics{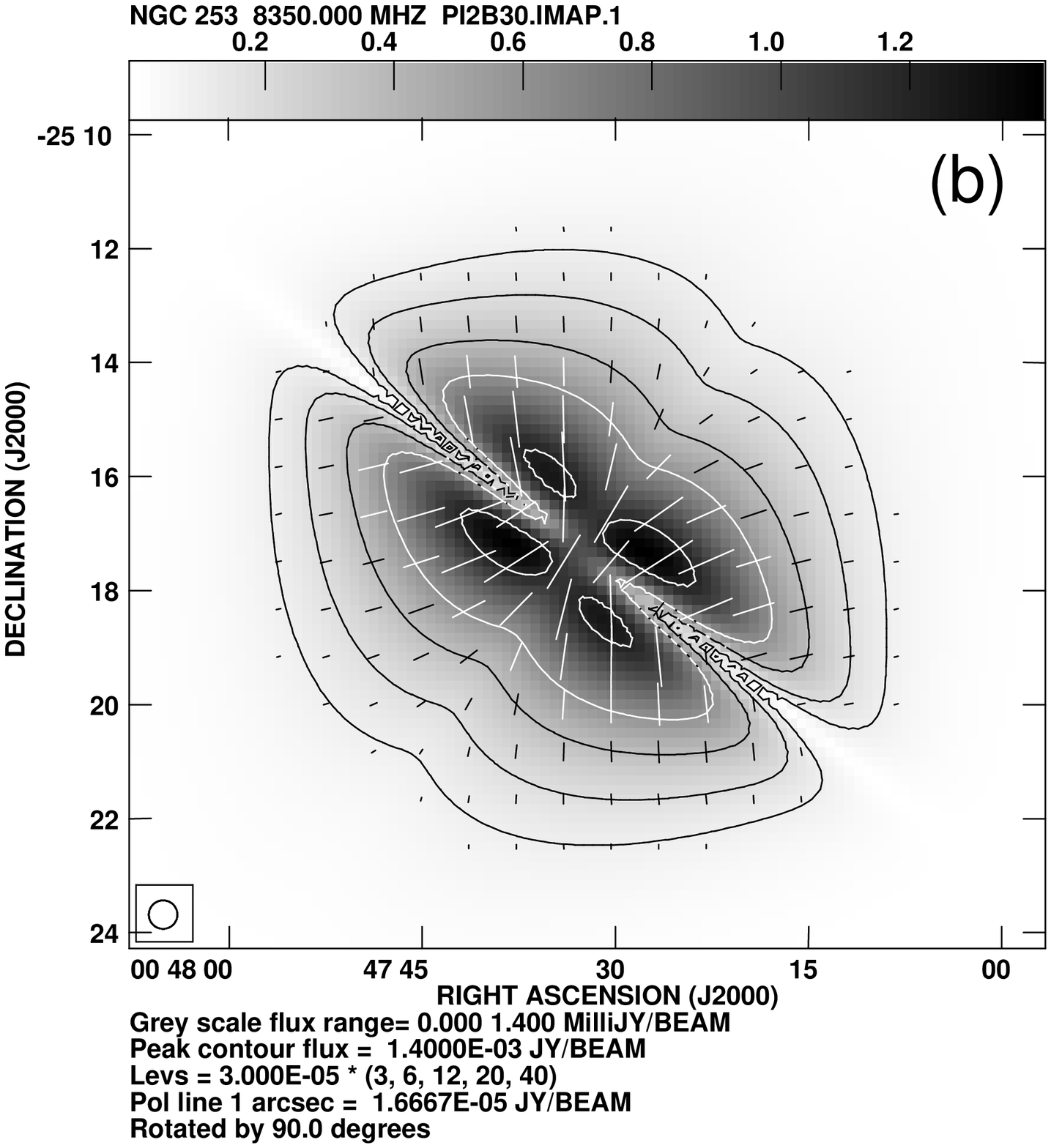}}
  \end{minipage}
  \begin{minipage}[b]{0.25\textwidth}
    \resizebox{\hsize}{!}{\includegraphics{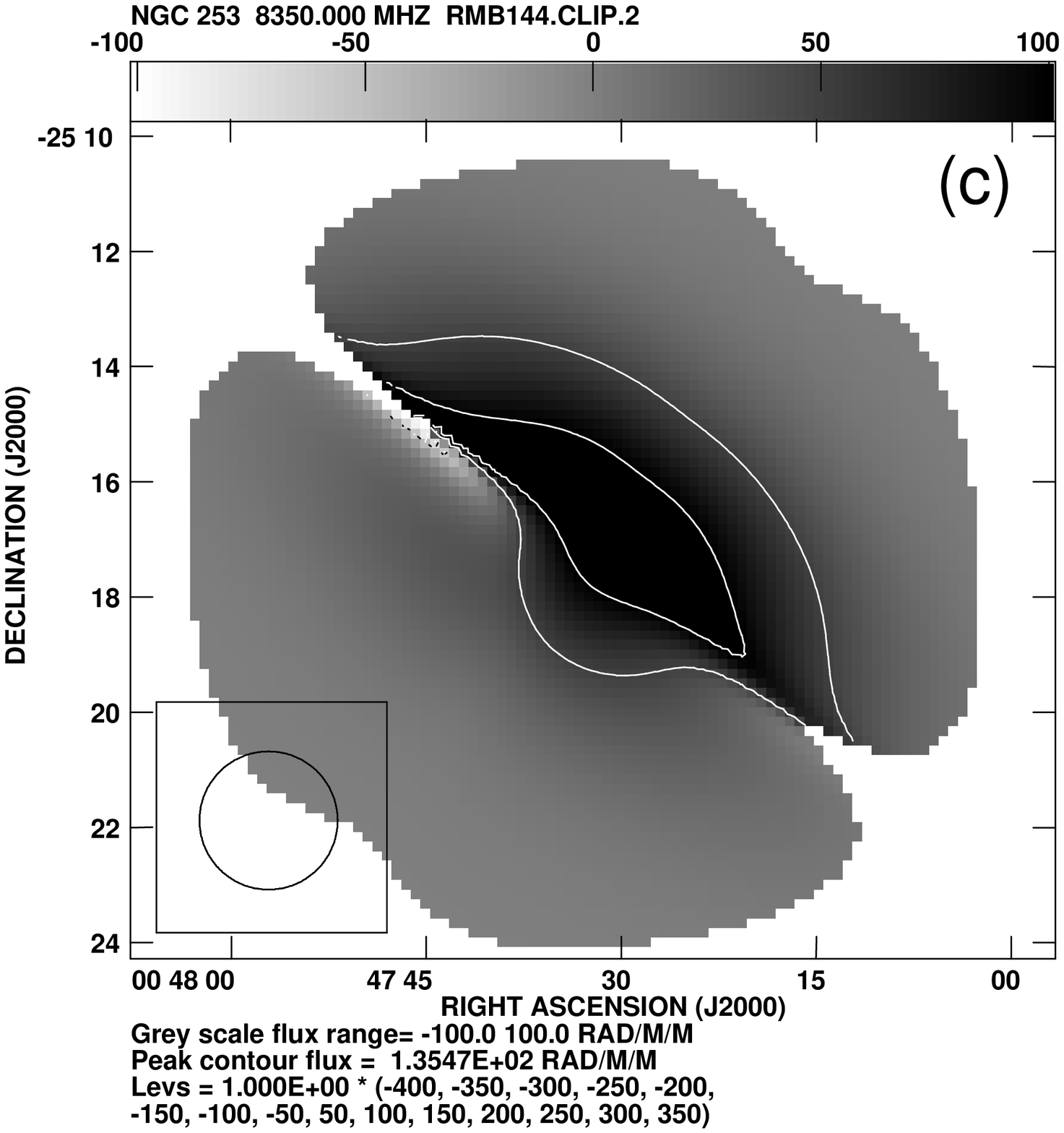}}
  \end{minipage}
  \begin{minipage}[b]{0.24\textwidth}
    \centering\resizebox{0.8\hsize}{!}{\includegraphics{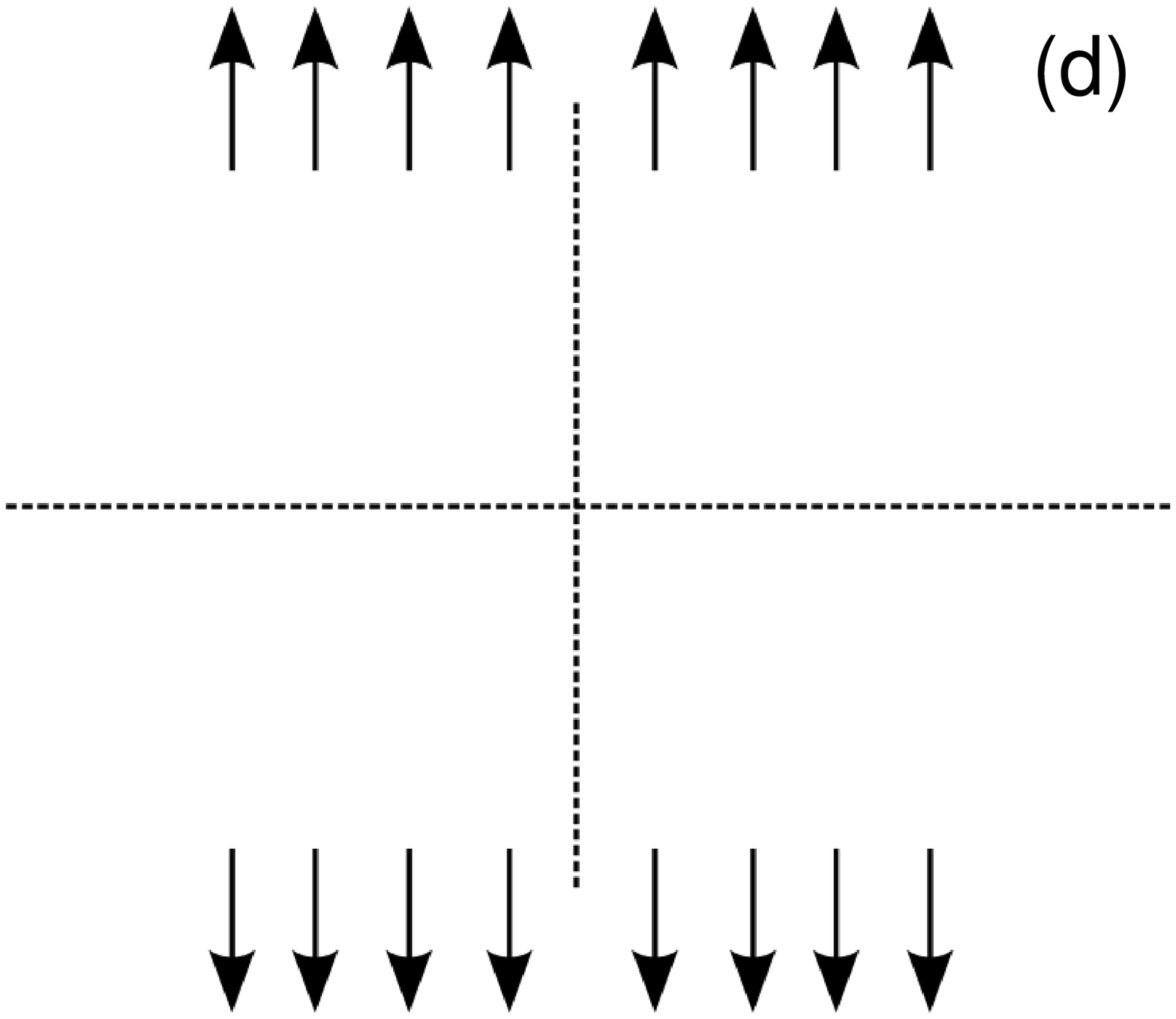}}
    \vspace{1.2cm}
  \end{minipage}
\caption{Even halo magnetic field. The halo field points away from the disk.}
\label{fig:halo_even}
\end{figure*}

\begin{figure*}[tbhp]
  \begin{minipage}[b]{0.25\textwidth}
    \resizebox{\hsize}{!}{\includegraphics{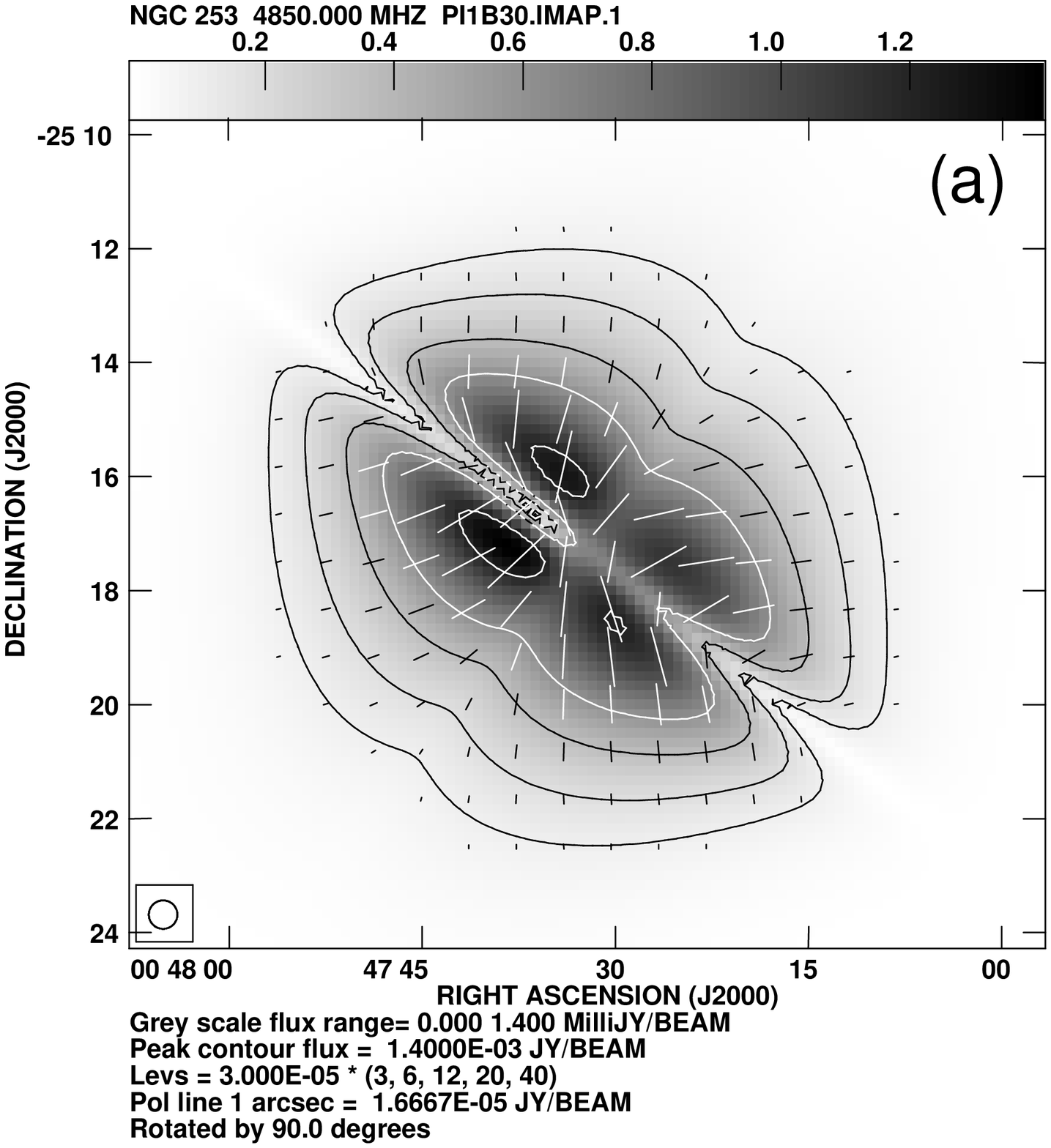}}
  \end{minipage}
  \begin{minipage}[b]{0.25\textwidth}
    \resizebox{\hsize}{!}{\includegraphics{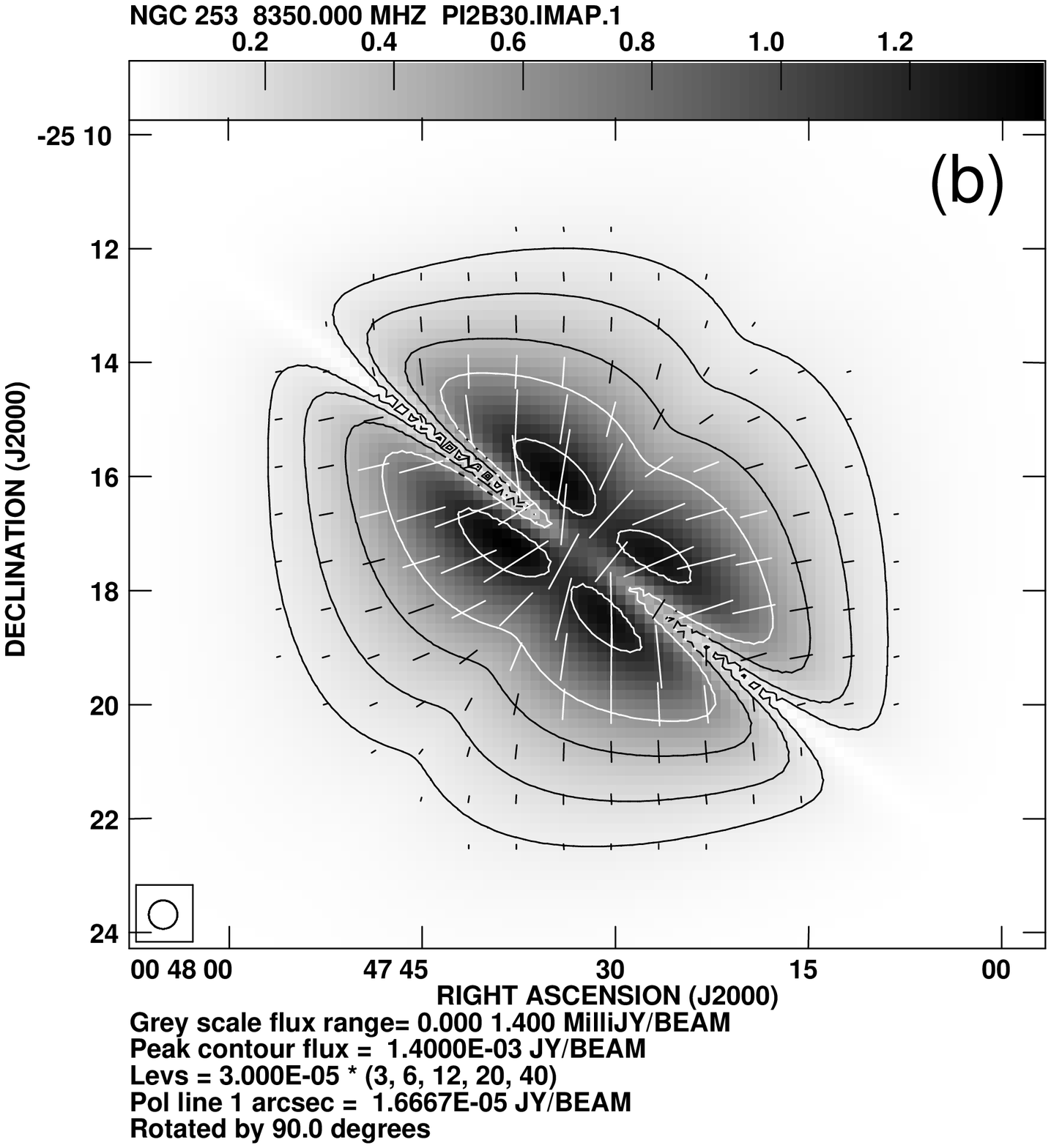}}
  \end{minipage}
  \begin{minipage}[b]{0.25\textwidth}
    \resizebox{\hsize}{!}{\includegraphics{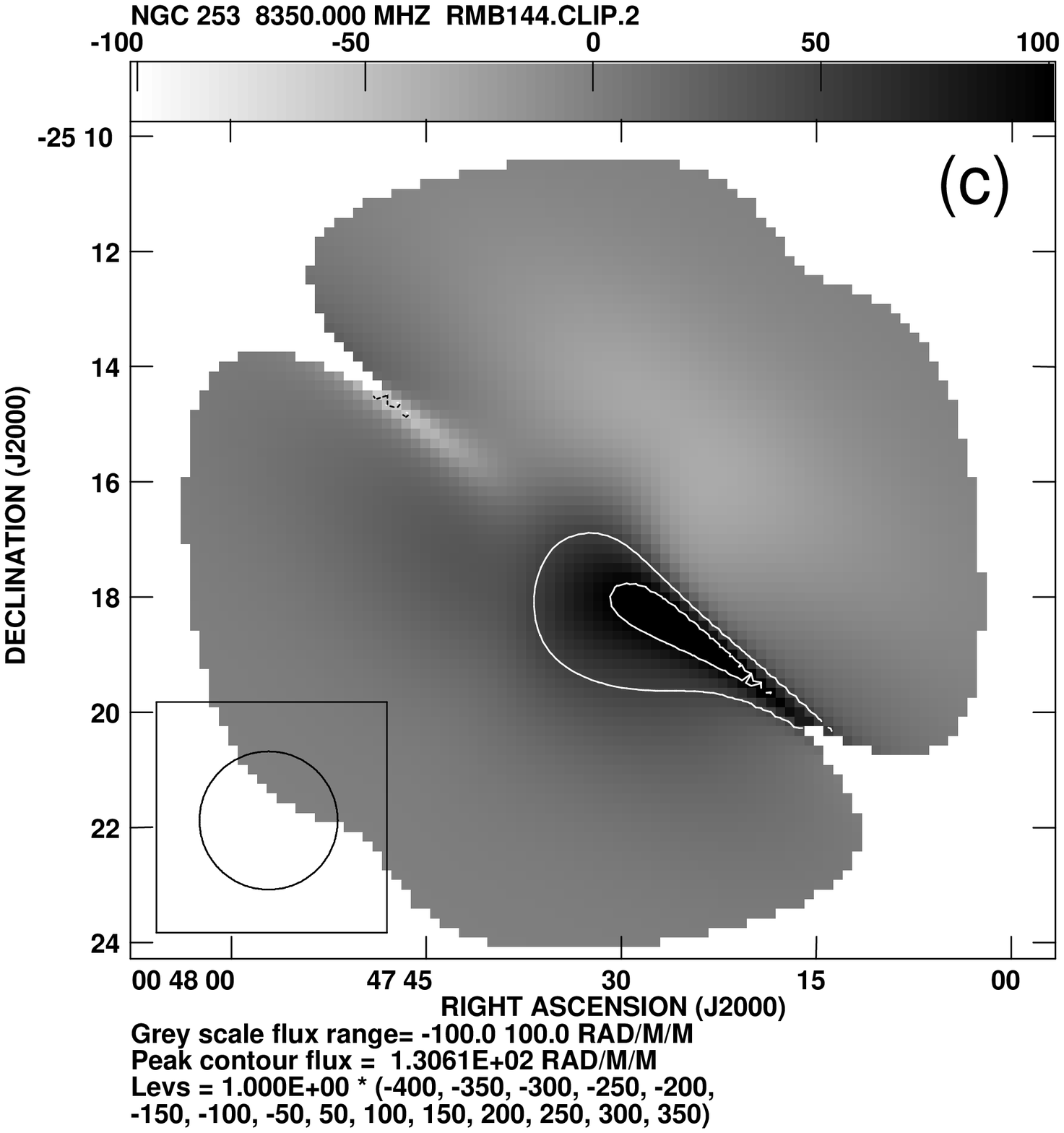}}
  \end{minipage}
  \begin{minipage}[b]{0.24\textwidth}
    \centering\resizebox{0.8\hsize}{!}{\includegraphics{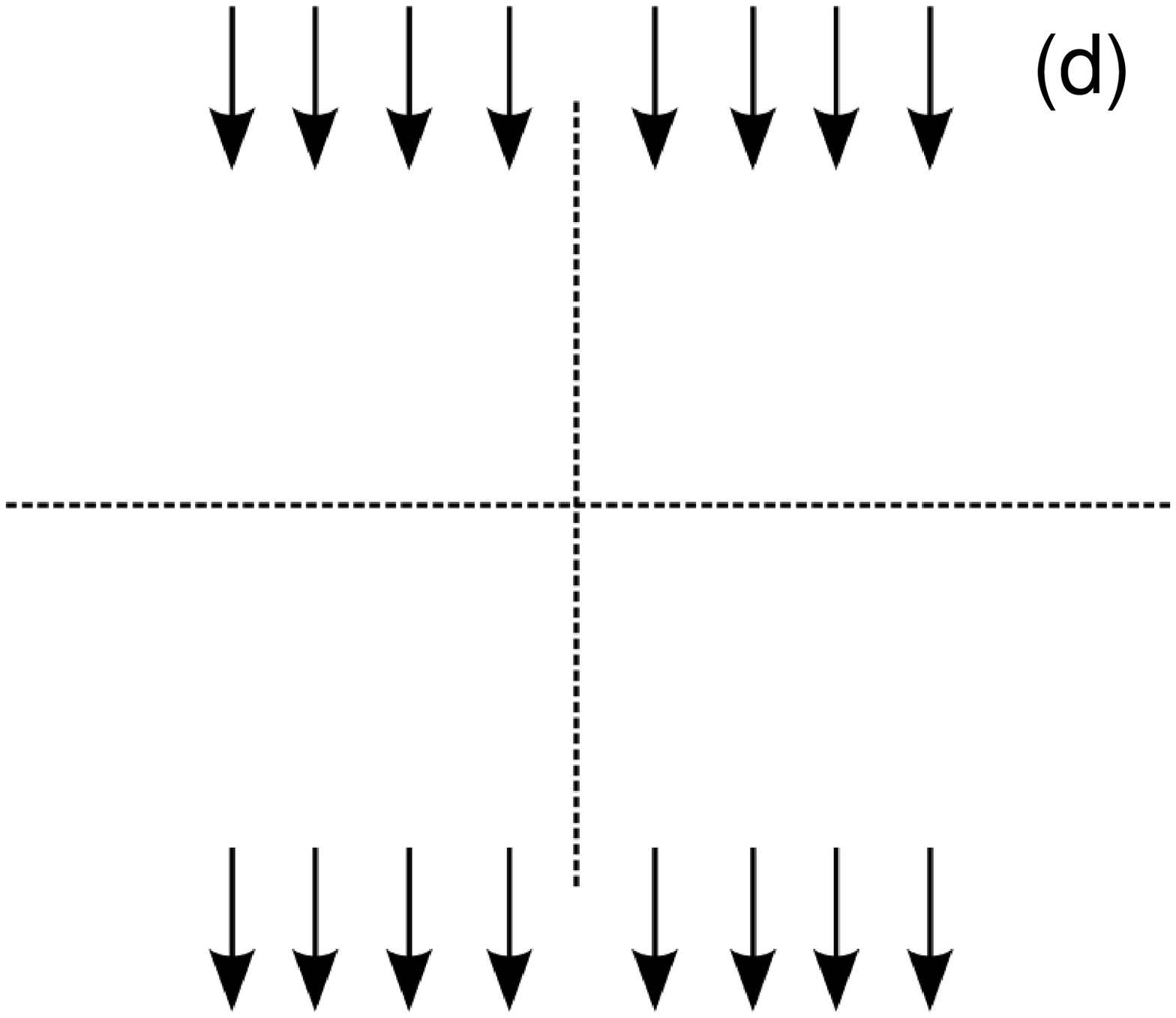}}
    \vspace{1.2cm}
  \end{minipage}
\caption{Odd halo magnetic field. The halo field points away from the disk in
  the southern halo.}
\label{fig:halo_odd}
\end{figure*}

\begin{figure*}[tbhp]
  \begin{minipage}[b]{0.25\textwidth}
    \resizebox{\hsize}{!}{\includegraphics{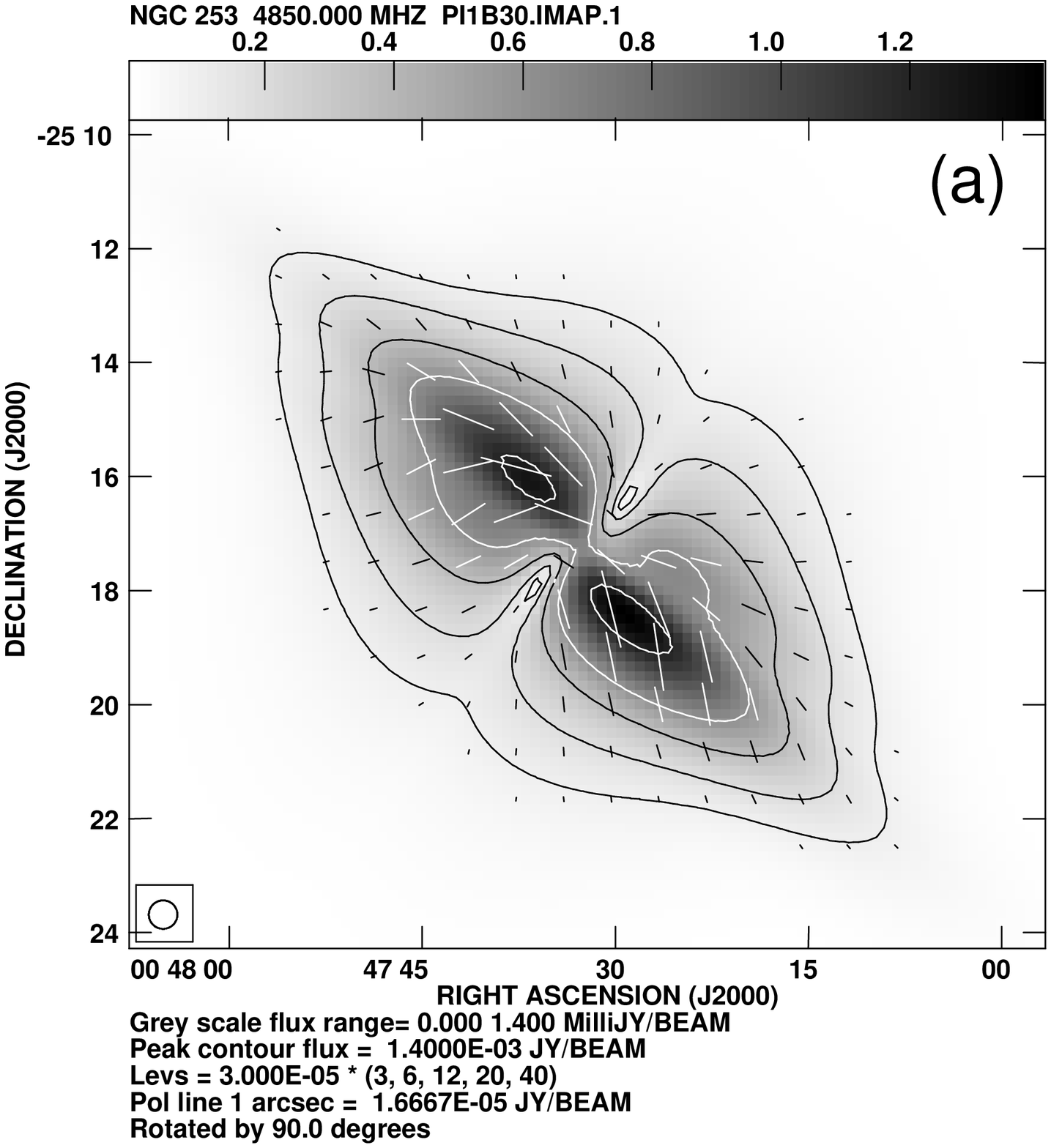}}
  \end{minipage}
  \begin{minipage}[b]{0.25\textwidth}
    \resizebox{\hsize}{!}{\includegraphics{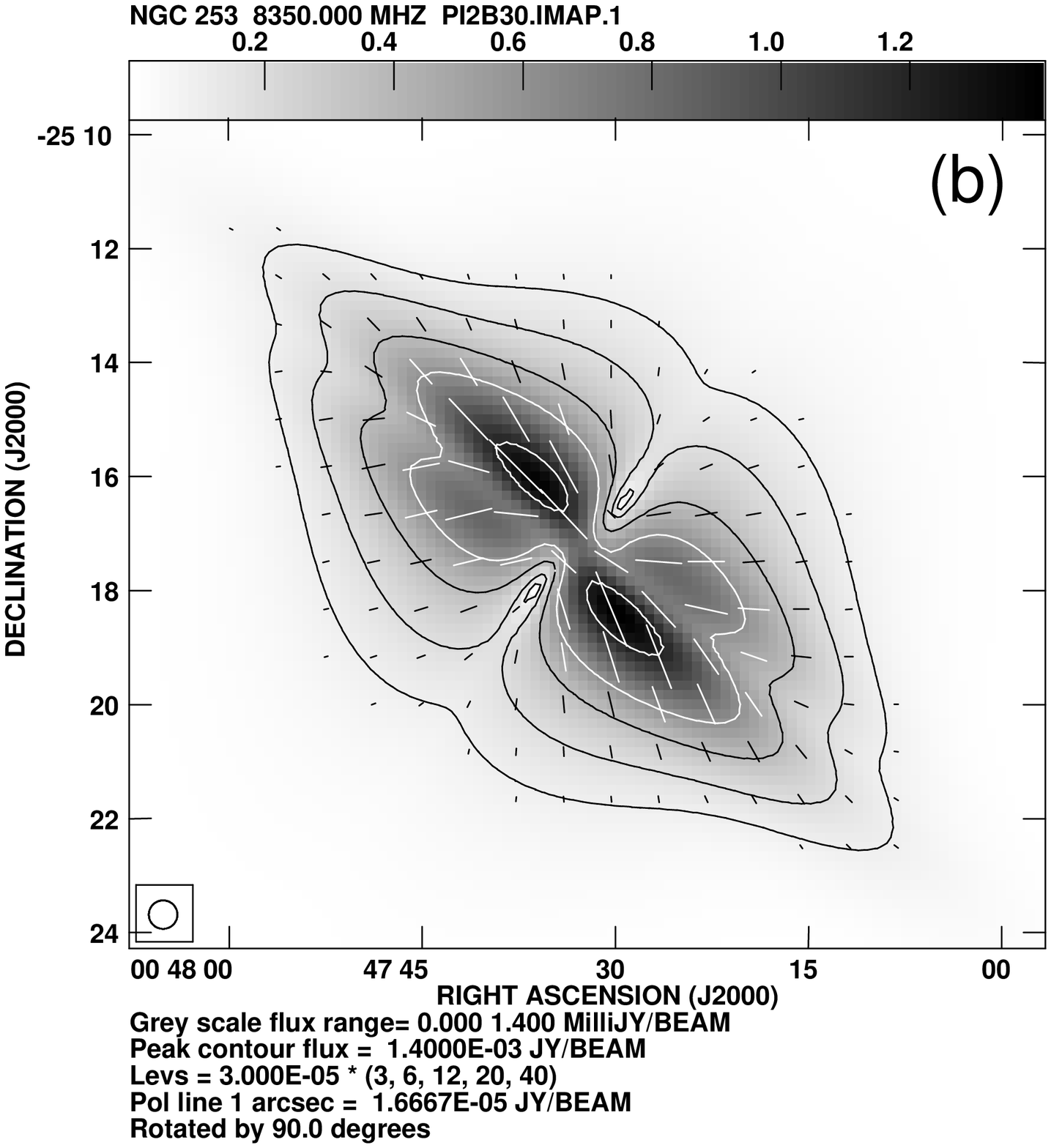}}
  \end{minipage}
  \begin{minipage}[b]{0.25\textwidth}
    \resizebox{\hsize}{!}{\includegraphics{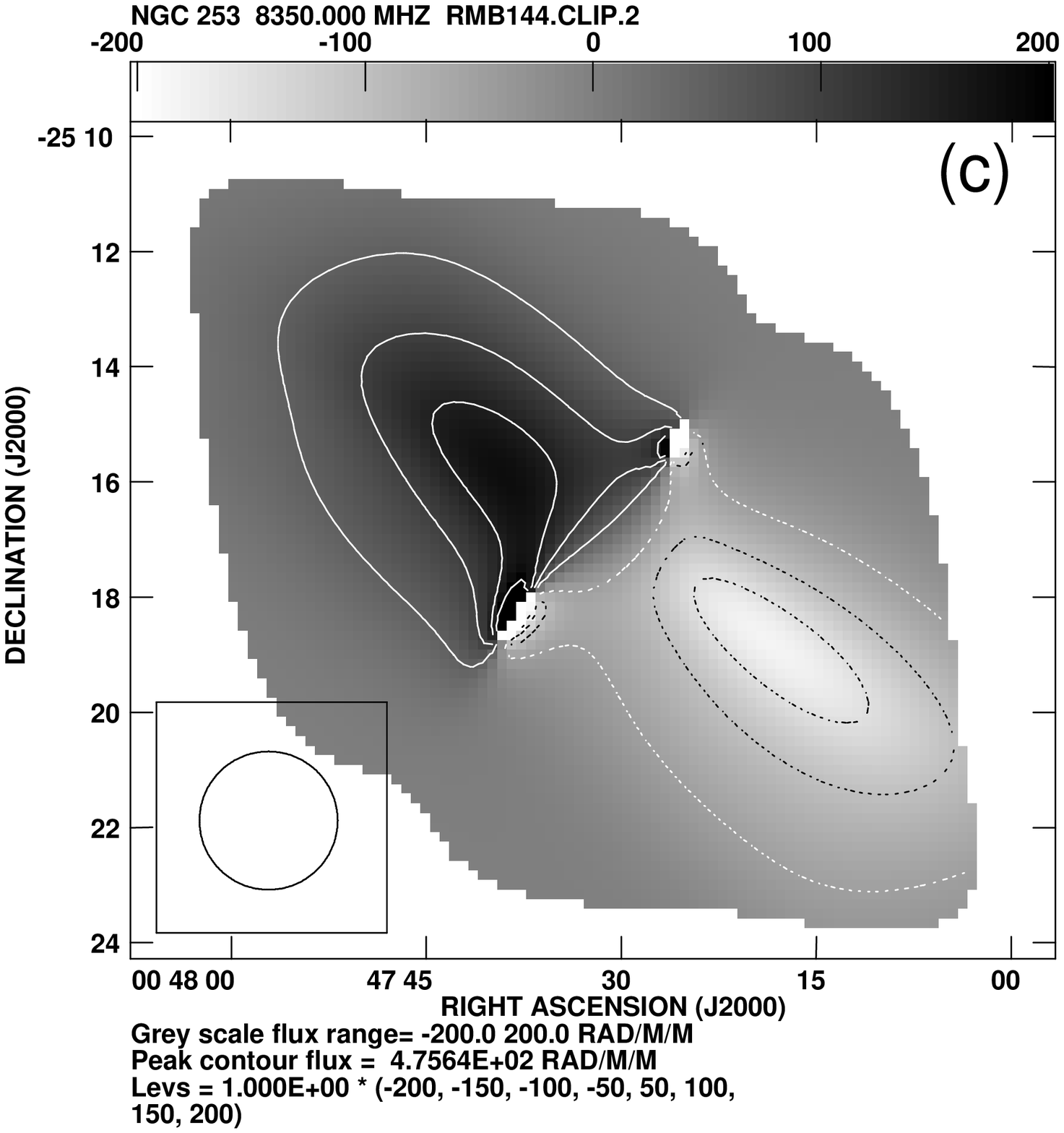}}
  \end{minipage}
  \begin{minipage}[b]{0.24\textwidth}
    \centering\resizebox{0.8\hsize}{!}{\includegraphics{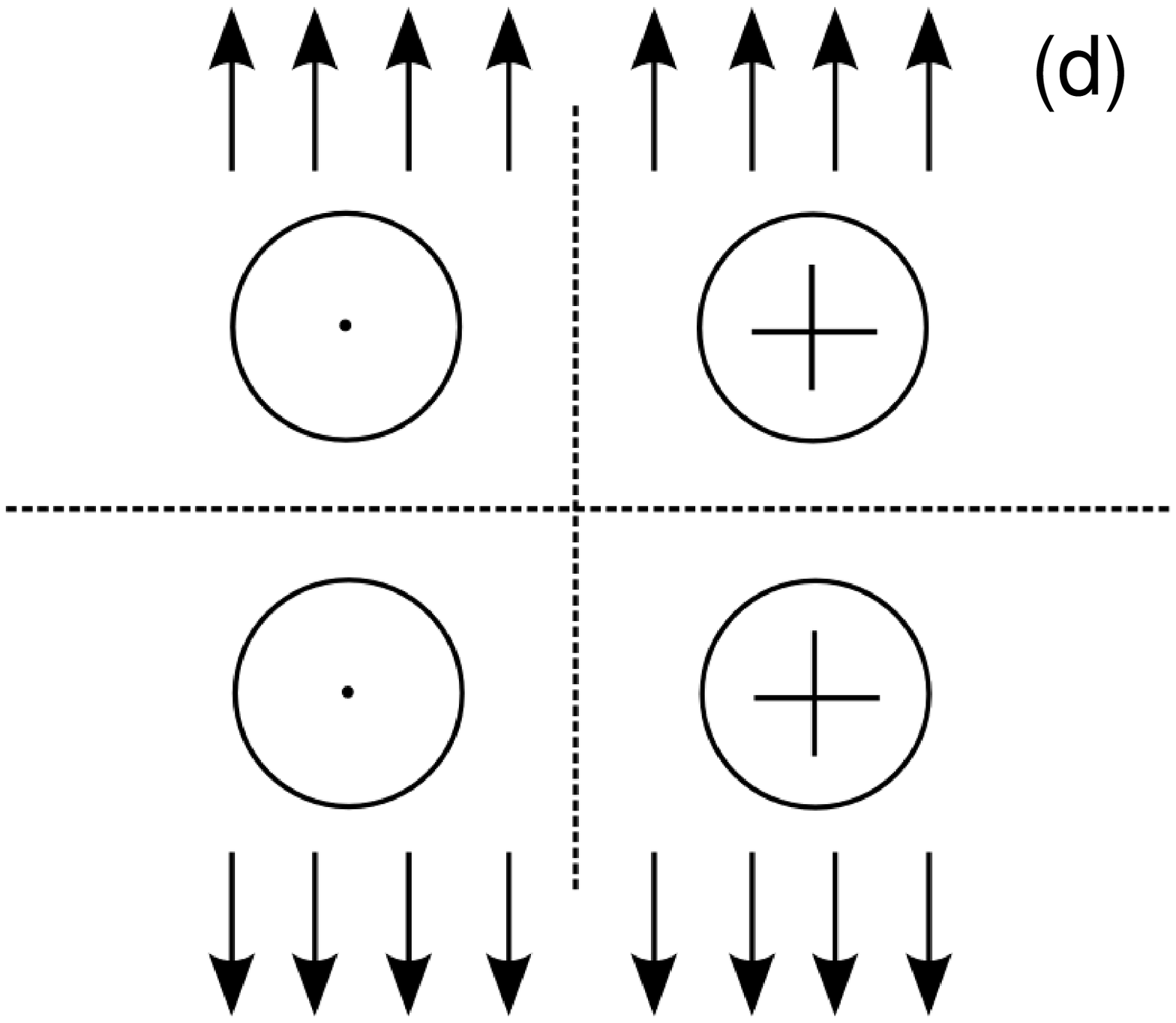}}
    \vspace{1.2cm}
  \end{minipage}
\caption{Even disk magnetic field and even halo magnetic field. The
  halo field points away from the disk.}
\label{fig:even_even}
\end{figure*}

\begin{figure*}[tbhp]
  \begin{minipage}[b]{0.25\textwidth}
    \resizebox{\hsize}{!}{\includegraphics{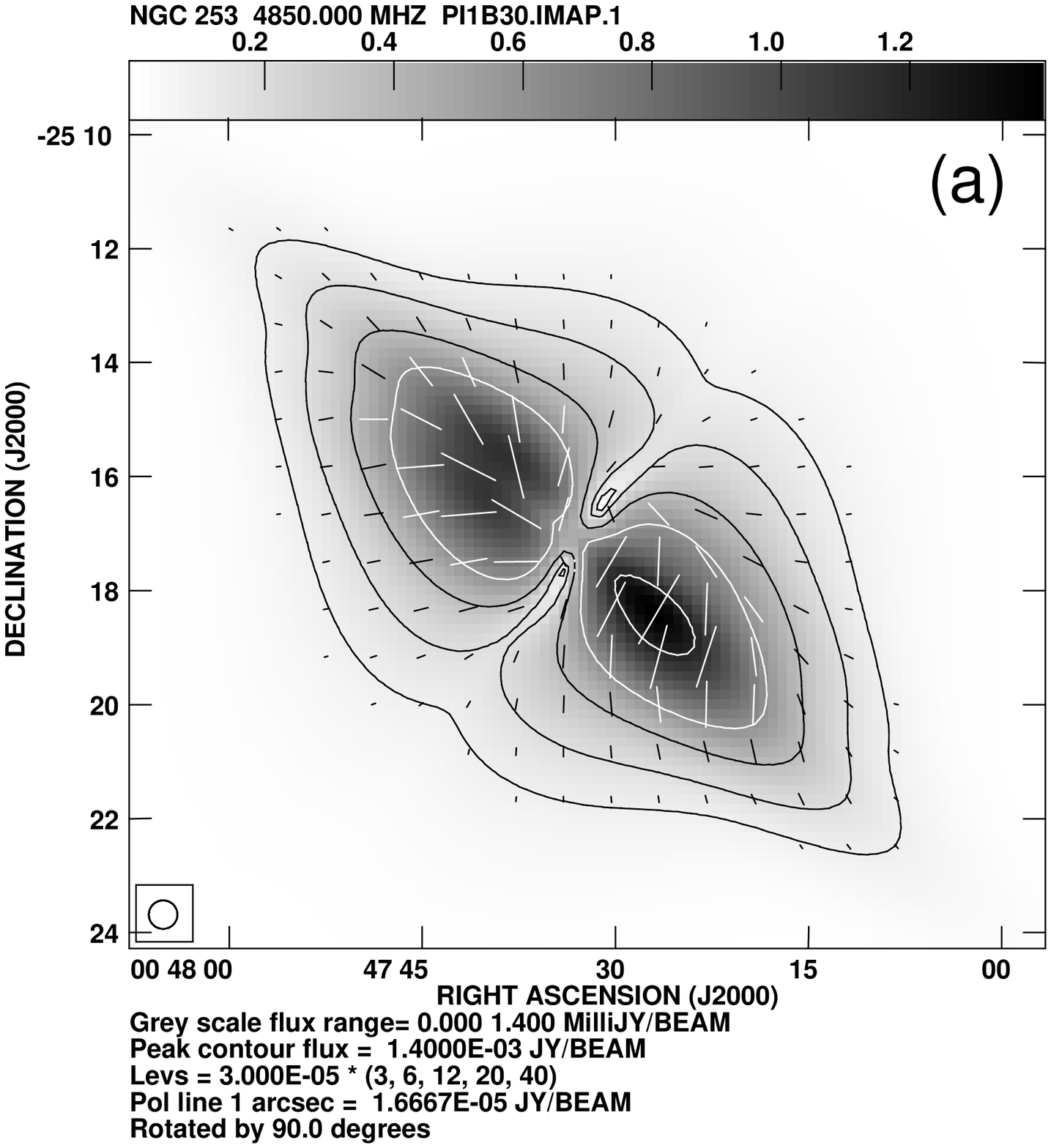}}
  \end{minipage}
  \begin{minipage}[b]{0.25\textwidth}
    \resizebox{\hsize}{!}{\includegraphics{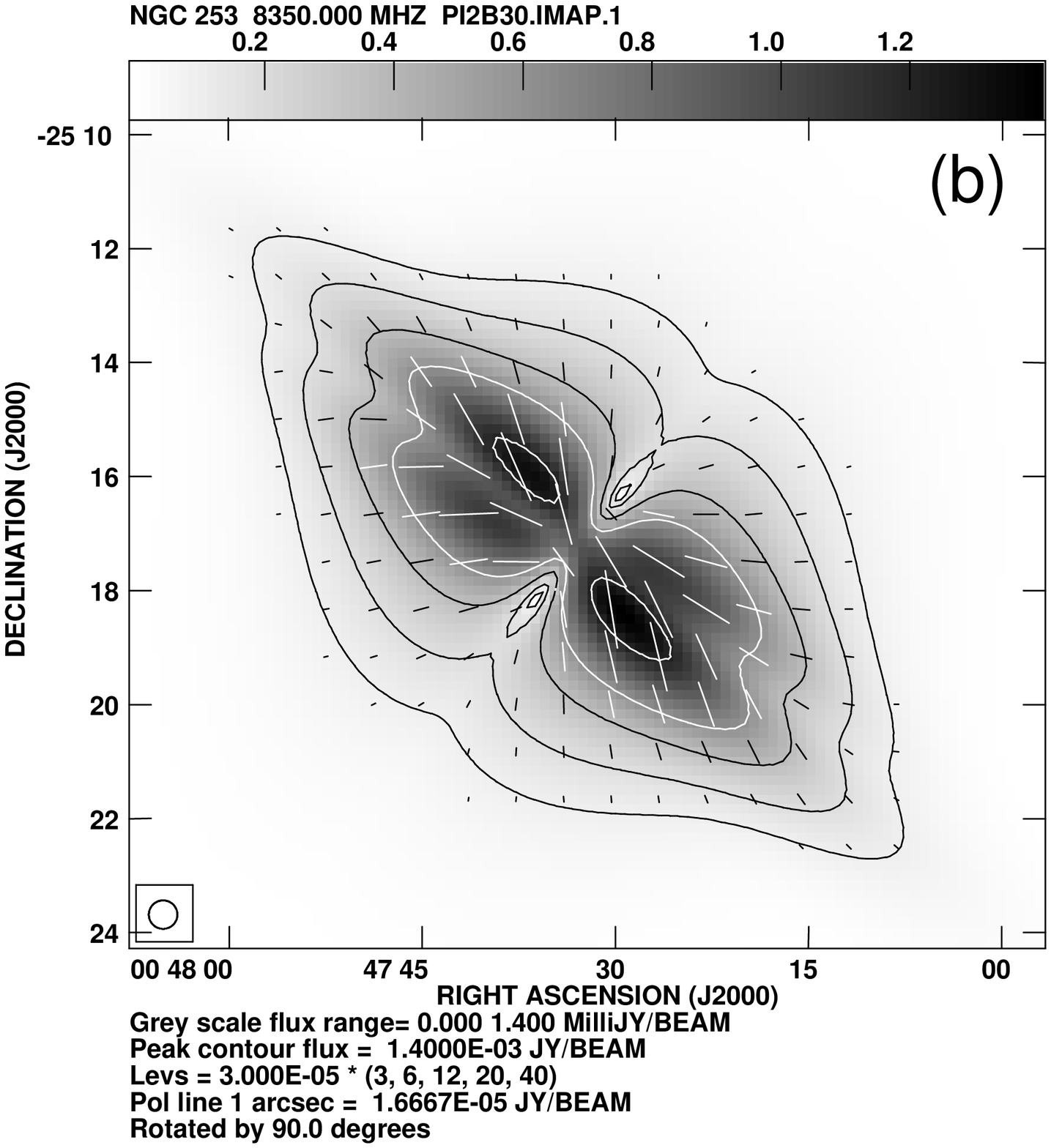}}
  \end{minipage}
  \begin{minipage}[b]{0.25\textwidth}
    \resizebox{\hsize}{!}{\includegraphics{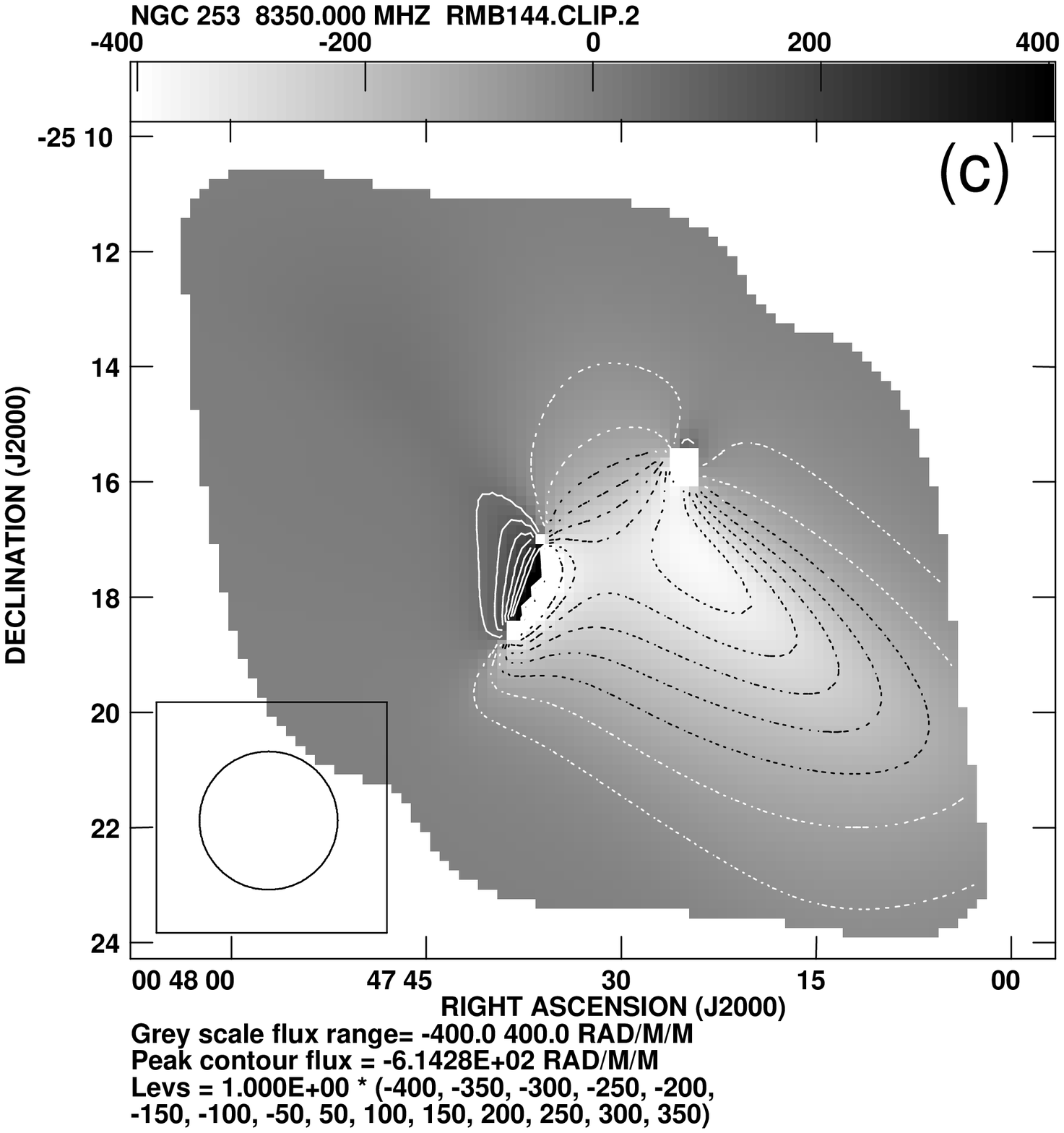}}
  \end{minipage}
  \begin{minipage}[b]{0.24\textwidth}
    \centering\resizebox{0.8\hsize}{!}{\includegraphics{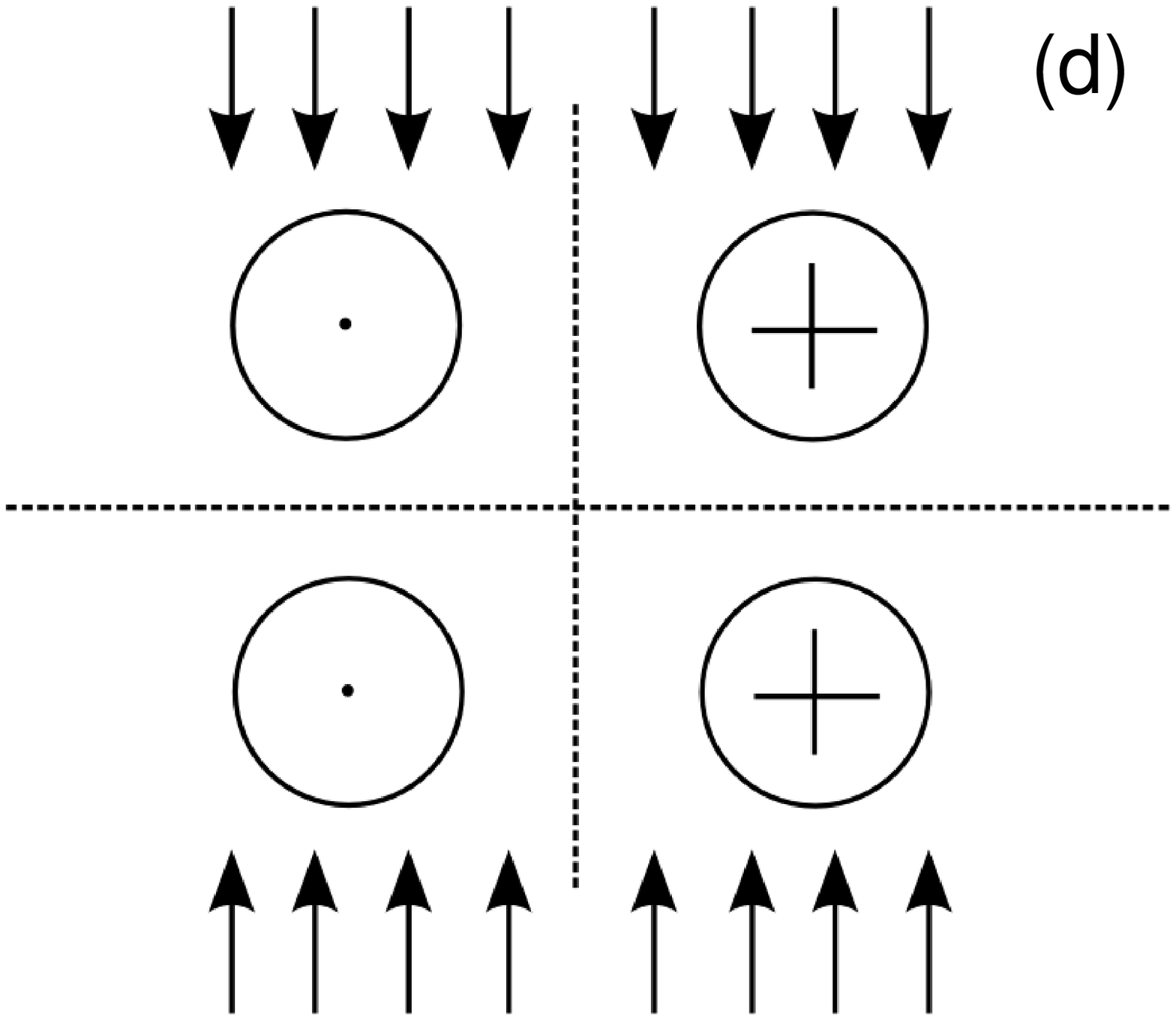}}
    \vspace{1.2cm}
  \end{minipage}
\caption{Even disk magnetic field and even halo magnetic field. The
  halo field points towards the disk.}
\label{fig:even_even_n}
\end{figure*}

\begin{figure*}[tbhp]
  \begin{minipage}[b]{0.25\textwidth}
    \resizebox{\hsize}{!}{\includegraphics{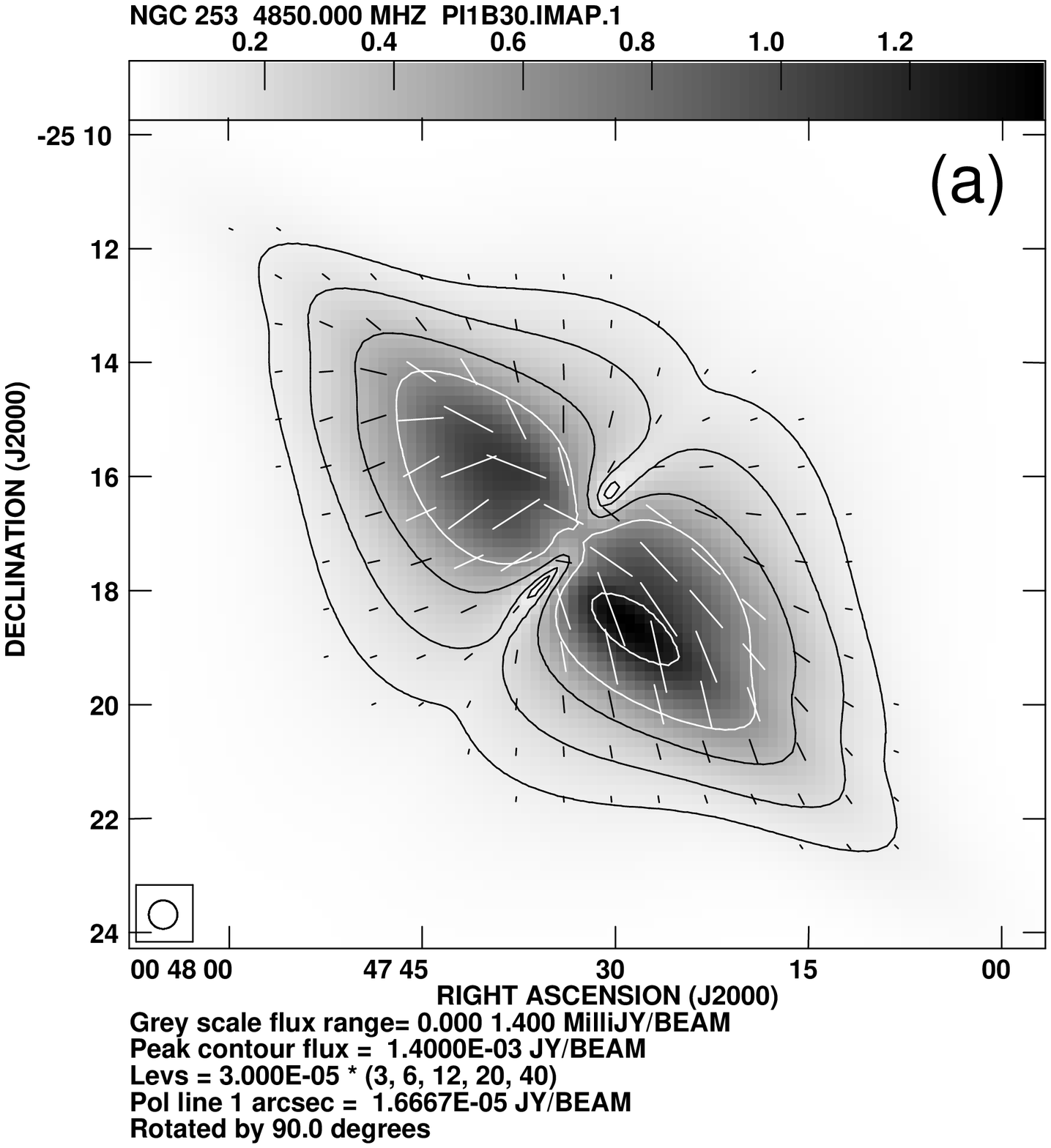}}
  \end{minipage}
  \begin{minipage}[b]{0.25\textwidth}
    \resizebox{\hsize}{!}{\includegraphics{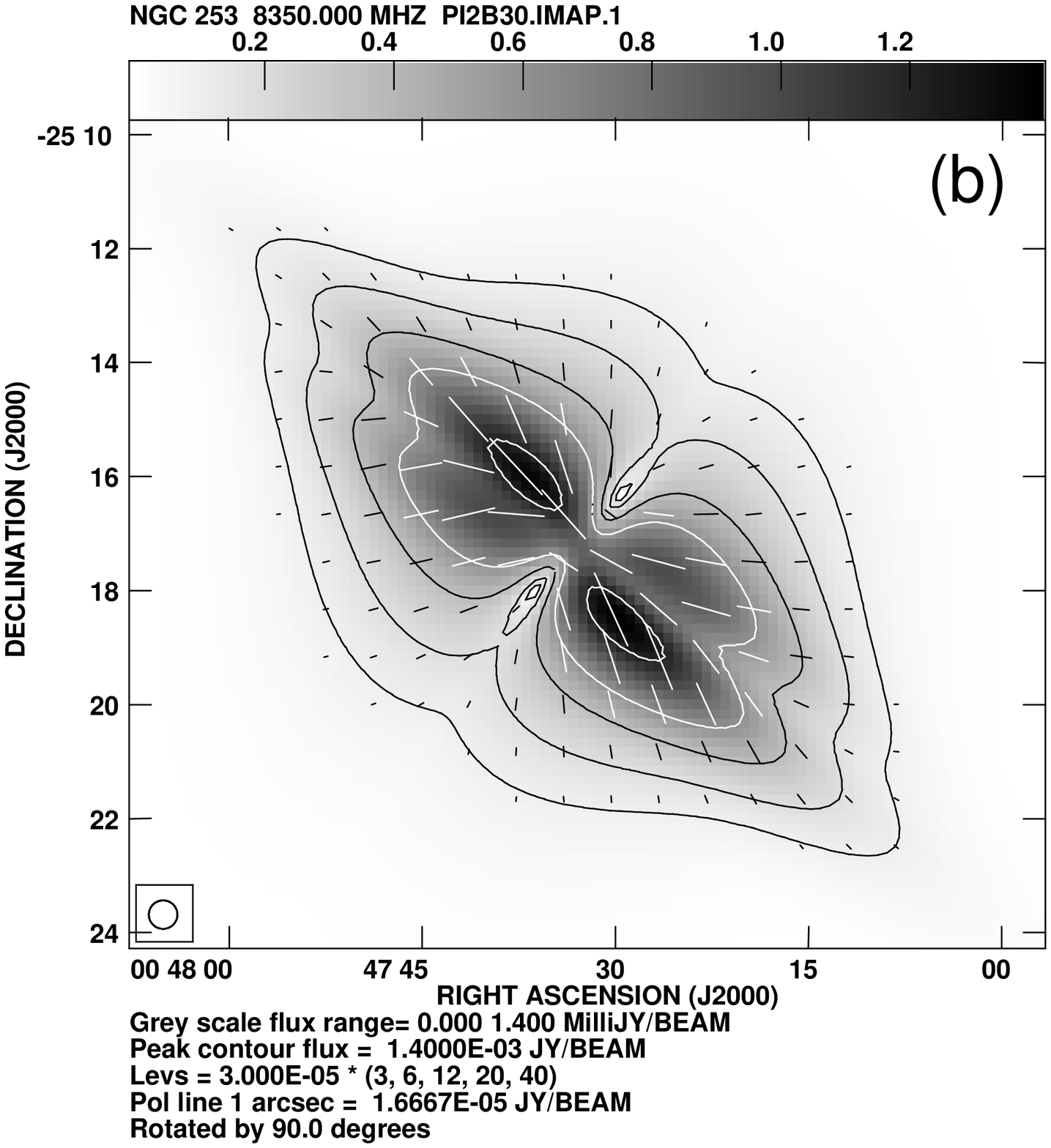}}
  \end{minipage}
  \begin{minipage}[b]{0.25\textwidth}
    \resizebox{\hsize}{!}{\includegraphics{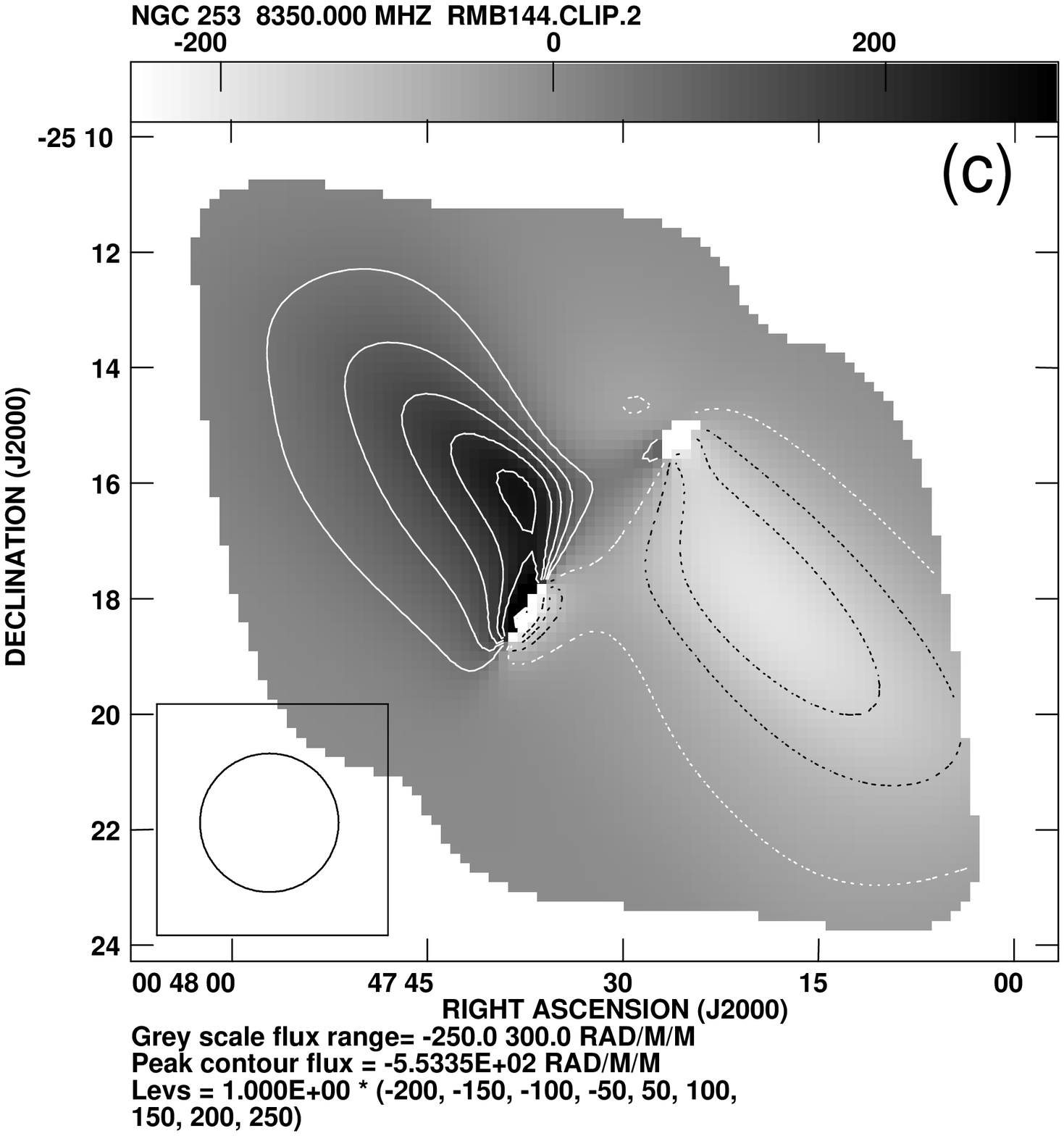}}
  \end{minipage}
  \begin{minipage}[b]{0.24\textwidth}
    \centering\resizebox{0.8\hsize}{!}{\includegraphics{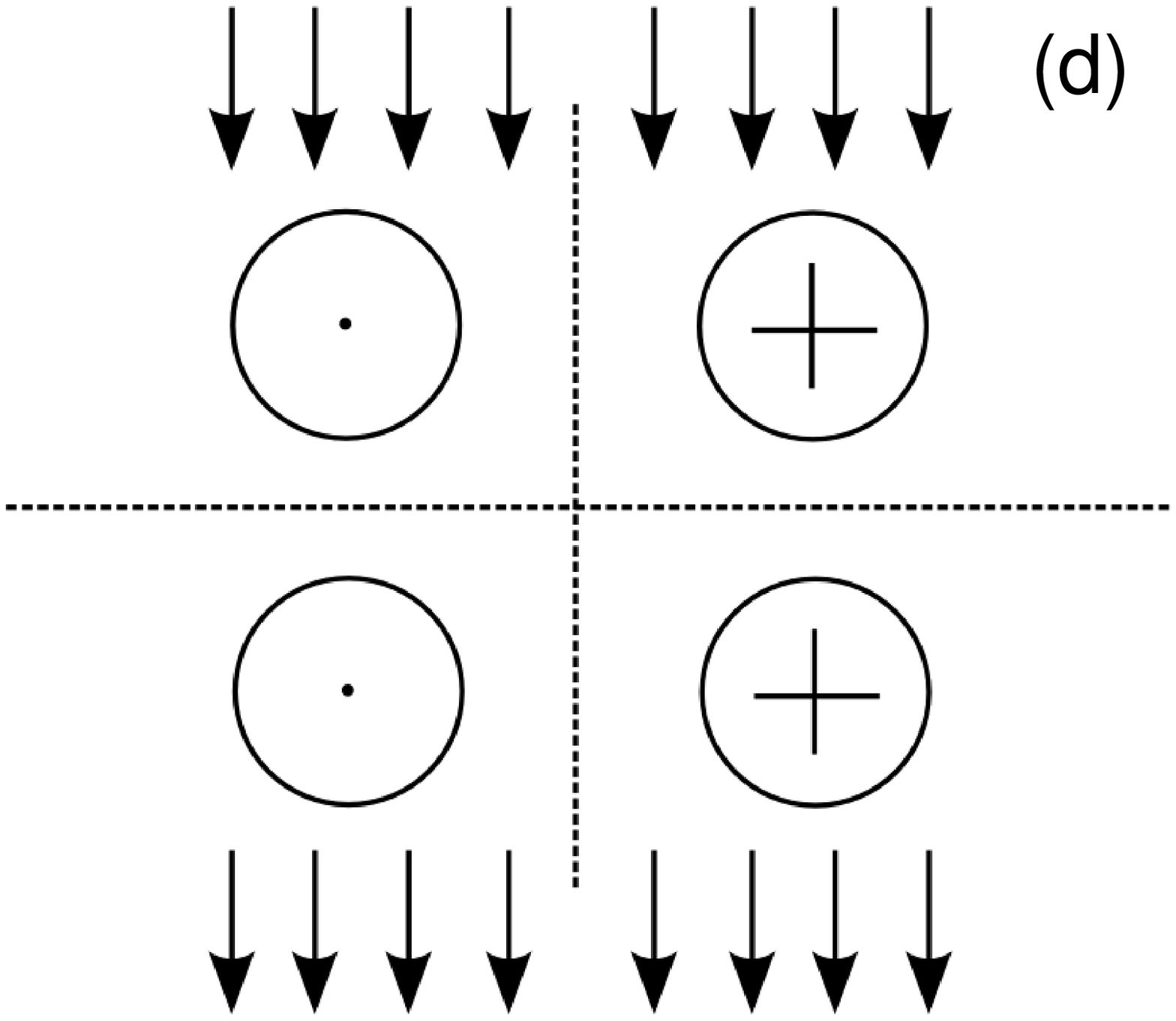}}
    \vspace{1.2cm}
  \end{minipage}
\caption{Even disk magnetic field and odd halo magnetic field. The
  halo field points away from the disk in the southern halo.}
\label{fig:even_odd}
\end{figure*}

\begin{figure*}[tbhp]
  \begin{minipage}[b]{0.25\textwidth}
    \resizebox{\hsize}{!}{\includegraphics{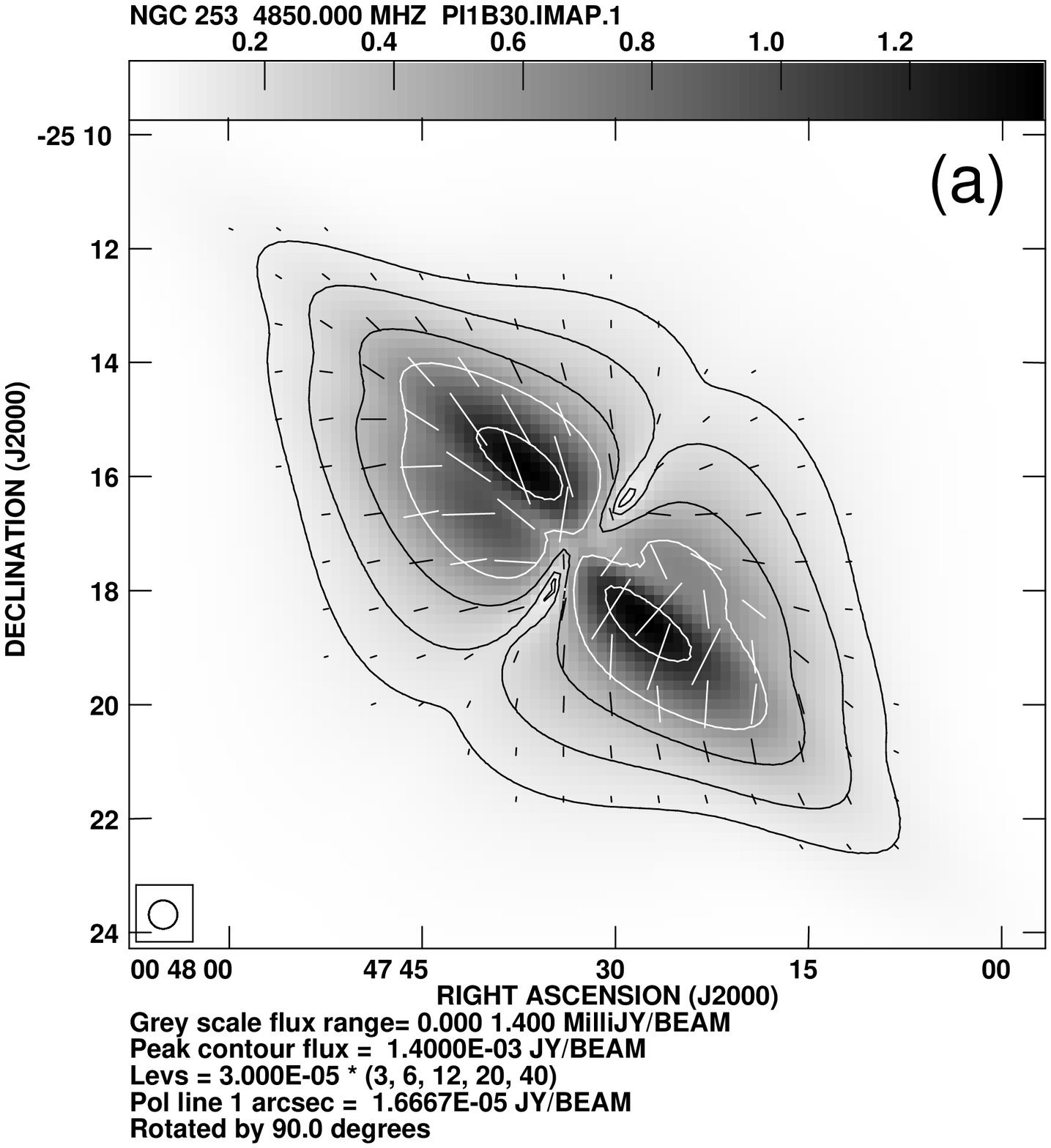}}
  \end{minipage}
  \begin{minipage}[b]{0.25\textwidth}
    \resizebox{\hsize}{!}{\includegraphics{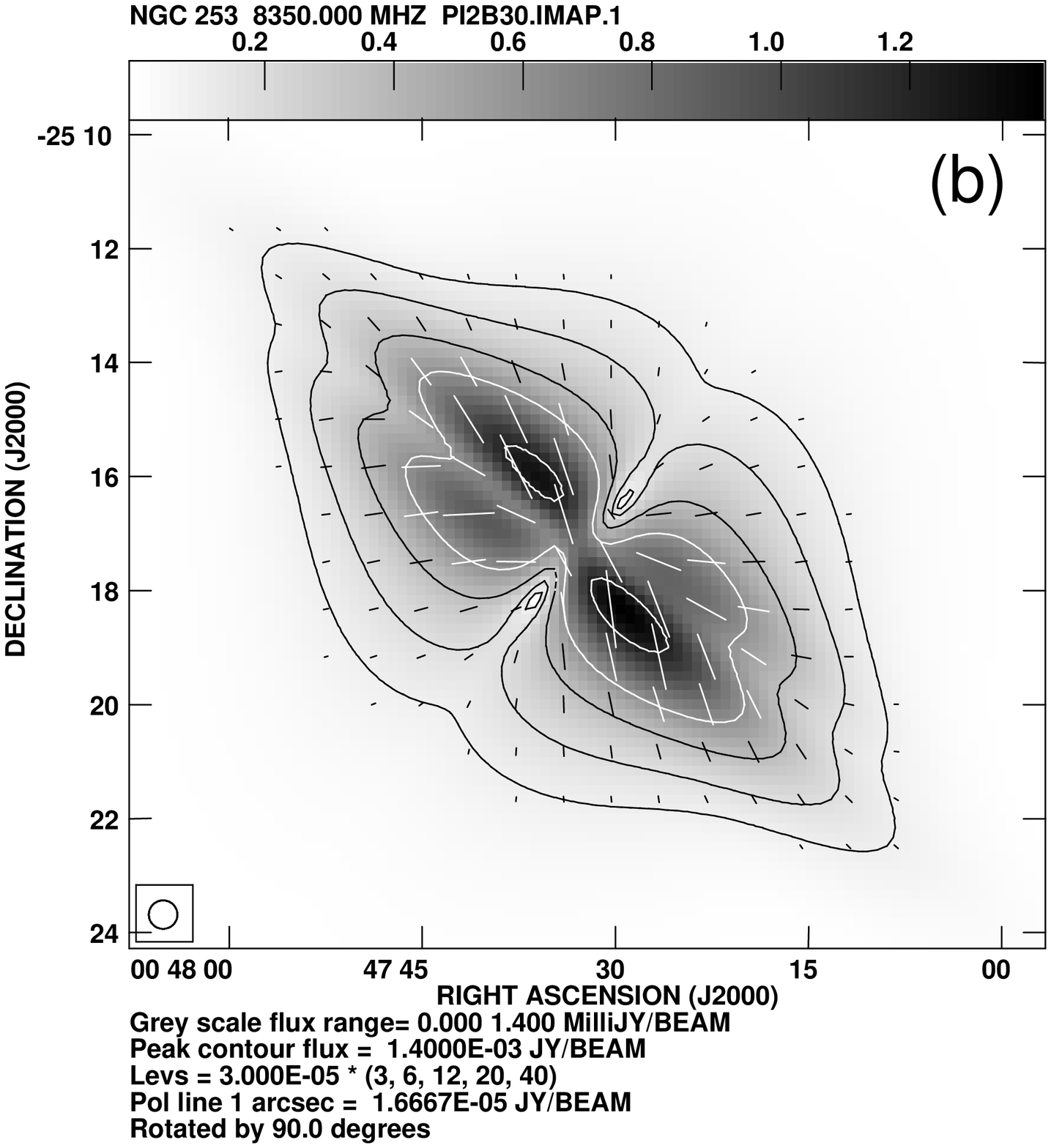}}
  \end{minipage}
  \begin{minipage}[b]{0.25\textwidth}
    \resizebox{\hsize}{!}{\includegraphics{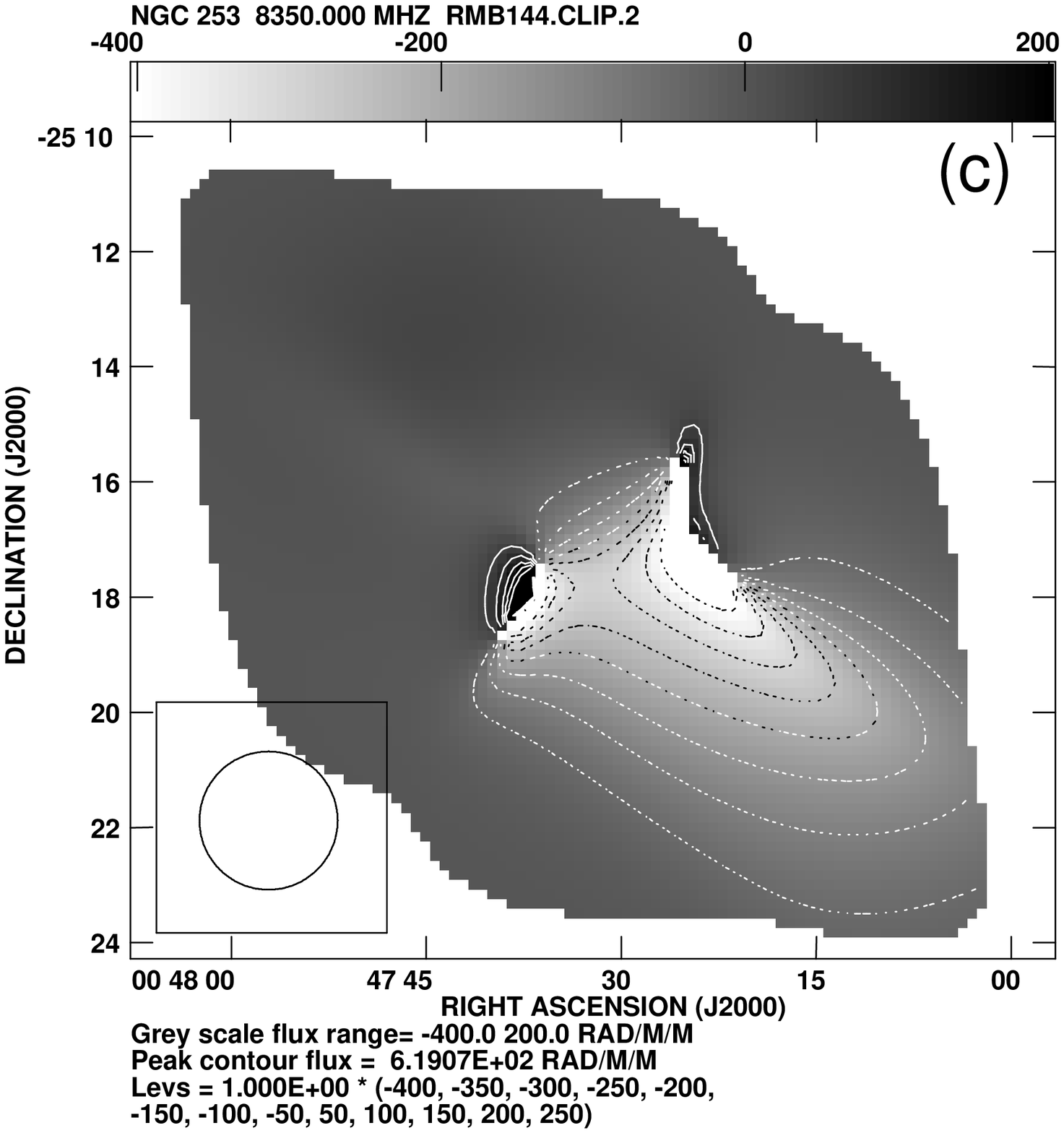}}
  \end{minipage}
  \begin{minipage}[b]{0.24\textwidth}
    \centering\resizebox{0.8\hsize}{!}{\includegraphics{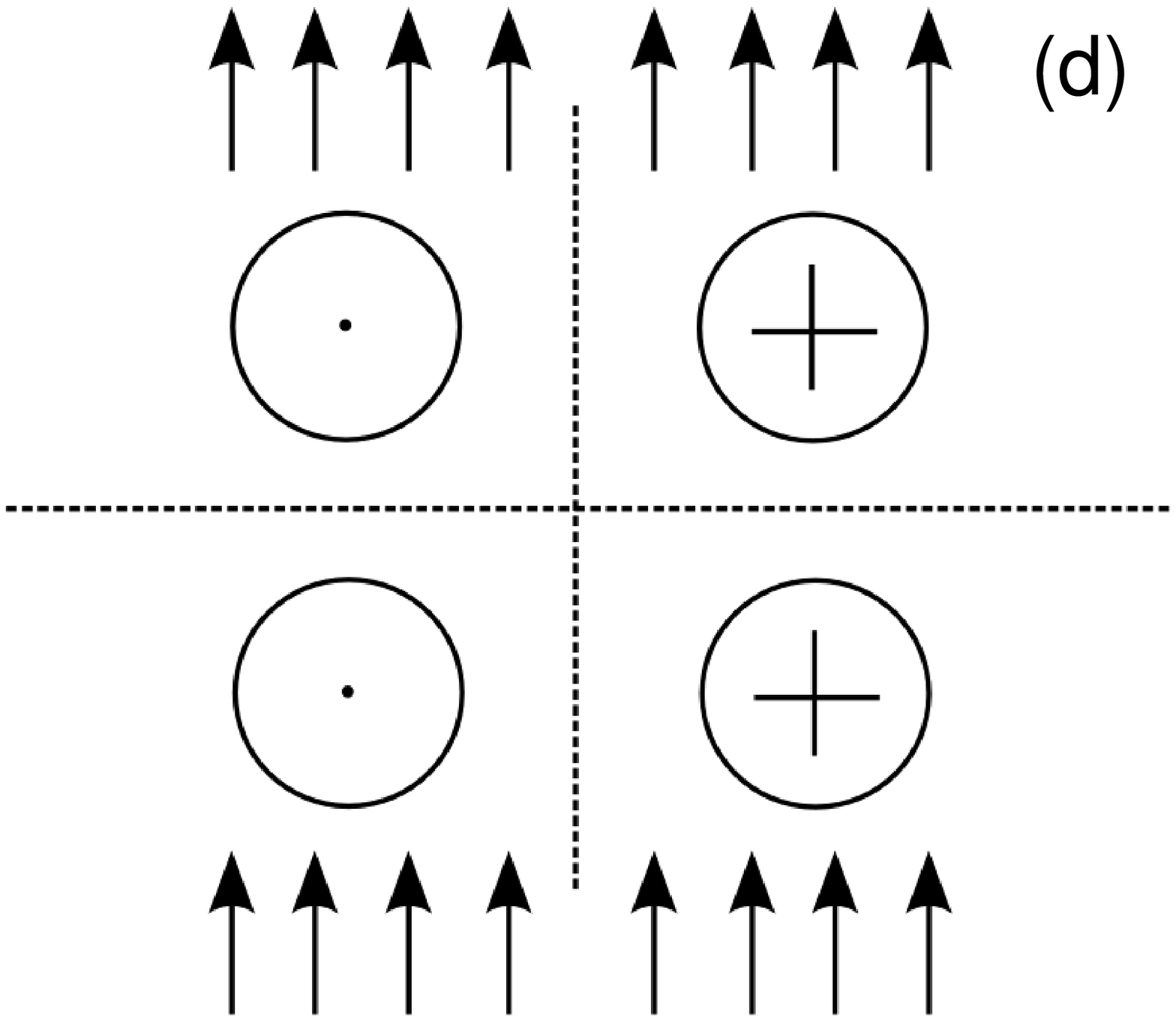}}
    \vspace{1.2cm}
  \end{minipage}
\caption{Even disk magnetic field and odd halo magnetic field. The
  halo field points towards the disk in the southern halo.}
\label{fig:even_odd_n}
\end{figure*}

\begin{figure*}[tbhp]
  \begin{minipage}[b]{0.25\textwidth}
    \resizebox{\hsize}{!}{\includegraphics{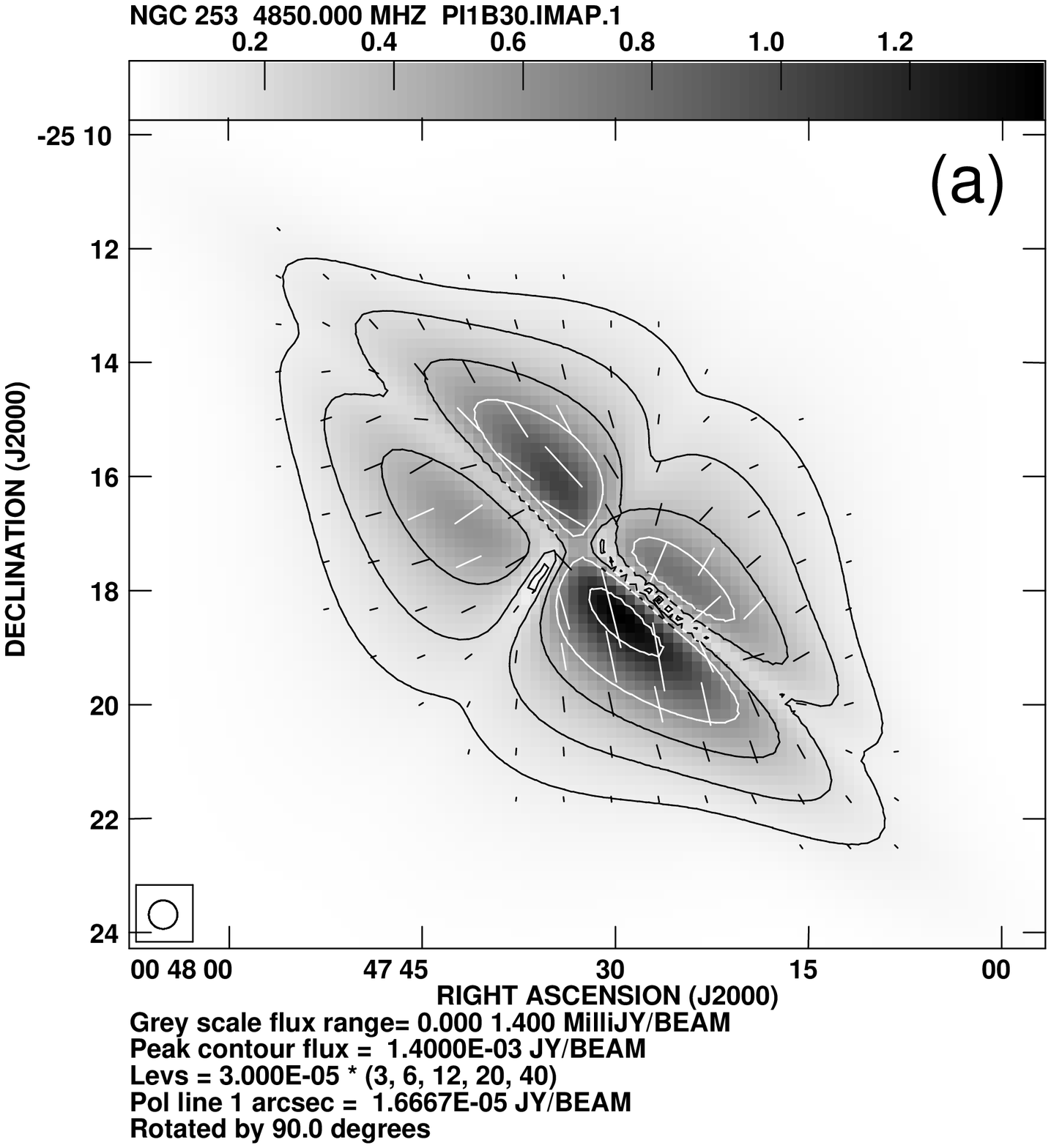}}
  \end{minipage}
  \begin{minipage}[b]{0.25\textwidth}
    \resizebox{\hsize}{!}{\includegraphics{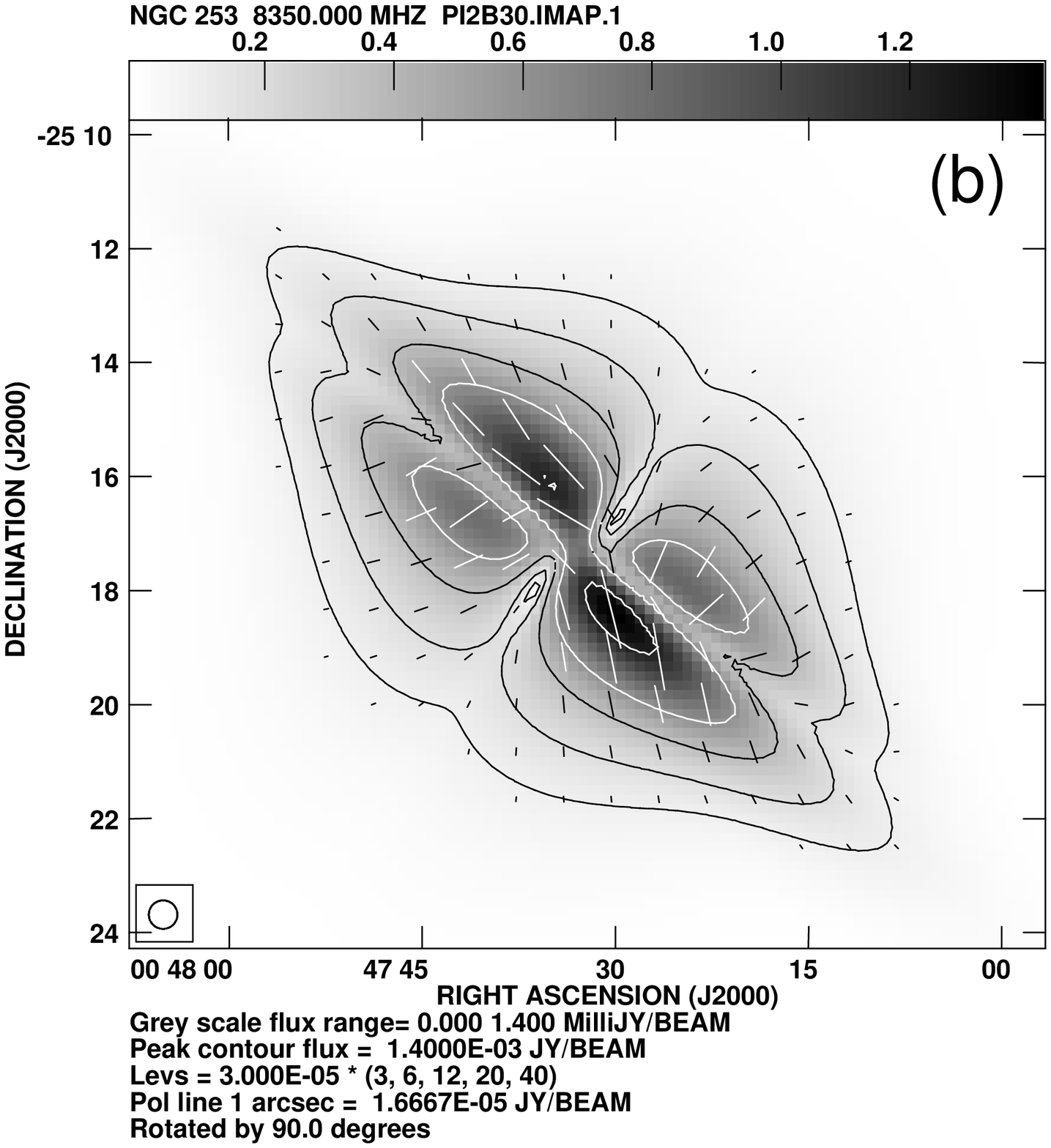}}
  \end{minipage}
  \begin{minipage}[b]{0.25\textwidth}
    \resizebox{\hsize}{!}{\includegraphics{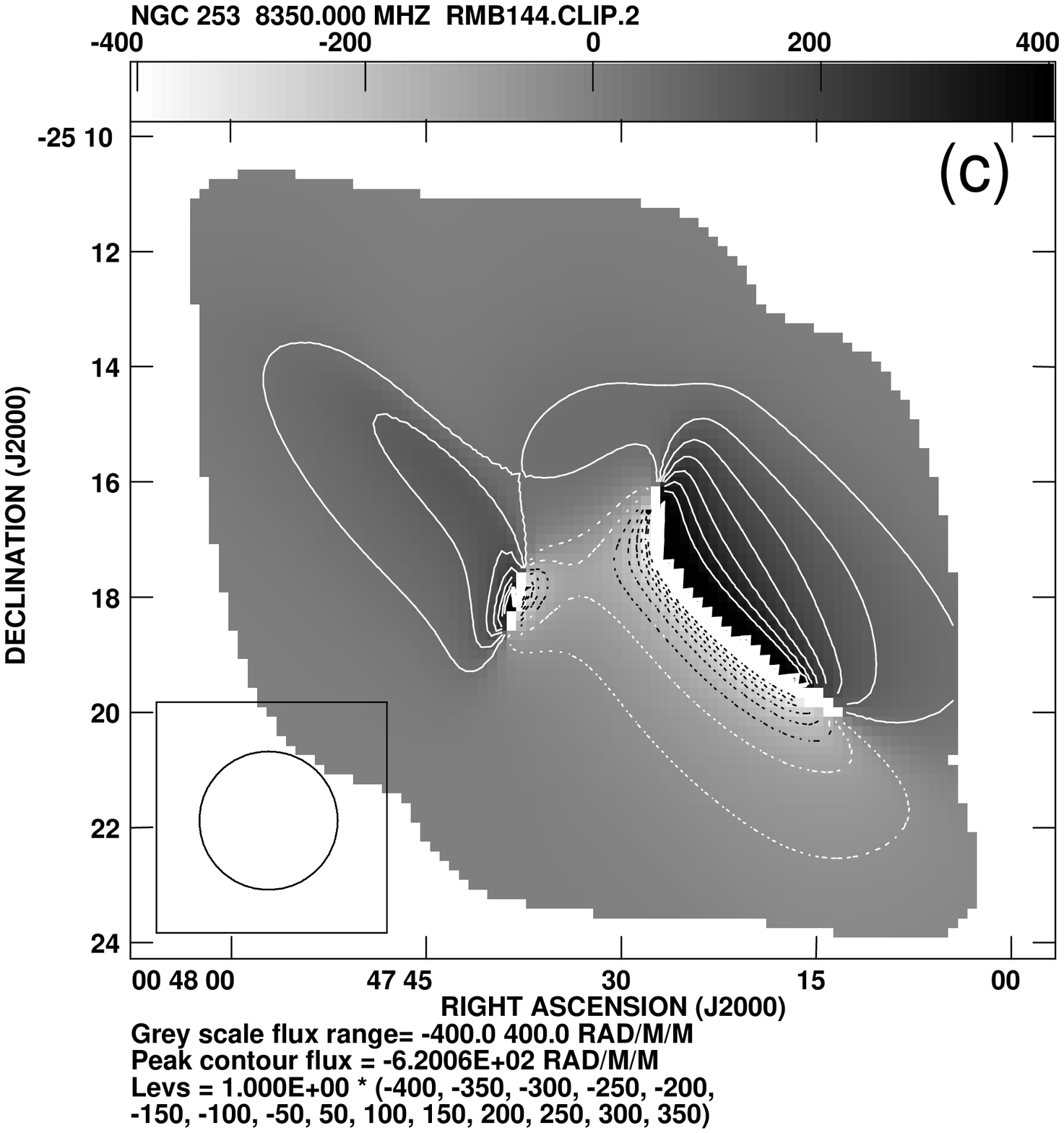}}
  \end{minipage}
  \begin{minipage}[b]{0.24\textwidth}
    \centering\resizebox{0.8\hsize}{!}{\includegraphics{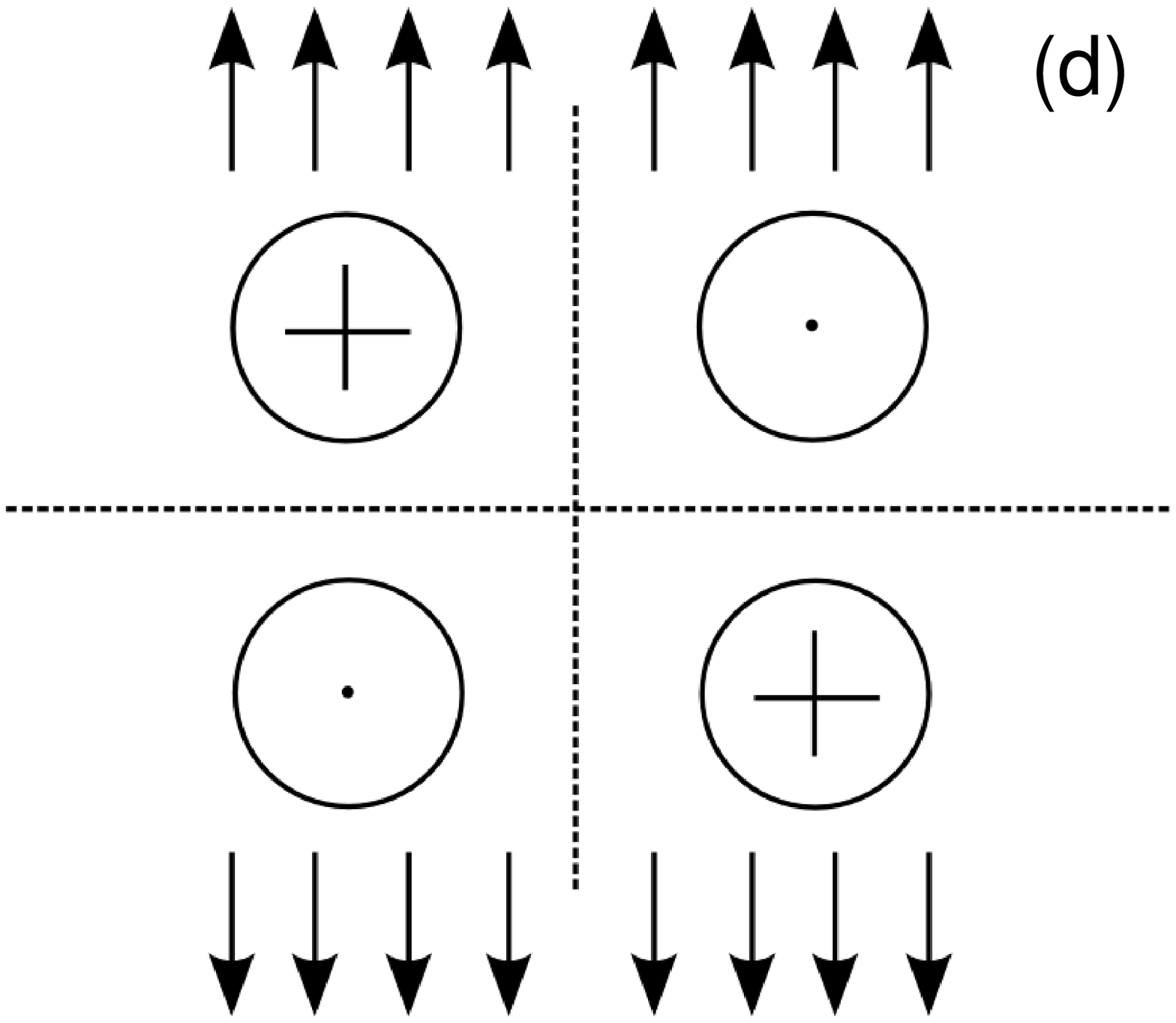}}
    \vspace{1.2cm}
  \end{minipage}
\caption{Odd disk magnetic field and even halo magnetic field. The
  halo field points away from the disk.}
\label{fig:odd_even}
\end{figure*}

\begin{figure*}[tbhp]
  \begin{minipage}[b]{0.25\textwidth}
    \resizebox{\hsize}{!}{\includegraphics{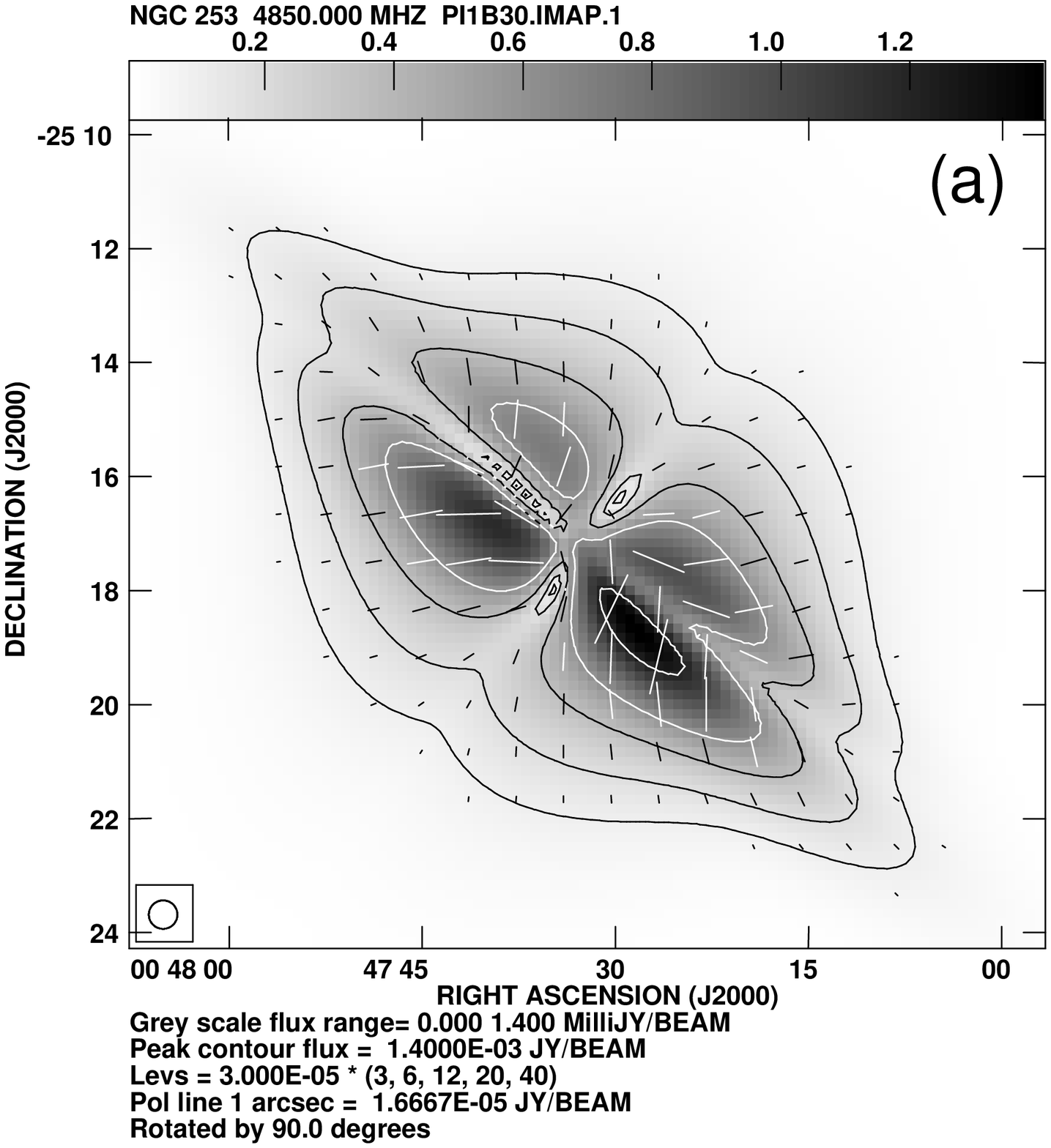}}
  \end{minipage}
  \begin{minipage}[b]{0.25\textwidth}
    \resizebox{\hsize}{!}{\includegraphics{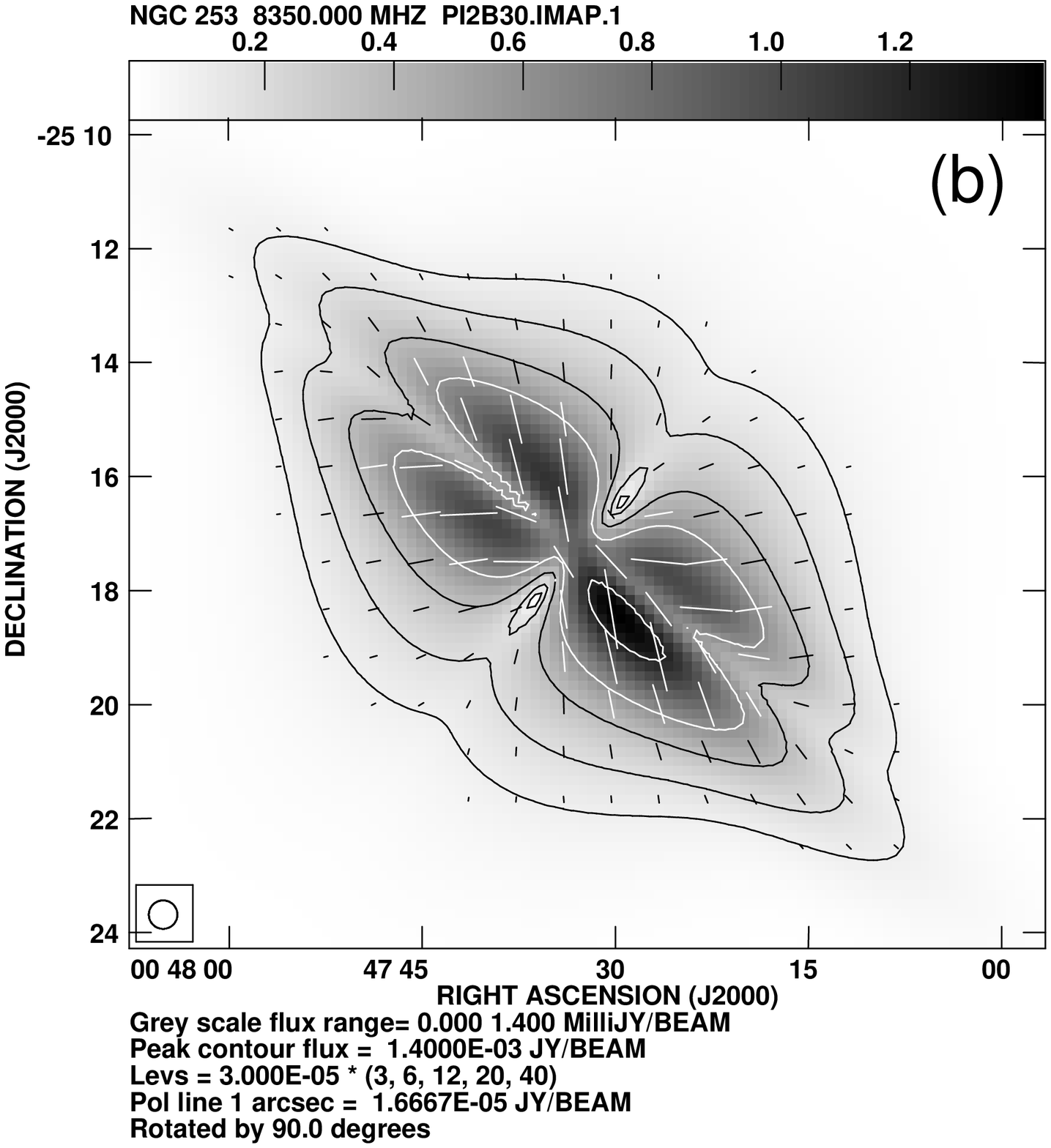}}
  \end{minipage}
  \begin{minipage}[b]{0.25\textwidth}
    \resizebox{\hsize}{!}{\includegraphics{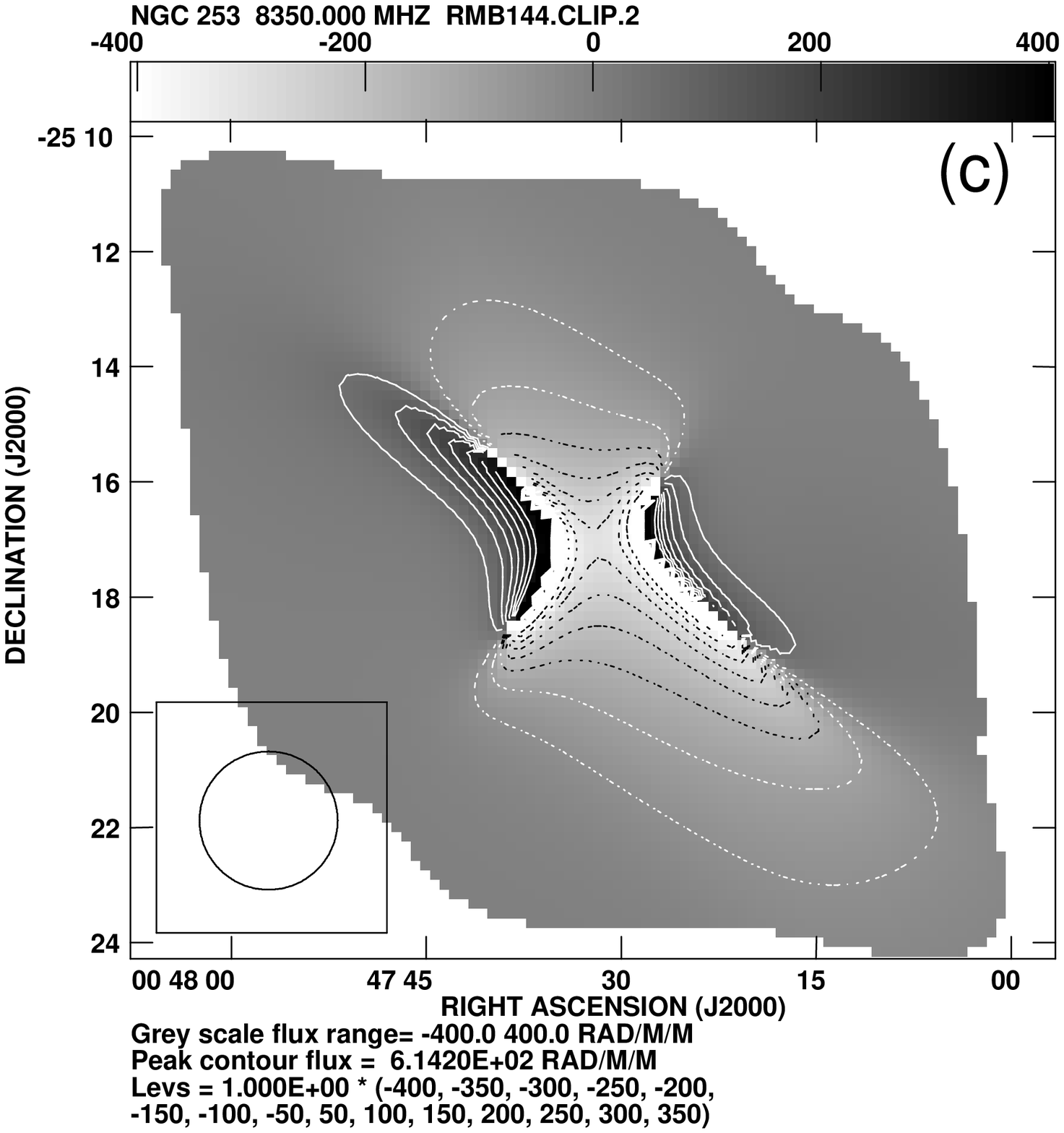}}
  \end{minipage}
  \begin{minipage}[b]{0.24\textwidth}
    \centering\resizebox{0.8\hsize}{!}{\includegraphics{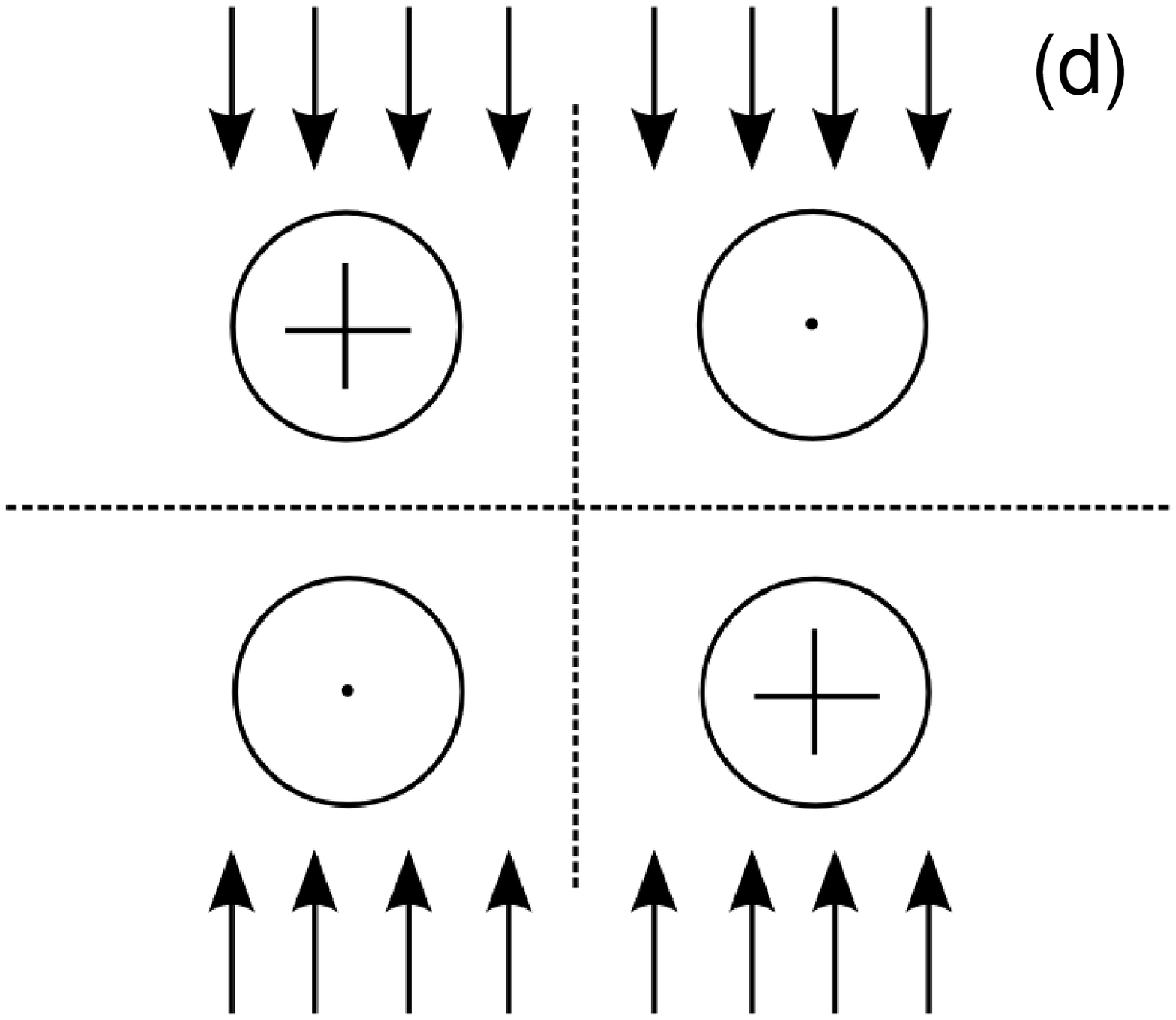}}
    \vspace{1.2cm}
  \end{minipage}
\caption{Odd disk magnetic field and even halo magnetic field. The
  halo field points towards the disk.}
\label{fig:odd_even_n}
\end{figure*}

\begin{figure*}[tbhp]
  \begin{minipage}[b]{0.25\textwidth}
    \resizebox{\hsize}{!}{\includegraphics{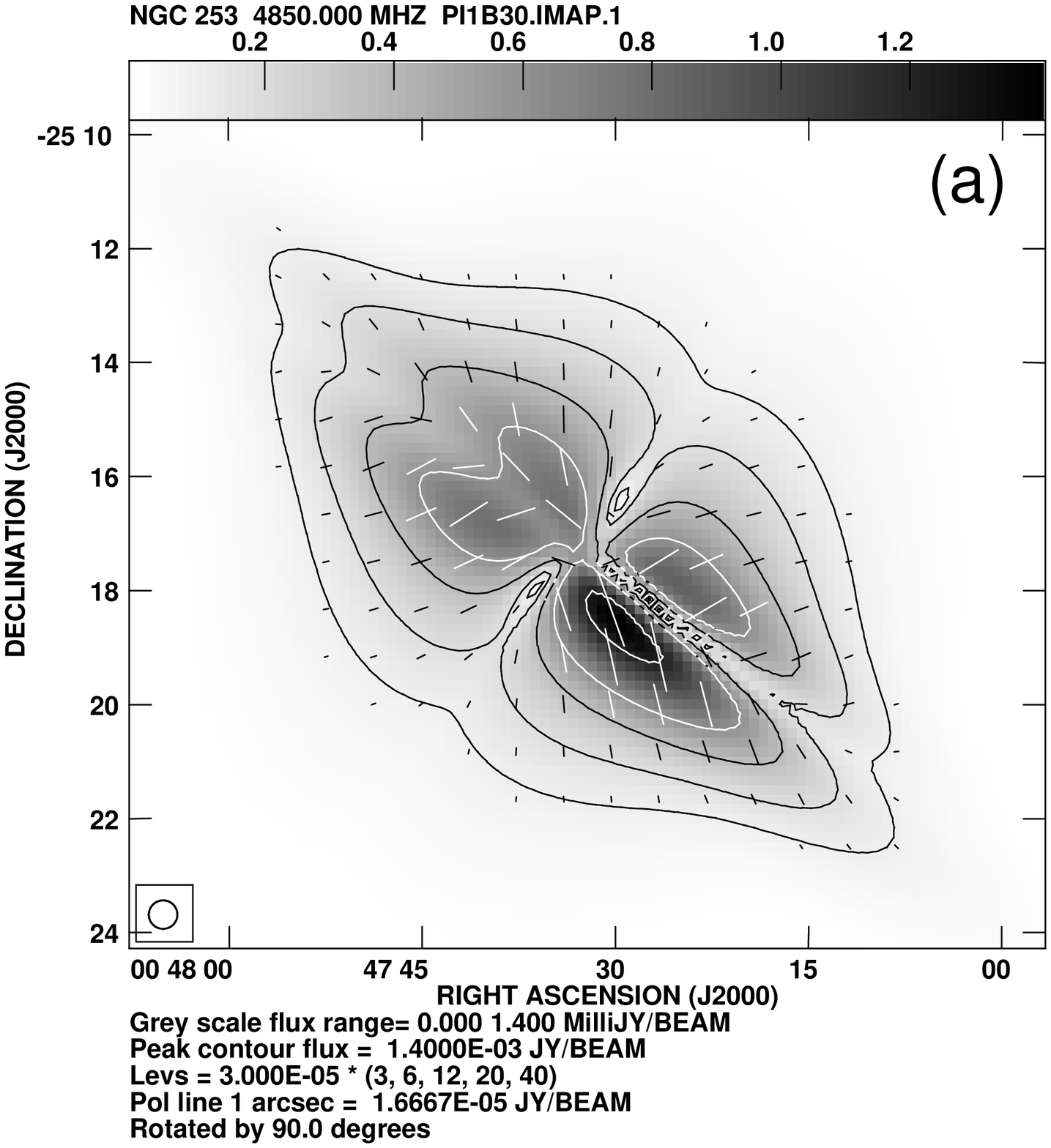}}
  \end{minipage}
  \begin{minipage}[b]{0.25\textwidth}
    \resizebox{\hsize}{!}{\includegraphics{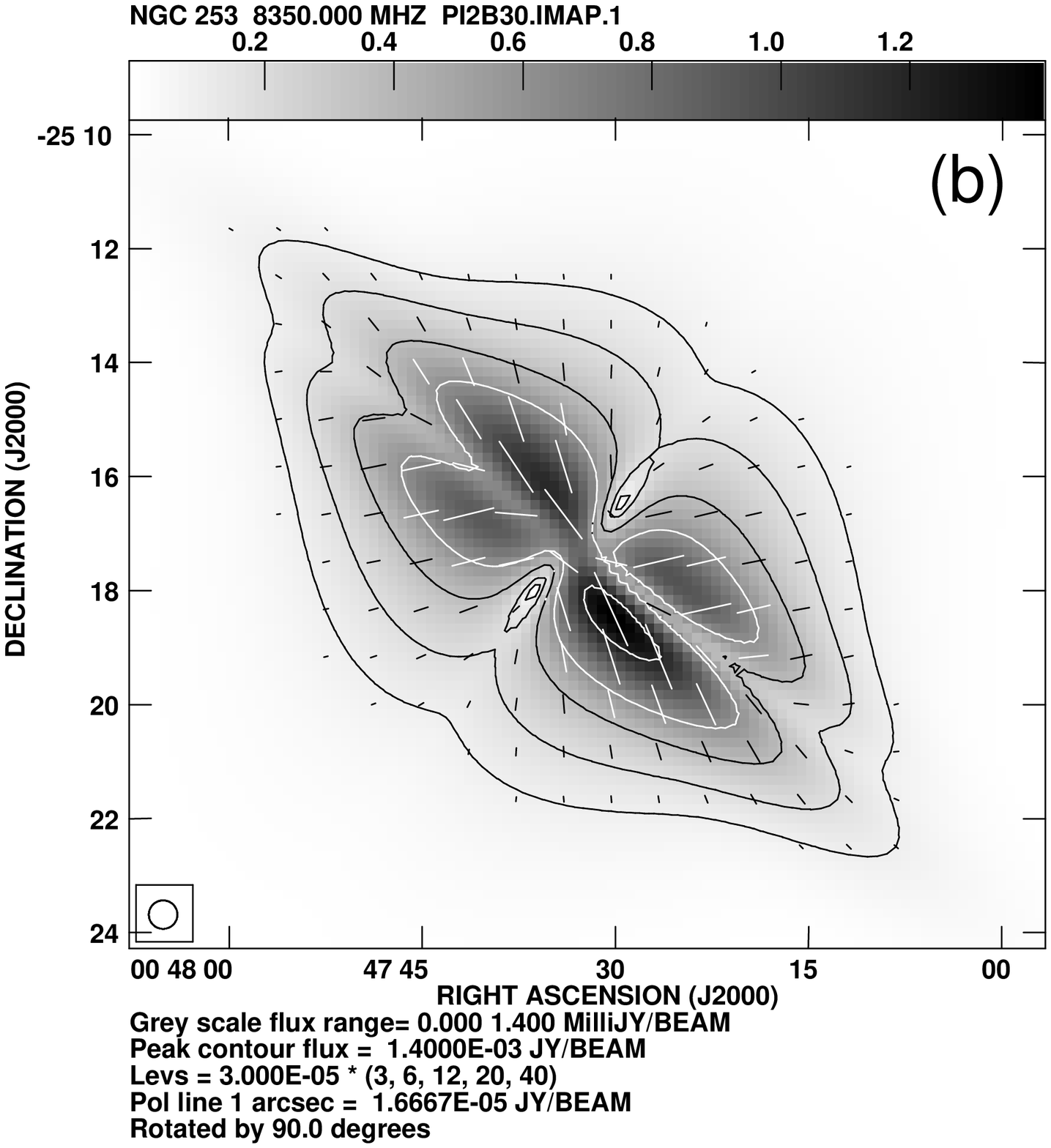}}
  \end{minipage}
  \begin{minipage}[b]{0.25\textwidth}
    \resizebox{\hsize}{!}{\includegraphics{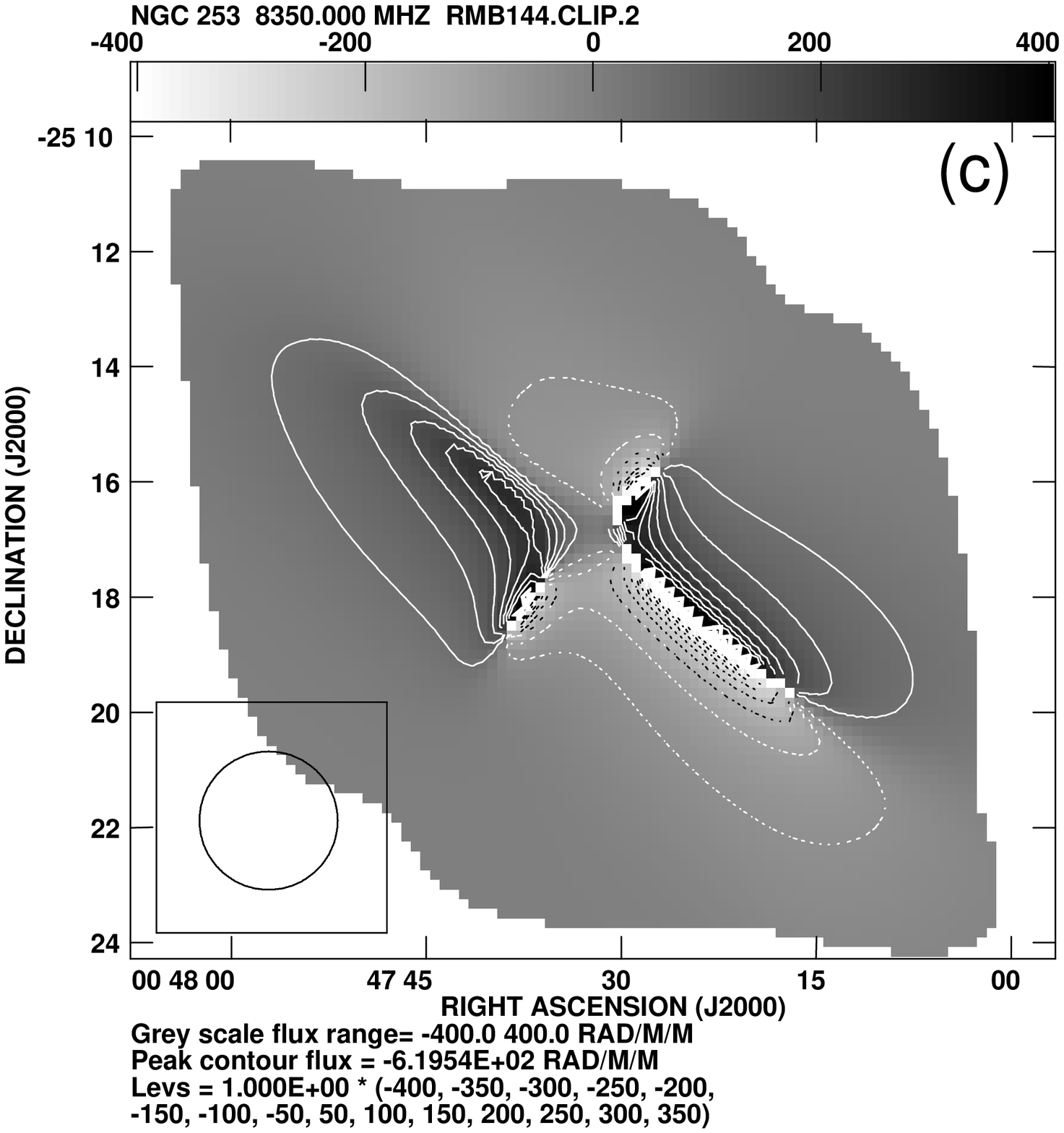}}
  \end{minipage}
  \begin{minipage}[b]{0.24\textwidth}
    \centering\resizebox{0.8\hsize}{!}{\includegraphics{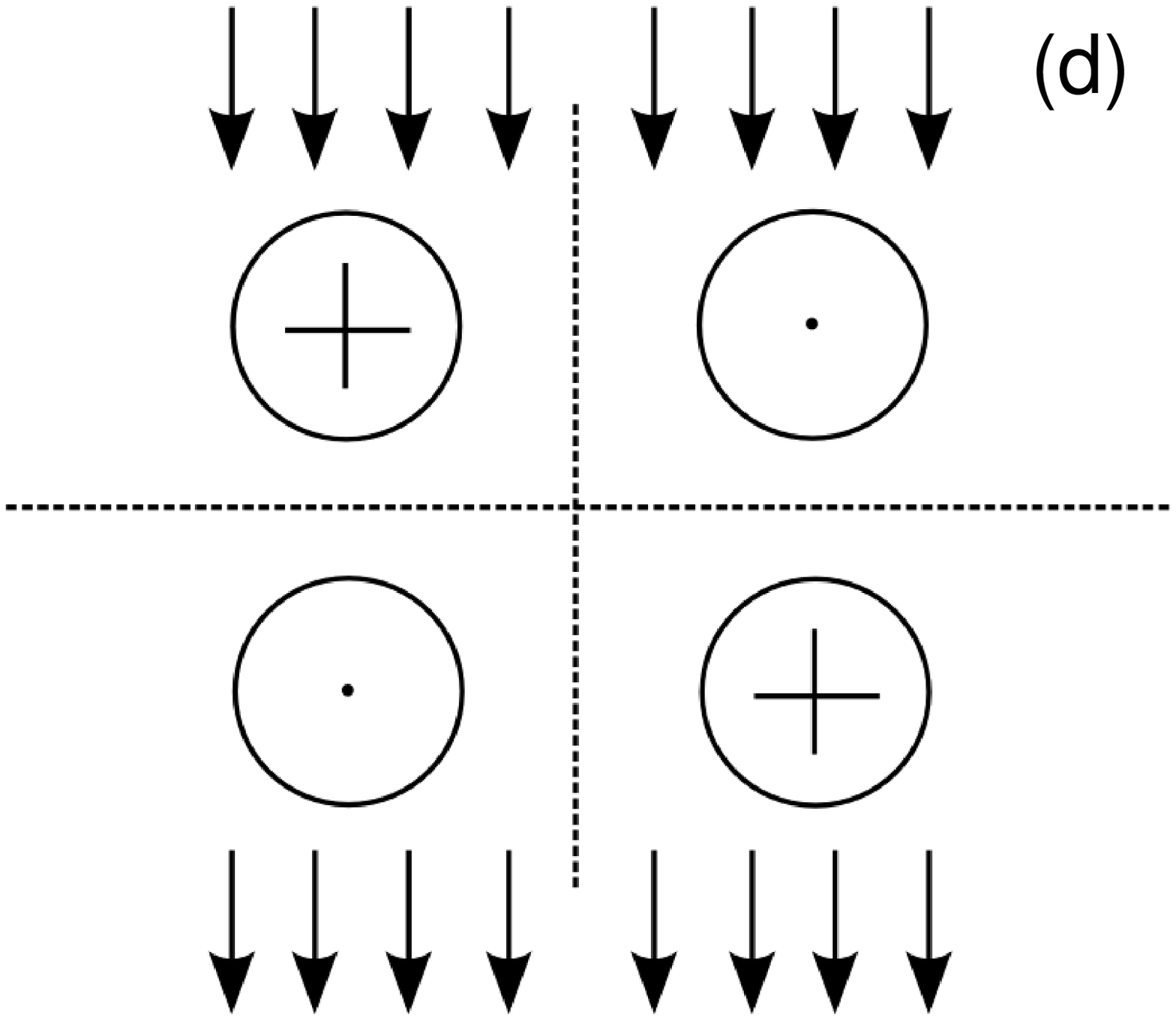}}
    \vspace{1.2cm}
  \end{minipage}
\caption{Odd disk magnetic field and odd halo magnetic field. The halo
field points away from the disk in the southern halo.}
\label{fig:odd_odd}
\end{figure*}

\begin{figure*}[tbhp]
  \begin{minipage}[b]{0.25\textwidth}
    \resizebox{\hsize}{!}{\includegraphics{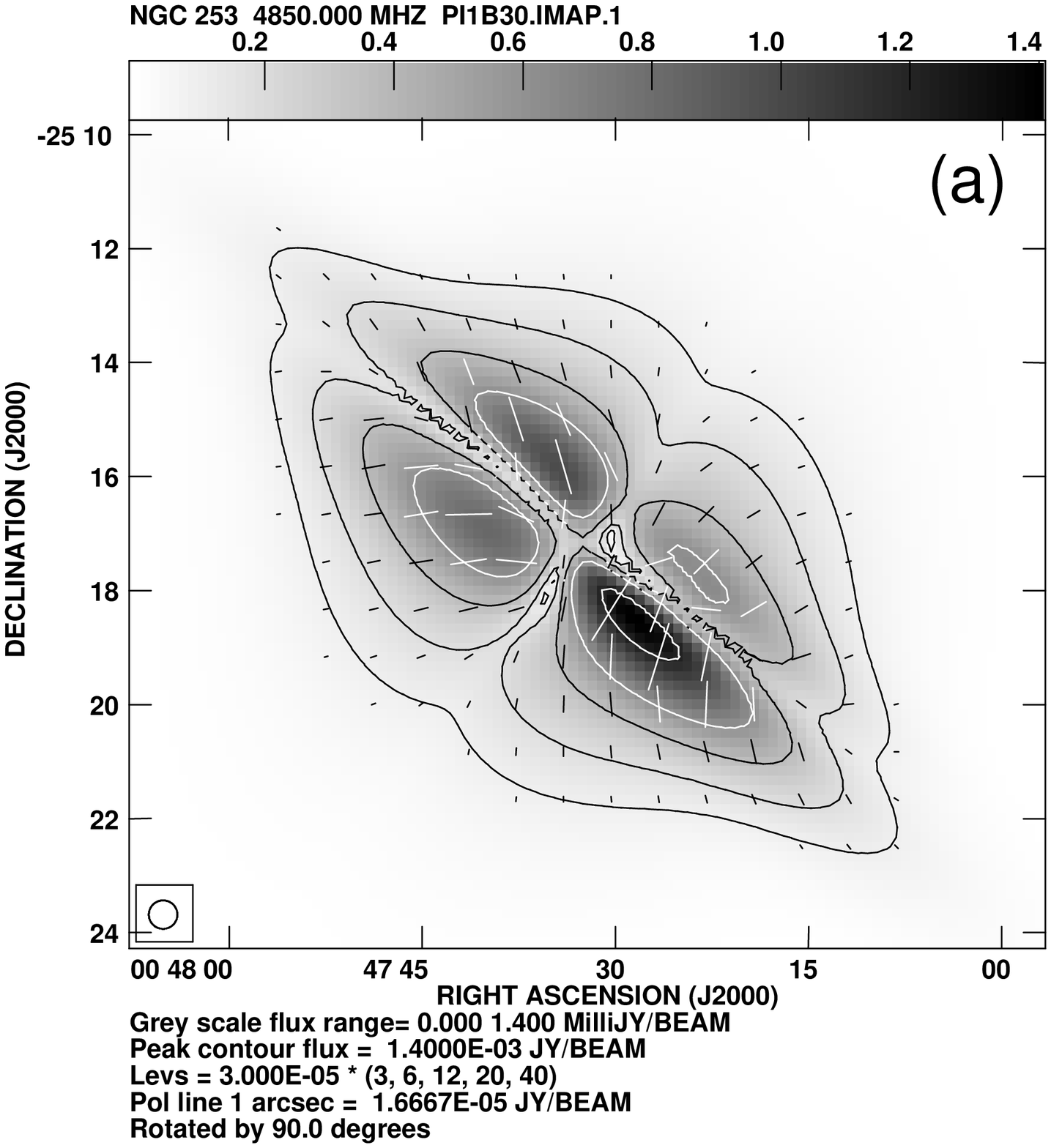}}
  \end{minipage}
  \begin{minipage}[b]{0.25\textwidth}
    \resizebox{\hsize}{!}{\includegraphics{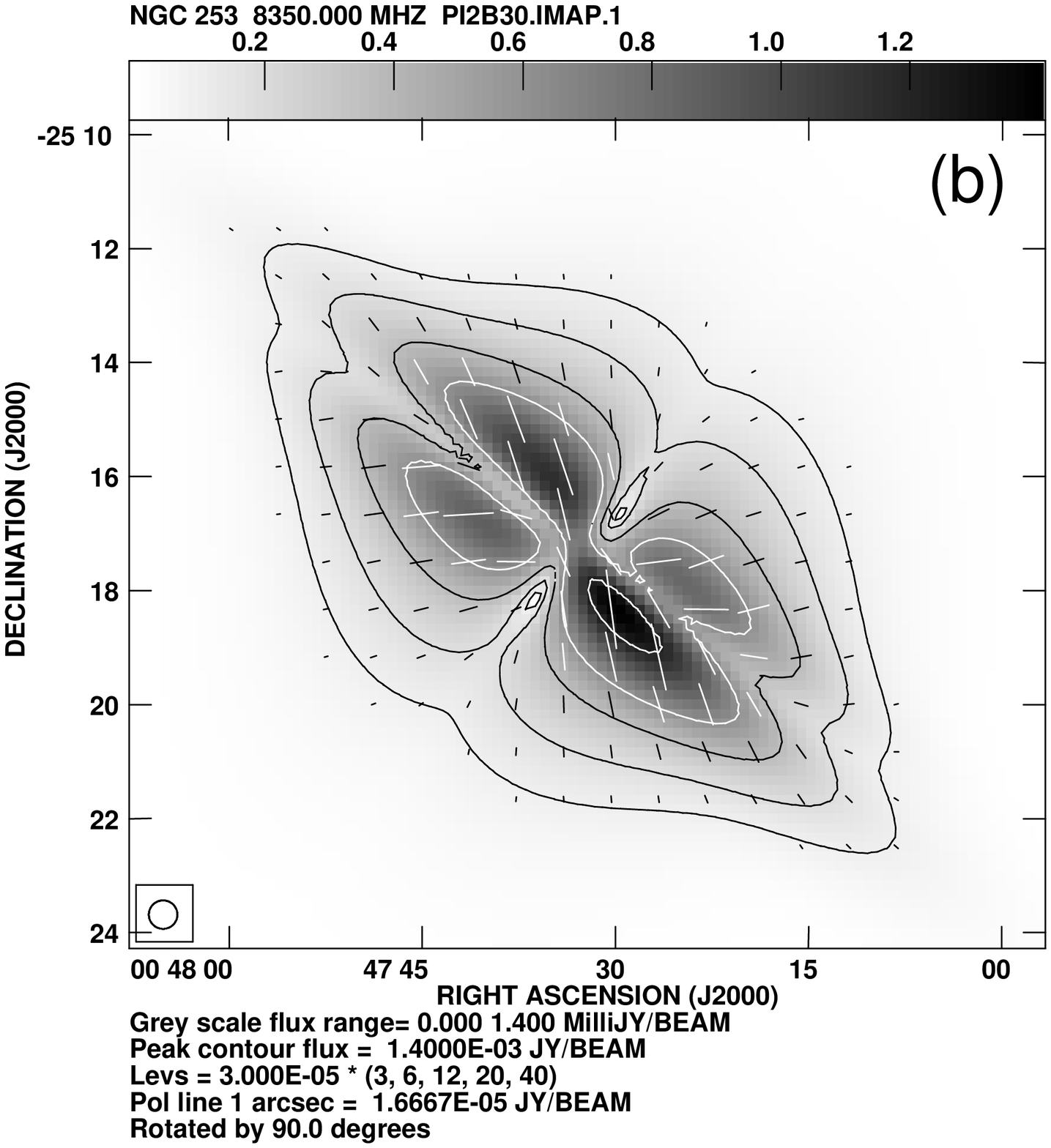}}
  \end{minipage}
  \begin{minipage}[b]{0.25\textwidth}
    \resizebox{\hsize}{!}{\includegraphics{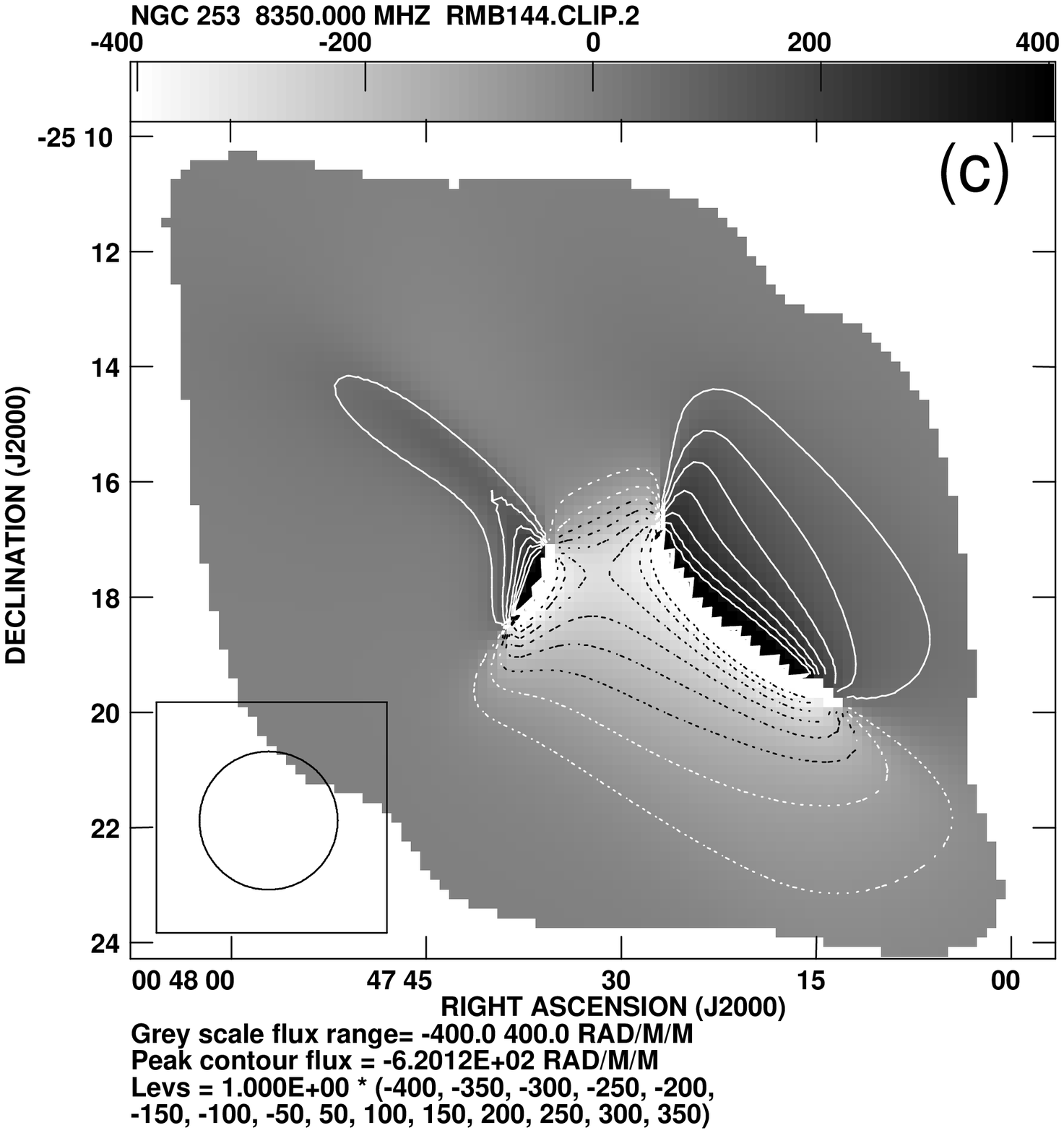}}
  \end{minipage}
  \begin{minipage}[b]{0.24\textwidth}
    \centering\resizebox{0.8\hsize}{!}{\includegraphics{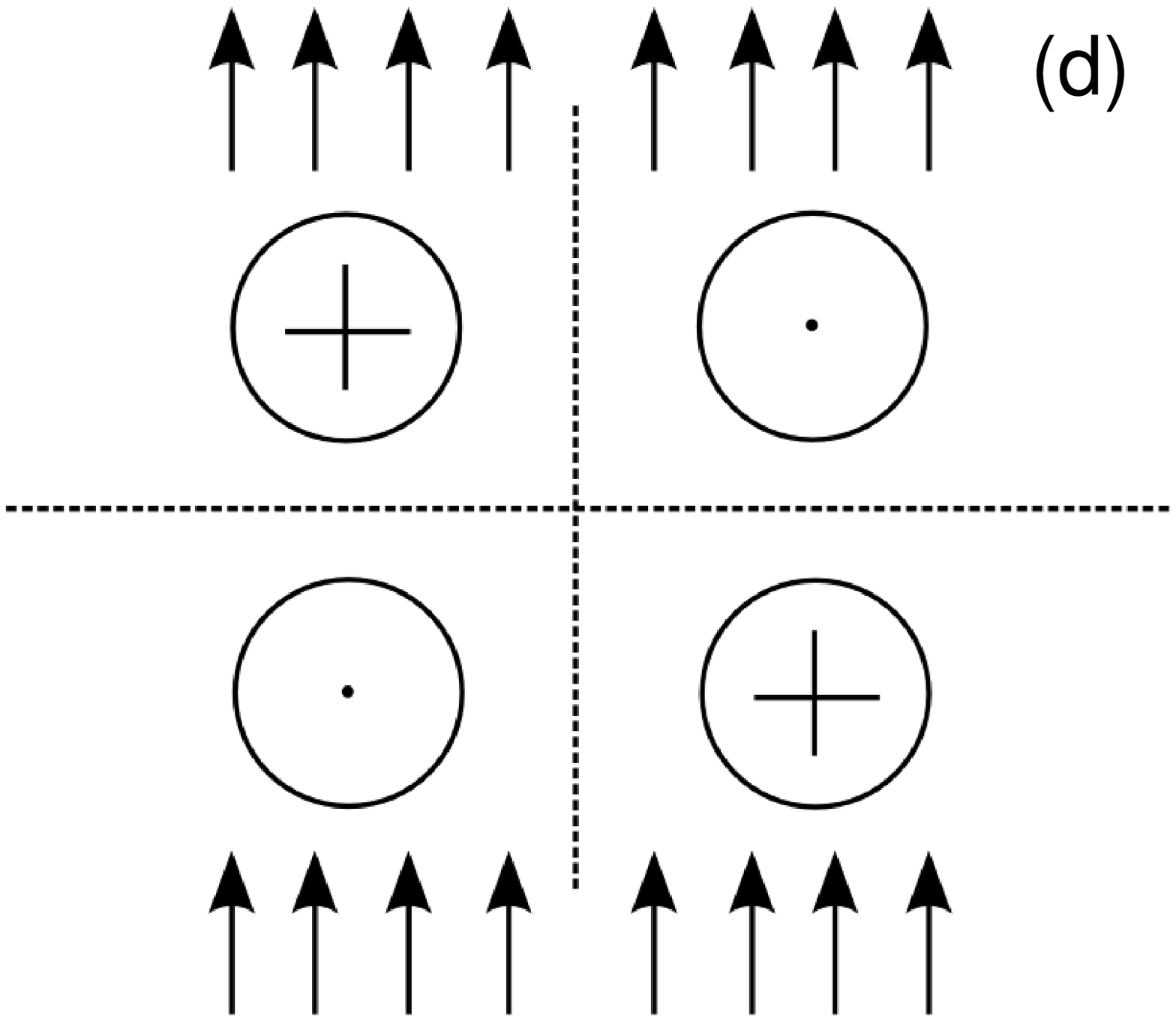}}
    \vspace{1.2cm}
  \end{minipage}
\caption{Odd disk magnetic field and odd halo magnetic field. The halo
field points towards the disk in the southern halo.}
\label{fig:odd_odd_n}
\end{figure*}

In order to find the best model for the magnetic field in NGC\,253 we studied
the expected polarized intensities and RMs of various possibilities of the
disk and halo magnetic field alone and combinations of them. The setup of the
models is described in Sect.\,\ref{subsec:mf_dynamo} with the parameters of
Table~\ref{tab:model_parameters}. In
Figs.\,\ref{fig:disk_even}-\ref{fig:odd_odd_n} we present the polarized
intensity and the magnetic field orientation (not corrected for Faraday
rotation) for $\lambda\lambda$ $6.2\,\rm cm$ (a) and $3.6\,\rm cm$ (b) at
$30\arcsec$ resolution. Note that the Faraday corrected magnetic field
orientation is identical for all models consisting both of the disk and halo
field (Figs.\,\ref{fig:even_even}-\ref{fig:odd_odd_n}). The difference in the
models is only in the \emph{direction} of the magnetic field that becomes
visible in the RM distribution (c) shown with $144\arcsec$
resolution. Sketches (d) show the direction of the magnetic field in the disk
and in the halo. Here, a ``$\cdot$'' indicates that the field points to the
observer and a ``$+$'' denotes a field pointing away.
Figures~\ref{fig:disk_even} and \ref{fig:disk_odd} show the disk magnetic field
alone for the even and odd case. They are similar but the odd field has a
smaller amplitude in the RM distribution. This can be understood as the
scaleheight of the disk is smaller than the projected minor axis. We see
therefore only one side of the disk (the southern one).
Figs.\,\ref{fig:halo_even} and \ref{fig:halo_odd} show the halo magnetic field
alone for the even and odd cases. Note the asymmetry of both the polarized
intensity and the RM. Because the magnetic field orientation changes along the
line-of-sight, there is some depolarization as the magnetic field vectors are
not parallel to each other. If the Faraday rotation has the same sense as the
rotation of the magnetic field, the depolarization is stronger. If they have
opposite senses, the depolarization is weaker. This can explain the asymmetry
although the magnetic field is axisymmetric.
Figures~\ref{fig:even_even}-\ref{fig:even_odd_n} show the models with the even
disk magnetic field. Among them Figs.\,\ref{fig:even_even} and
\ref{fig:even_odd} are the models whose RM distribution agrees best with the
observations. The two models only differ in the direction of the northern halo
field.  The even halo magnetic field (Fig.\,\ref{fig:even_even}) is our
best-fit model. Figs.\,\ref{fig:odd_even}-\ref{fig:odd_odd_n} show the models
for an odd disk magnetic field. They all have strong gradients in the RM which
are not observed.

\end{document}